\documentclass[preprint,12pt,authoryear]{elsarticle}

\usepackage{amssymb}
\usepackage{amsmath} 

\usepackage{placeins}
\usepackage{bm}
\usepackage{color}
\usepackage{graphicx}
\usepackage{braket}

\newcommand{\parenth}[1]{\ensuremath{\left(#1\right)}}

\journal{Physics Reports}

\begin{document}

\begin{frontmatter}



\title{Squeezed states of light \\ and their applications in laser interferometers}

\author{Roman Schnabel}
\address{Institut f\"ur Laserphysik, Zentrum f\"ur Optische Quantentechnologien,\\  Universit\"at Hamburg, 
Luruper Chaussee 149, 22761 Hamburg, Germany}

\begin{abstract} 
According to quantum theory the energy exchange between physical systems is quantized. As a direct consequence, measurement sensitivities are fundamentally limited by quantization noise, or just `quantum noise' in short. Furthermore, Heisenberg's Uncertainty Principle demands measurement back-action for some observables of a system if they are measured repeatedly.
In both respects, squeezed states are of high interest since they show a `squeezed' uncertainty, which can be used to improve the sensitivity of measurement devices beyond the usual quantum noise limits including those impacted by quantum back-action noise. Squeezed states of light can be produced with nonlinear optics, and a large variety of proof-of-principle experiments were performed in past decades.  
As an actual application, squeezed light has now been used for several years  to improve the measurement sensitivity of GEO\,600 -- a laser interferometer built for the detection of gravitational waves. Given this success, squeezed light is likely to significantly contribute to the new field of gravitational-wave astronomy.
This Review revisits the concept of squeezed states and two-mode squeezed states of light, with a focus on experimental observations.  
The distinct properties of squeezed states displayed in quadrature phase-space as well as in the photon number representation are described. 
The role of the light's quantum noise in laser interferometers is summarized and the actual application of squeezed states in these measurement devices is reviewed.\\
\end{abstract}


\end{frontmatter}

\tableofcontents
\newpage

\section{Introduction} \label{Sec:1} 
Laser interferometers are used to monitor small changes in refractive indices, rotations, or surface displacements such as mechanical vibrations. They transfer a differential phase change between two light beams into a changing power of the output light, which is photo-electrically detected, for example by a photo diode. The light is produced in a lasing process that usually aims for a coherent (Glauber) state. In practice, laser light is often in a mixture of coherent states producing excess noise in the interferometric measurement. 
But even if the laser light is in a (pure) coherent state its detection is associated with noise, usually called `shot-noise'. This arises from the quantisation of the electro-magnetic field, which, for a coherent state, results in Poissonian counting statistics of mutually independent photons. 
\begin{figure}[ht]
  \vspace{-2mm}
  \includegraphics[width=11.1cm]{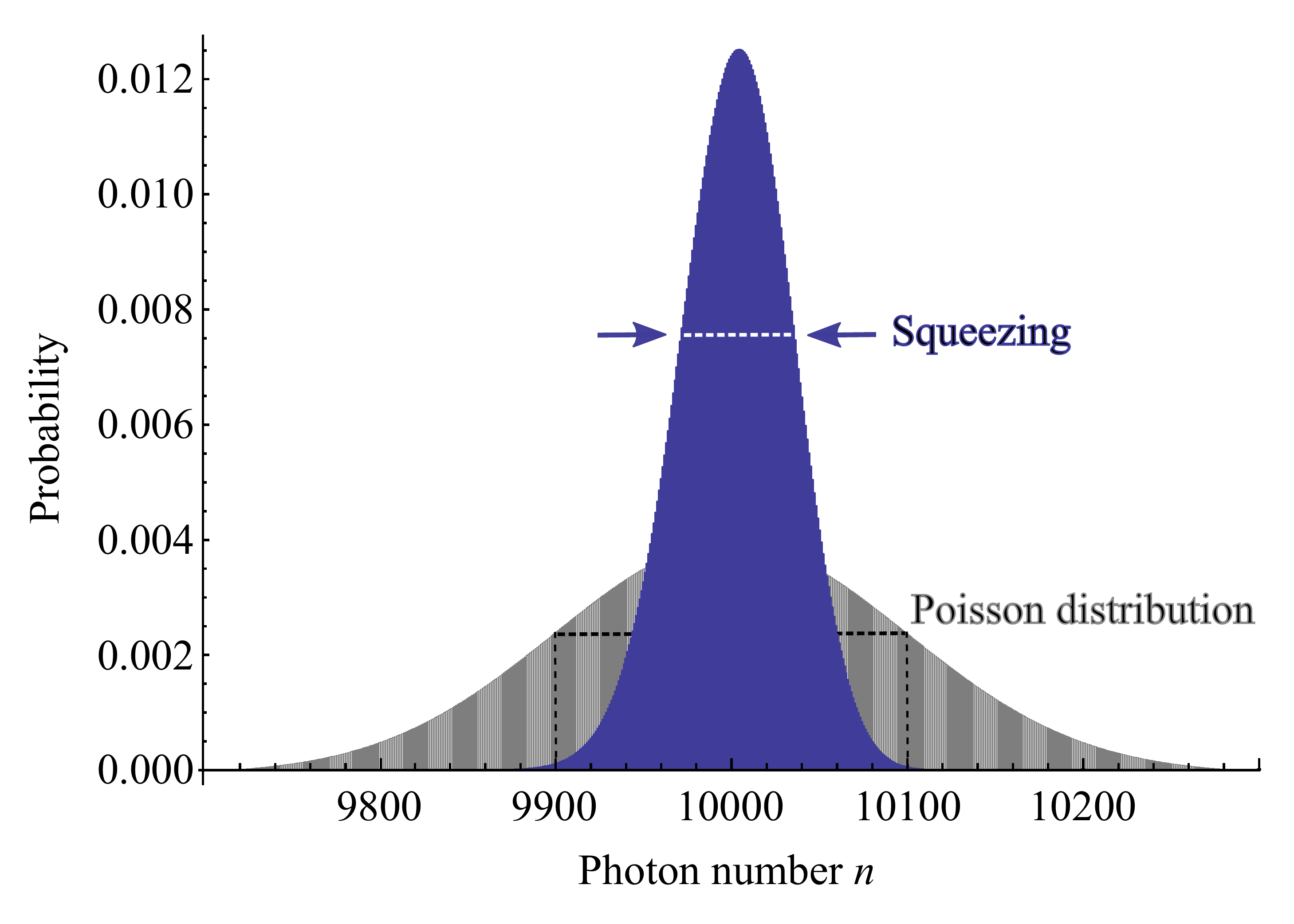}
  \vspace{-2mm}
 \caption{
\textbf{Poissonian and squeezed photon statistics} -- The upper boundary of each area represents the probability distribution of detected photon number $n$,  when performing a large number of measurements on an ensemble of identical states having an average photon number of  $\overline{n} = |\alpha|^2 = 10,\!000$, where $\alpha$ is the coherent field excitation, or `displacement'. The broader curve shows the `Poissonian' distribution, which describes the counting statistic of mutually independent particles, i.e.~those of the coherent state. Due to the large value of $\alpha$, the distribution is almost Gaussian with a standard deviation of $\pm \sqrt{\overline{n}}$. The narrow curve corresponds to the equally displaced 10\,dB squeezed state, which obviously has a `sub-Poissonian' photon statistic. Note that squeezed states with small or even without any coherent excitation (squeezed vacuum states) exhibit quite different photon statistics -- see Fig.\,\ref{fig:13} for example.  
} \vspace{-3mm}
\label{fig:1}
\end{figure}

If the coherent state is highly excited and thus the average number of photons $\overline{n}$ per detection interval is large, the Poissonian distribution can be approximated by a Gaussian distribution with a standard deviation of $\pm \sqrt{\overline{n}}$. 
During the past decades \emph{squeezed states of light} have attracted a lot of attention because they can exhibit less quantum noise than a coherent state of the same coherent excitation, i.e.~they can show sub-Poissonian counting statistic, see Fig.\,\ref{fig:1}.

\begin{figure}[ht]
  \vspace{2mm}
  \includegraphics[width=13.6cm]{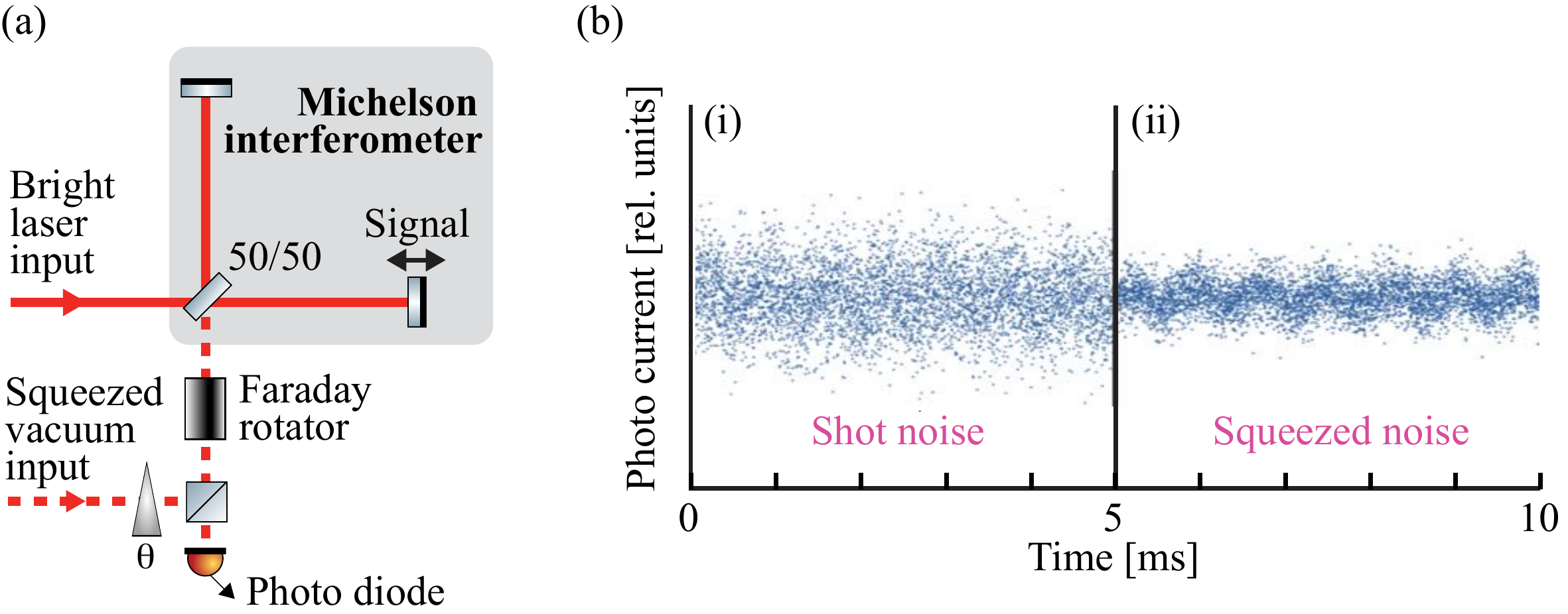}
  \vspace{-1mm}
 \caption{
\textbf{Squeezed-light enhanced Michelson interferometer} -- 
(a) In addition to the conventional operation of a Michelson laser interferometer with bright coherent light, a broadband squeezed-vacuum field is injected into the signal output port and overlapped with the bright interferometer mode. The interferometer is operated close to a dark fringe, such that most of the bright coherent light as well as most of the squeezed vacuum are back-reflected from the Michelson interferometer, respectively. Due to interference with the broadband squeezed vacuum, the interferometer's output light on the photo diode shows reduced variance in the photon number statistic, as shown in Fig.\,\ref{fig:1}. Overlapping the two light fields is possible with theoretically zero loss by the combination of a Faraday rotator and a polarizing beam splitter (PBS). A signal is produced by modulating the relative arm length.  
(b) Simulated data for photo diode measurements. Without squeezing (i), the signal of the laser interferometer is not visible. With squeezing (ii), the shot noise is reduced and, here, a sinusoidal signal visible.  
}
\label{fig:2}
\end{figure}

Squeezed states belong to the class of `non-classical' states, which are considered to be at the heart of quantum mechanics. 
These states are defined as those that cannot be described as a mixture of coherent states. In this case, their Glauber-Sudarshan $P$-functions [\cite{Sudarshan1963,Glauber1963}] do not correspond to (classical) probability density functions, i.e.~they are not positive-valued functions. As a `classical' example, the $P$-function of a coherent state corresponds to a $\delta$-function. 

But the question remains what property of coherent states justifies the name `classical', even though coherent states are quantum states and show quantum uncertainties. My answer to this question is the following. All experiments which only involve coherent states and mixtures of them allow for a description that uses a combination of classical pictures. As we will see below, this description swaps between two different classical pictures and is thus not truly classical but semi-classical. (A more precise description of the nature of coherent states uses the term `semi-classical'.)

Let us consider a laser interferometer that uses light in a coherent state. Firstly, the light beam is split in two halves by a beam splitter. The two beams travel along different paths and are subsequently overlapped on a beam splitter \emph{where they interfere exactly as classical waves would do}. The electric fields superimpose, thereby producing the phenomenon of interference. Up to this point there is no reason to argue light might be composed of particles. \\
Secondly, the new (still coherent) beams that result from the interference are absorbed, for instance by a photo-electric detector. \emph{In the case of coherent states the detection process can be perfectly described in the classical particle picture in which the particles appear independently from each other in a truly random fashion}, yielding the aforementioned Poisson statistic. During the detection process, no wave feature of the light is present. Let us have a closer look:  A truly random (`spontaneous') event is an event that has not been triggered by anything in the past. This allows us to make a clear cut between the first part of the experiment, described by the classical wave picture, and the second part of the experiment, described by the classical particle picture. Both `worlds'  are disconnected. The subsequent application of two classical pictures is not truly classical, but `semi-classical'. It is indeed the observation that the photons occur individually with truly random statistics that allows this semi-classical description. In the case of a mixture of coherent states the photon statistics are super-Poissonian, which can be understood as a mixture of different Poissonian distributions. In the case of a slowly changing coherent state the mean value $\overline{n}$ depends on time. In all these cases, the semi-classical description is appropriate. Let me point out that in this very reasonable description photons do not exist before they are detected, e.g.~absorbed.
Further note, that the famous double-slit experiment with coherent states also allows for the same semi-classical description.\\

For squeezed states [\cite{Yuen1976,Walls1983}] the situation is different. As before, the interference can be fully described by the classical wave picture. The result of the detection process, however, is different from that of mutually independent random events. It is also different from any super-Poissonian statistics that could be produced by mixing an arbitrary number of different and/or time-dependent Poissonian distributions. 
Instead, the squeezed probability distribution in Fig.\,\ref{fig:1} suggests that the probability of detecting a photon decreases with the more photons that are already detected in the same time interval over which a single measurement is integrated.  
From this observation, one must conclude that the photons do not individually appear in a random fashion upon detection. 
There must be `quantum' correlations between the photons. These correlations must existed \emph{before} detection, since there is no interaction between the photons during their detection. Pre-existing correlations between detected photons seem to imply that the photons themselves existed before detection, i.e.~at times when interference occurred. 
In a semi-classical description, however, photons are classical particles and cannot interfere, for instance on a beam splitter. At this point, the semi-classical picture breaks down. Squeezed states are therefor `nonclassical'.\\
The failure of the semi-classical model described above generally certifies nonclassicality.\\   

Squeezed states are usually not characterized by counting their photons, but by measuring canonical continuous-variable phase-space observables. 
Measurements are performed, as usual, on an ensemble of identical states, and quasi-probability density functions are calculated from the data. The Glauber-Sudarshan $P$-function is the quasi-probability density distribution over coherent states. If the $P$-function of a state is entirely positive, the state is a coherent state or a (classical) mixture of coherent states. The state is considered as semi-classical. If the $P$-function is not a positive-valued function, the state cannot be expressed as a (classical) mixture of coherent states and is thus nonclassical [\cite{Gerry2005,Vogel2006}]. A non-positive-valued $P$-function is the sufficient and necessary condition for the failure of the semi-classical model. 
The Wigner function is the quasi-probability phase-space representation over the canonical continuous-variable phase-space observables themselves  [\cite{Gerry2005}]. 
The Wigner functions of squeezed states are entirely positive. Although subject to discussion, this fact does not mean that squeezed states are less nonclassical than Fock states or cat states, which not only have a nonclassical $P$-function but also a partially negative Wigner function. 
(A cat state is a quantum superposition of two macroscopically distinct states [\cite{Monroe2002}], referring to Schr\"odinger's-cat gedanken experiment [\cite{Schroedinger1935}]). 
In practice, squeezed states can even be regarded as superior nonclassical states because they represent the only nonclassical state that has been produced in a steady state fashion.\\ 
In almost all experiments so far, the generation of Fock states and cat states involves a probabilistic event, such as the detection of a photon in another beam path, to herald these states. 
In fact, squeezed states provide the nonclassical resource for the probabilistic preparation of Fock states as well as cat states. But only the squeezed states themselves show a nonclassical effect in a stationary way: 
Limited only by the time duration and the frequency span of the mode that is in a squeezed state, the squeezing effect can be continuously observed independently of the time when the measurement is performed, and also independently of the measurement integration time.  
This fact is of great importance for applications of squeezed states in measurement devices since a squeezed-light-enhanced measurement remains unconditional and the effective measurement time is not reduced. \\

In past decades, squeezed states of light were used in many proof-of-principle experiments to research their potential for improving the sensitivity  of laser interferometers [\cite{Grangier1987,Xiao1987,McKenzie2002,Vahlbruch2005,Goda2008,Taylor2013}] or the performance of imaging beyond the shot-noise limit [\cite{Lugiato2002,Treps2003}], both accompanied by a huge number of theoretical works. Potential applications in secure optical communication (quantum key distribution) were also proposed, and proof-of-principle experiments demonstrated [\cite{Ralph1999,Furrer2012,Gehring2015}]. 
This review restricts itself to the improvement of laser interferometers, since only here has the application of squeezed light gone beyond proof-of-principle. The gravitational-wave detector (GWD) GEO\,600 has operated with squeezed light now for more than seven years, starting in 2010 [\cite{LSC2011,Grote2013}]. 
GEO\,600 is a 600\,m long Michelson laser interferometer built for the detection of gravitational waves. These waves are audio-band and sub-audio-band changes of space-time curvature originating from cosmic events such as the merger of neutron stars or black holes, as detected recently [\cite{Abbott2016}]. In GWDs such as GEO\,600 [\cite{Dooley2016}], Advanced LIGO [\cite{Aasi2015}], Advanced Virgo [\cite{Acernese2015}], and KAGRA [\cite{Aso2013}], conventional laser technology has been pushed to extremes over the past decades. Noise spectral densities normalized to space-time strain of less than $10^{-23}$\,Hz$^{-1/2}$ have been measured [\cite{Abbott2016}]. 
Progress will continue and, based on the successful application in GEO\,600, squeezed light is now widely accepted to provide a new additional technology to contribute to the new field of gravitational-wave astronomy. It was also successfully tested in one of the LIGO detectors in 2013 [\cite{LSC2013}] and is an integral part of the European design study for the 10\,km Einstein-Telescope [\cite{Punturo2010}].

GEO\,600 has already taken several years of `squeezed' observational data, which has increased its sensitivity at signal frequencies above 500\,Hz. With the implementation of a squeezed light source in GEO\,600, the application of nonclassical states in metrology has been pushed beyond merely proof-of-principle.\\ 

`Two-mode squeezed states' show a squeezed uncertainty in at least one \emph{joint} continuous variable of two subsystems `A' and `B'. Examples of joint variables are differences and sums of phase-space observables of A and B. Two-mode squeezed states not only belong to the class of nonclassical states but, due to their bi-partite character, also to the class of `inseparable' or `entangled' states. They are the ideal states to demonstrate the Einstein-Podolsky-Rosen paradox [\cite{Einstein1935}], as first achieved in [\cite{Ou1992}]. Apart from fundamental research on quantum mechanics, recent proof-of-principle experiments demonstrated their usefulness in interferometric measurements that go beyond the application of simple squeezed states [\cite{Steinlechner2013b, AstM2016}]. This experiment is the final topic of this review.\\ 

\section{Observations on light fields in squeezed states} \label{Sec:2} 
Generally there are two different kinds of observables that can be subject of a measurement performed on a quantum system. The first kind is associated with the system's wave property. In optics, it corresponds to the electric field strength at a given phase angle $\vartheta$. The according (dimensionless) operators are called the \emph{quadrature amplitudes} $\hat X^\vartheta$ and have a continuous spectrum of eigenvalues. Quadrature amplitudes are measured in very good approximation with a balanced homodyne detector using the interference with a bright local oscillator beam, see Fig.\,\ref{fig:3}\,(a). In practice, any measurement of $\hat X^\vartheta$ integrates over some sideband (Fourier) spectrum within the angular frequencies $\Omega \pm \Delta \Omega/2$. The sideband information always needs to be quoted. A straight forward but rather untypical way is by adding subscripts, which leads to $\hat X^\vartheta_{\Omega,\Delta \Omega}$. \emph{The classical analogue of the quadrature amplitude operator $\hat X^\vartheta_{\Omega,\Delta \Omega}$ is the modulation depth of the optical field  at modulation phase angle $\vartheta$ and at angular modulation frequency $\Omega$ measured over the band $\Delta \Omega < 2 \Omega$}. The uncertainties of the state's quadrature amplitudes at different phases $\vartheta$ are limited by a Heisenberg uncertainty relation, see section \ref{Sec:3}. The second kind of measurement is associated with the system's particle property and is given by the \emph{photon number} operator $\hat n$ associated with a measuring time interval $\Delta t$. Its precise measurement requires a photon counter, ideally with single photon resolution. The measurement result obviously has a discrete spectrum. Continuous as well as discrete observables are usually subject to quantum uncertainties and thus quantum noise.\\
Usually, the measurement's integration time and frequency band actually define the physical system that is characterized. In quantum optics experiments, the interrogated physical system is called a `mode'.

\subsection{Definition of a `single mode'} \label{Ssec:2.1} 
Let us define a light field, or generally any quantum system, to be a \emph{single} mode if it corresponds to the `smallest entity of a wave'. In this case its spectral and temporal distributions as well as waist size and divergence are at their Fourier limits and all other properties such as optical axis, waist position and polarization are well defined. 
For instance, a linearly polarized  longitudinal resonance of an optical standing-wave cavity defines such a single mode if the cavity finesse is high and transversal modes are non-degenerate. The complete photo-electrical detection of a cavity mode, however, is not straight forward. Most quantum optical experiments are instead performed on \emph{propagating} light. In this case single modes are defined by spatial filters and by temporal-spectral measurement windows, both being at the Fourier limit. Examples for single modes are a laser pulse and a spectral/temporal cutout from a continuous observation of a quasi-monochromatic continuous-wave light beam in the spatial TEM$_{00}$ mode, both at the Fourier limits.

In classical physics the only remaining free parameter of a given single mode is its excitation energy. In quantum physics the situation is different.  For a given energy a single mode can be in many different quantum states, which differ in their quantum statistics. 
Examples are coherent states, number (Fock) states, and squeezed states.

\subsection{Observations on squeezed states using a single PIN photo-diode} \label{Ssec:2.2}

An ideal PIN photo-diode absorbs the full energy of a light mode and produces one photo electron for every absorbed photon energy. 
It uses the internal photo-electric effect inside a semiconductor such as silicon or InGaAs. In contrast to avalanche photo-diodes, PIN photo-diodes operate with unity gain. `PIN' stands for `positive', `intrinsic' and `negative' and is describing the doping of the semiconductor layers.
A PIN photo-diode is optimally suited for the continuous monitoring of a rather bright light field of up to several tens of milliwatts. 
An example is the photo-diode in the output port of a gravitational-wave detector, as shown in Fig.\,\ref{fig:2}\,(a). 
The prominent wavelength of 1064\,nm, which is emitted by Nd:YAG lasers, has an optical frequency of $\nu = \omega / (2 \pi) = 2.82 \cdot 10^{14}$\,Hz. 
The period of the field oscillation is a few femtoseconds and cannot be directly resolved with photo-electric detectors. However, \emph{variations} of the electric field around the averaged optical field oscillation on longer time-scales can be resolved. 
Applying an electronic bandpass filter at the sideband angular frequency $\Omega \pm \Delta \Omega/2$ to the photo voltage provides information about the `depth of the light's amplitude modulation', which is also called the `amplitude of the amplitude quadrature'. It can also slowly vary in time and reads 
\begin{equation}
\hat X^{\vartheta=0^\circ}_{\Omega,\Delta \Omega} (t) \equiv \hat X_{\Omega,\Delta \Omega} (t) \equiv \hat X \, .
\label{eq:X}
\end{equation}
The subscript is usually skipped, as it is done with the time dependence, as indicated on the right. 
Applying the electronic bandpass filter in fact defines the \emph{mode} of the light being detected. The structure of the definition in Eq.\,(\ref{eq:X}) forms the basis of interferometric signals and quantum noise, also in the semi-classical case of coherent states. 
Lets take an example. In the recent observation of gravitational waves [Fig.~1, bottom row, in \cite{Abbott2016}], the time-frequency representation of the gravitational-wave signal corresponded to the amplitude quadrature amplitude $\hat X_{\Omega,\Delta \Omega} (t)$ of the interferometer output light. 
Note that a larger value of $\Delta \Omega$ allows for changes of the quadrature amplitude on shorter time scales.\\

If the light field's `modulation mode' does not contain any quanta, simply because there are no photons that have a frequency difference of  $\pm \Omega$ with respect to the carrier, it is in its ground state. 
In this case `vacuum noise' is observed, which originates from the ground state uncertainty. Since the vacuum noise only becomes measurable as a beat with a bright light field, it can also be seen as the carrier's band-path filtered shot noise. 
A modulation mode in a displaced vacuum state (a coherent state) corresponds to nonzero coherent modulation.\\ 
The measured level of the vacuum noise generally depends on the power of the bright carrier light and on the electronic amplification. In any case, it provides the reference for certifying `squeezing'. Observations using a single PIN photo-diode require an independent measurement to quantify vacuum noise. A necessary condition is that attenuating the total field's light power results in the same attenuation of the measured $X_{\Omega,\Delta \Omega}$ values. If they show a stronger attenuation, a coherent modulation or thermal noise might be present. If they show a weaker attenuation the photo-diode and its electronics might be saturated.\\

Fig.\,\ref{fig:2}\,(b) illustrates how a broadband squeezed field improves the measurement of an amplitude modulation in time domain, based on a PIN photo-diode. Shown is a simulated time sequence of $\hat X_{\Omega, \Delta\Omega}$-data sampled from the photoelectric voltage. In this simulation all sideband frequencies from zero (DC) to the cutoff frequency of the detector electronics ($\Omega_{\rm cut}$) are included ($\Omega=\Delta\Omega / 2 = \Omega_{\rm cut}/2$). No additional band pass filter is applied making it a maximally broadband detection.
Although the data in Fig.\,\ref{fig:2}\,(b,i) contains a classical amplitude modulation of the detected light, this signal is not visible due to random noise, here representing shot noise.  Fig.\,\ref{fig:2}\,(b,ii) shows the same situation but with shot noise that is squeezed over the full detection band. The quantum uncertainty of the modulation depth is squeezed and the classical signal becomes visible. 
 
It needs to be noted that a single PIN photo-diode can only measure the amplitude of the amplitude quadrature $\hat X_{\Omega,\Delta \Omega}(t)$, but not the non-commuting observable, the `amplitude of the phase quadrature' 
\begin{equation}
\hat X^{\vartheta=90^\circ}_{\Omega,\Delta \Omega} (t) \equiv \hat Y_{\Omega,\Delta \Omega} (t) \equiv \hat Y \, .
\label{eq:Y}
\end{equation}
For values that are small compared to the field strength of the bright field, the quantity $Y$ approximately describes the bright field's `phase modulation depth'. \\

\subsection{Observations  on squeezed states using a balanced homodyne detector} \label{Ssec:2.3}  

\begin{figure}[ht]
  \vspace{3mm}
  \includegraphics[width=13.6cm]{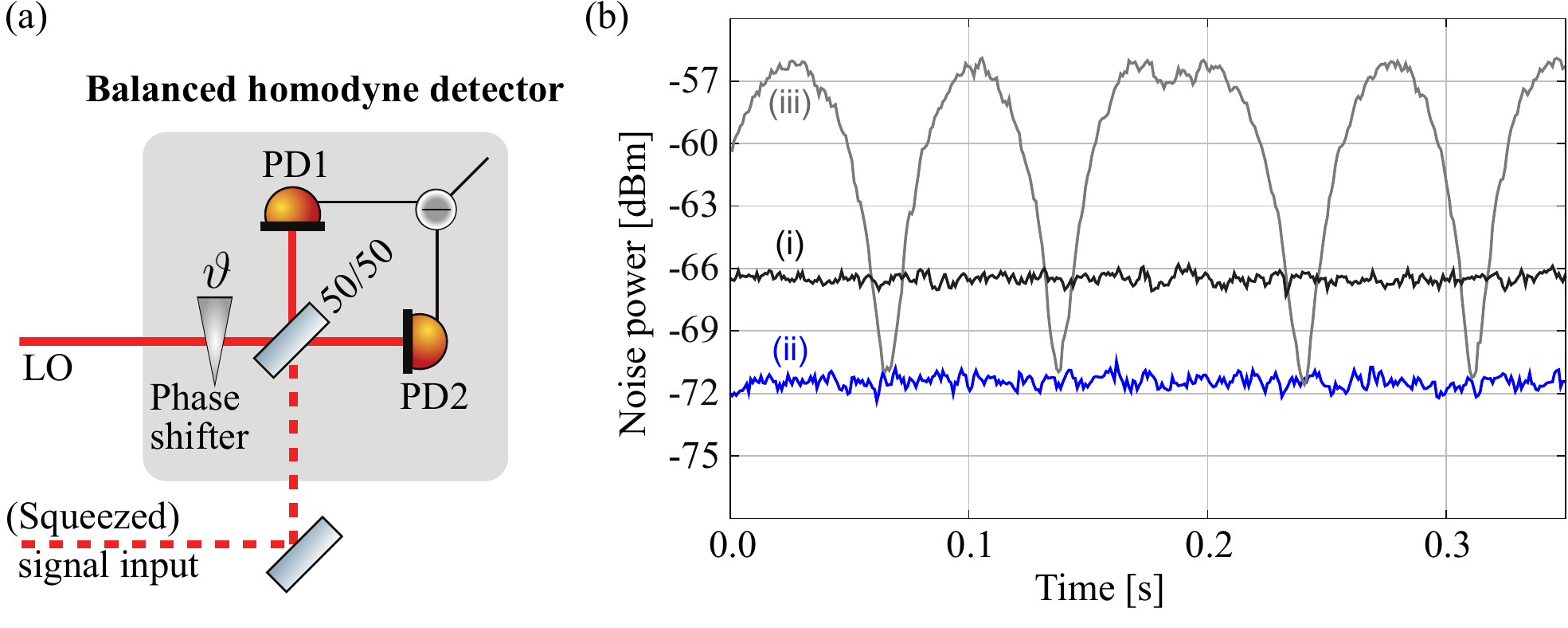} 
  \vspace{-1mm}
 \caption{{\textbf{Balanced homodyne detection (BHD)}} -- (a) Setup. The quadrature at choosable angle $\vartheta$ of the signal field is measured by overlapping the latter with a local oscillator (LO) field of the same mode parameters on a balanced beam splitter and recording the difference voltage from two PIN photo-diodes as shown. In order to meet the BHD approximation, the LO needs to be much more intense than the signal field. A close to perfect mode overlap between LO and  signal input field is crucial. For a non-perfect overlap, the detector measures the input state with unwanted contributions of the vacuum state. 
(b) Noise power measurements (i) on an electronically amplified and band-pass filtered quadrature amplitude of the vacuum field ($\hat X^{\rm{vac}}_{\Omega, \Delta \Omega}$) (signal input  blocked),  (ii) on a squeezed quadrature ($\hat X^{\rm{sqz}}_{\Omega, \Delta \Omega}$) of a squeezed vacuum state ($\vartheta =: 0$), and (iii) on respective quadratures of the same squeezed state where the phase angle $\vartheta$ was continuously shifted by changing the optical path length of the LO. The measurement data shows about 5\,dB of squeezing and was first published in [\cite{Chelkowski2007}]. $\Omega / 2\pi = 5$\,MHz, $\Delta \Omega / 2\pi = 100$\,kHz. 
}
\label{fig:3}
\end{figure}

In contrast to a single PIN photo diode, a balanced homodyne detector (BHD) is suitable to measure the quantum statistic of all types of modulations, i.e.~for all angles $\vartheta$. Such a detector consists of two identical PIN photo-diodes, a balanced beam splitter, and an external homodyne local oscillator field that is much brighter than the signal beam and that has an adjustable phase. The signal beam corresponds to the squeezed field, which in many experiments is in a squeezed vacuum field, having an optical power that usually corresponds to just a few photons per mode.  
The two beams are overlapped on the balanced beam splitter with close to perfect mode matching and the two interference outputs are focussed onto the photo diodes, see Fig.\,\ref{fig:3} (left). The electric output signal of the BHD is the difference of the photo diode voltages.
The LO takes over the role of the carrier light field, but with the possibility to choose the phase shift $\vartheta$. This way, eigenvalues of $\hat X$, $\hat Y$ or $\hat X^\vartheta$ can be measured, where the latter is given by the following linear combination of the first two 
\begin{equation}
\hat X^\vartheta (t) = \rm{cos} (\vartheta)\, \hat X (t) +  \rm{sin} (\vartheta)\, \hat Y (t) \, .
\label{eq:quadrature}
\end{equation}
If the modulation depths of signal and local oscillator beams are weak compared to their coherent amplitudes $|\alpha|$ and $|\alpha_{\rm LO}|$, the output voltage of a BHD corresponds to eigenvalues of the following operator
\begin{equation}
\hat V (t) \propto  2 \rm{cos}(\vartheta) \, |\alpha_{LO}| |\alpha| + |\alpha_{LO}|\, \hat X^\vartheta (t) + |\alpha| \, \hat X^\vartheta_{LO} (t) \, .
\label{eq:BHD}
\end{equation}
The `homodyne approximation' further involves $|\alpha_{\rm LO}|  \gg |\alpha| $ such that the term on the right can be neglected even if the local oscillator shows some classical quadrature excitation. 
The output voltage of a BHD is usually spectrally analysed or at least spectrally filtered, which removes the DC part, in full analogy to a single photo diode (see previous subsection).
Sampling the filtered voltage provides eigenvalues proportional to the generalized quadrature amplitude in Eq.\,(\ref{eq:quadrature}) 
\begin{equation}
\hat V^{\rm BHD}_{\Omega,\Delta \Omega} (t) \propto   |\alpha_{LO}| \, \hat X^\vartheta_{\Omega,\Delta \Omega} (t) \, .
\label{eq:BHD}
\end{equation}
~\\

Fig.\,\ref{fig:3}\,(a) shows the setup of a balanced homodyne detector for the characterization of squeezed states.
Setting $\vartheta = 0^\circ$, eigenvalues of the amplitude modulation depths can be sampled from the photo voltage according to Eq.\,(\ref{eq:BHD}). 
Setting  $\vartheta =90^\circ$, eigenvalues of the phase modulation depths are measured.
The data's expectation values $\langle \hat X^\vartheta \rangle$ provide the coherent displacement of the squeezed state. The data's variances
\begin{equation}
\Delta^2 \hat X^\vartheta \equiv \langle (\hat X^\vartheta)^2 \rangle - \langle \hat X^\vartheta \rangle^2 \, 
\label{eq:VAR}
\end{equation}
provide the state's (quantum) noise. 
A pure squeezed state as well as a squeezed state that experienced photon loss have Gaussian quantum statistics and are thus fully described by the expectation values and variances (first and second moments) of two orthogonal quadratures, but only if one quadrature reflects the lowest quadrature variance. 

In most experiments with squeezed light, the photo electric voltage according to Eq.\,(\ref{eq:BHD}) is not sampled with a data aquisition system, but the signal is directly fed into a spectrum analyser measuring the noise power of the voltage. If the expectation value $\langle \hat X^\vartheta \rangle$ is zero, the noise power is proportional to the variance $\Delta^2 \hat X^\vartheta$ in Eq.\,(\ref{eq:VAR}). The reference for quantifying the squeeze factor is measured by blocking the (squeezed) signal field in Fig.\,\ref{fig:3}\,(a).  
The measured vacuum noise level corresponds to the LO's (electronically amplified) shot noise level.

Traces (ii) and (iii) in Fig.\,\ref{fig:3} (b) show measured noise powers of the modulation mode ($\Omega/2\pi = 5$\,MHz, $\Delta\Omega/2\pi = 100$\,kHz) being in a squeezed vacuum state. (i) is proportional to the variance of the ground state uncertainty $\Delta^2 \hat X^{\rm vac}_{\Omega,\Delta\Omega}$. (ii) is proportional to the quantum noise variance of the squeezed quadrature amplitude $\Delta^2 \hat X^{\rm sqz}_{\Omega,\Delta\Omega}$. (iii) is proportional to the quantum noise variance of the quadrature amplitude with scanned phase $\Delta^2 \hat X_{\Omega,\Delta\Omega}(\vartheta (t))$.\\

To fully characterize a quantum state, i.e.~to do quantum state tomography [\cite{Vogel1989}], a BHD is a prerequisite. 
But also interfero\-metric measurements with balanced homodyne detectors instead of single PIN photo-diodes have several advantages. A correctly implemented BHD readily provides the vacuum noise level, when the signal beam is blocked. With a BHD, the optimum operating point of the interferometer is precisely at a dark fringe. If a perfect dark fringe can practically be achieved, amplitude noise of the laser does not couple into the signal port. If the interferometer has balanced arm length also frequency noise of the laser then does not couple into the signal port. Some quantum non-demolition schemes with the prospect of evading quantum radiation pressure noise require the detection of a non-canonical quadrature angle [\cite{Jaekel1990,Kimble2001}]. Here, the adjustable phase of a BHD provides a straight forward approach. The experimental exploration of BHDs for gravitational-wave detectors only has started recently [\cite{SteinlechnerS2015}].\\
\begin{figure}[ht!!!!!!!!!!!!!!!!!!!!!!!!!!!!!!!!!]
  \vspace{0mm}
  \includegraphics[width=12.4cm]{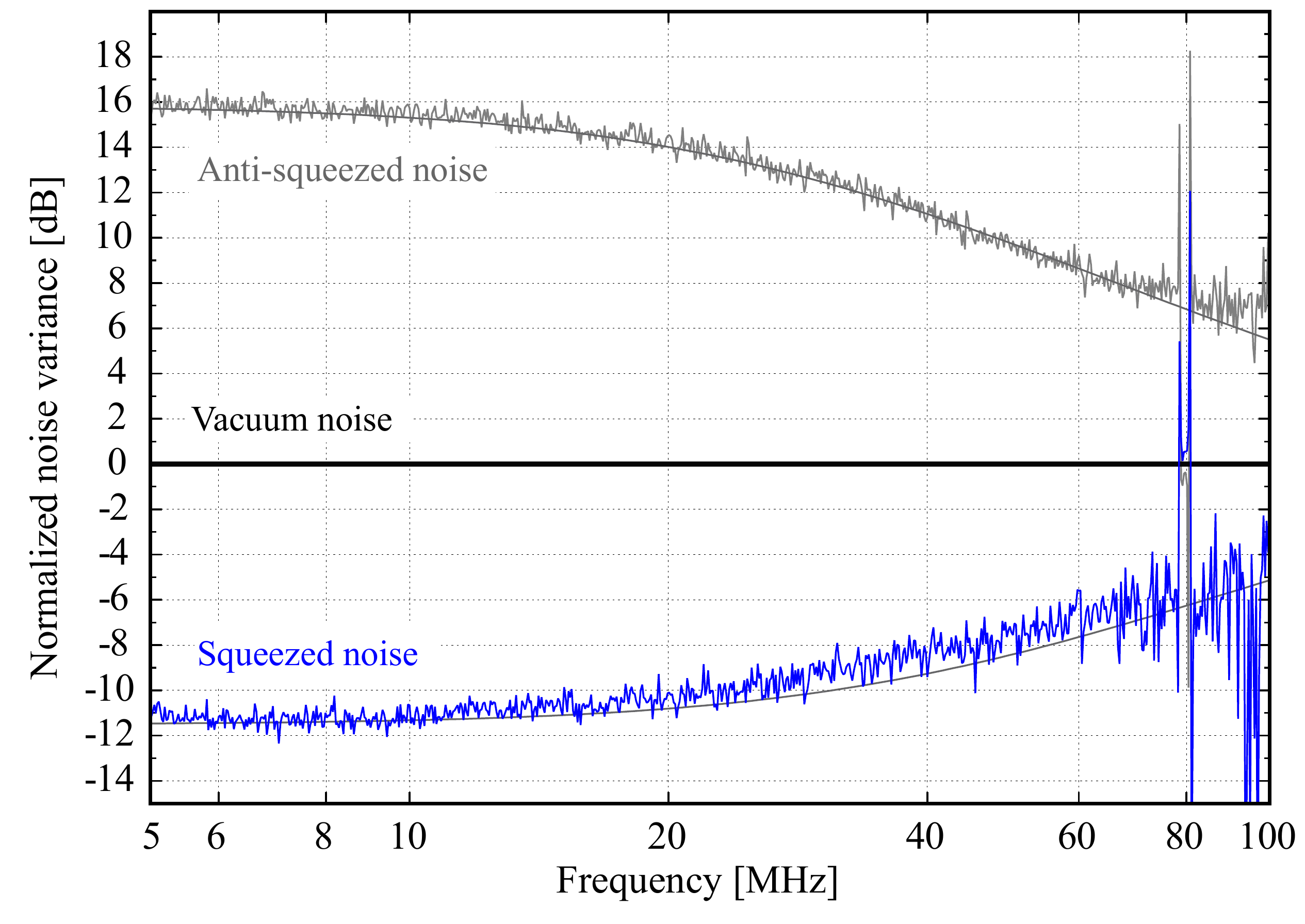} 
  \vspace{-1mm}
\caption{{\textbf{Spectrum of quadrature amplitude variances}} --  Shown are the quantum noise properties of a large number of modulation modes having a resolution bandwidth of $\Delta\Omega/ (2 \pi)=1$\,MHz. For all traces the balanced homodyne detector output was analysed with a spectrum analyser. Squeezing of $\hat X_{\Omega, \Delta \Omega}$ (bottom trace) and anti-squeezing of $\hat Y_{\Omega, \Delta \Omega}$ (top trace) versus $f = \Omega / (2 \pi)$ are shown relative to the vacuum noise variance. The spectrum below 5\,MHz is not shown since it contained less squeezing due to laser relaxation oscillation in the carrier field as well as disturbances from back-scattered light [\cite{Vahlbruch2007}]. Disturbances at frequencies above 70\,MHz originated from relatively large detector dark noise, which was subtracted from all traces shown here. The thin line represents a theoretical model that takes into account for the linewidth of the squeezing cavity. The data was first presented in in Ref.~[\cite{Mehmet2010}].
}
\label{fig:4}
\end{figure}
~\\

A light field can be analysed with respect to many different modulation frequencies $\Omega$. The result constitutes a \emph{spectrum} [\cite{Breitenbach1998}], where, in principle, every modulation mode can be in a different quantum state. 
Fig.\,\ref{fig:4} shows spectra of squeezed states from 5\,MHz to 100\,MHz with $\Delta\Omega/2\pi = 1$\,MHz. The lower curve shows the spectrum of the most strongly squeezed variances, in this case the variances of $\hat X_{\Omega,\Delta\Omega}$. The upper spectrum shows the variance in the orthogonal quadrature amplitude ($\hat Y_{\Omega,\Delta\Omega}$). All variances are normalized to those of the corresponding vacuum state. The squeeze factor reduces towards higher frequencies due to the linewidth of the squeezing cavity. The anti-squeezing is always higher than the absolute value of the squeezing due to Heisenberg's uncertainty relation and due to the presence of optical loss. The curves do not represent \emph{pure} squeezed states but \emph{mixed} squeezed states, with a significant contribution from vacuum states, due to optical loss. 
Pure squeezed states can only be produced by making the influence of all decoherence processes negligible.

The choice of the resolution bandwidth (RBW, $\Delta \Omega$) during data taking and processing defines the spectral-temporal modulation modes, including their number within the detected spectrum. For any setting of the RBW, the quantum mechanical properties of the quadrature amplitudes $\hat X_{\Omega,\Delta \Omega}$ and $\hat Y_{\Omega,\Delta \Omega}$ [\cite{Caves1985}] fully correspond to those introduced for quadratures in standard text books, and which are  reviewed in Sec.\,\ref{Sec:3}.\\

\FloatBarrier

\subsection{Observations on two-mode squeezed states using balanced homodyne detectors} \label{Ssec:2.4}

Two-mode squeezed states are composed of two subsystems `A' and `B' and are bi-partite entangled states with a Gaussian quantum statistic. 
To avoid conflicts with different usage of the term `mode', they can synonymously be named `bipartite Gaussian entangled states' or `bipartite squeezed states', which will be mainly used in this Review.
In the same way multi-partite Gaussian entangled states  correspond to multi-partite squeezed states. 

\begin{figure}[ht]
  \vspace{-1mm}
  \includegraphics[width=10.6cm]{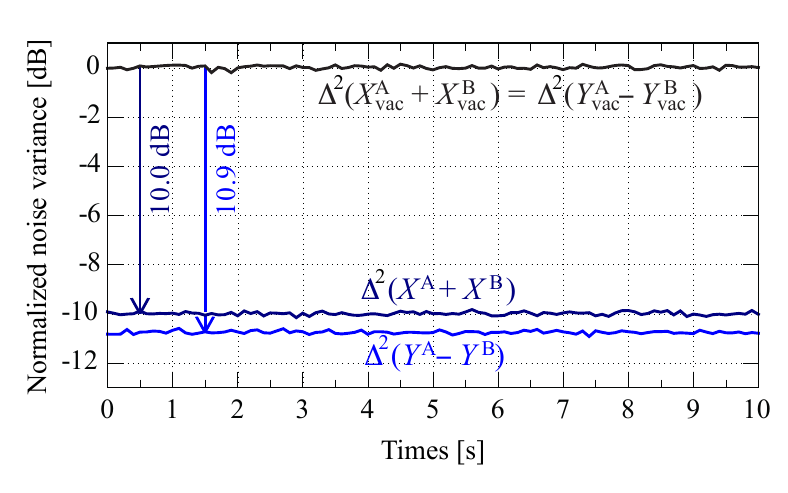} 
  \vspace{-3mm}
\caption{{\textbf{Two-mode squeezing measurement}} -- For this measurement the outputs of two balanced homodyne detectors are added or subtracted and the variances (noise powers) of the results recorded. The upper trace was measured with modes `A' and `B'  being in their ground states. This measurement served as a reference level. Strong two-mode squeezing was observed as shown by the lower two traces. The sideband frequency was $\Omega/ (2 \pi) = 8$\,MHz and the resolution bandwidth was $\Delta\Omega/ (2 \pi) = 200$\,kHz. The measurement results were first published in Ref.~[\cite{Eberle2013}].
}
\label{fig:5}
\end{figure}

The measurement observables that prove or disprove the bi-partite squeezing property are $\hat X^A_{\Omega, \Delta \Omega} - \hat X^B_{\Omega, \Delta \Omega}$ and $\hat Y^A_{\Omega, \Delta \Omega} + \hat Y^B_{\Omega, \Delta \Omega}$, where the minus and plus signs may be swapped. Bi-partite squeezed states are precisely those states that were discussed by Einstein, Podolsky, and Rosen (EPR) in their seminal paper [\cite{Einstein1935}]. 
Fig.\,\ref{fig:5} shows a measurement result on bi-partite squeezed light [\cite{Eberle2013}]. The variances of both joined observables are squeezed as shown in the two lower traces. They were {recorded} consecutively by adding or subtracting the outputs of two balanced homodyne detectors. But by interfering the subsystems on a beam splitter one could even measure both joined observables \emph{simultaneously}. This possibility is correctly described in quantum theory since their commutator is zero. 
 
The so-called EPR paradox arises as follows. If we either measure $\hat X^A_{\Omega, \Delta \Omega}$ and $\hat X^B_{\Omega, \Delta \Omega}$ or $\hat Y^A_{\Omega, \Delta \Omega}$ and $\hat Y^B_{\Omega, \Delta \Omega}$ it is obvious from the data in Fig.\,\ref{fig:5} that we can always predict the measurement result at subsystem `B' when knowing the result at subsystem `A'. This seems to suggest that \emph{both} quantities at `B' are precisely defined simultaneously \emph{before} the measurement on `A', which contradicts the rigorous (and correct) interpretation of their non-zero commutator that they are not precisely defined simultaneously.

To solve this paradox, EPR conjectured that the wavefunction, as defined by quantum theory, does not provide the full information.
This led to a discussion of whether hidden variables existed that needed to be included in a complete theory of quantum mechanics (see also Bell [\cite{Bell1966}]). The experimentally observed violation of Bell's inequality [\cite{Bell1964,Aspect1981,Giustina2013,Hensen2015}], however, ruled out the existence of (local) hidden variables. 

Based on that, the EPR paradox needs to be solved in a different way. Contrary to what EPR assumed, it \emph{is} in fact possible to predict the value of an arbitrary observable of a physical system $A$ with certainty via a measurement on system $B$, although this observable was \emph{not} defined before the measurement. 
Without any interaction, a measurement on subsystem `A' not only creates `reality' of e.g.~$\hat X^A_{\Omega, \Delta \Omega}$; simultaneously, `reality' is also created regarding the observable $\hat X^B_{\Omega, \Delta \Omega}$ describing subsystem `B'. Here, the term `reality' has the meaning as defined by EPR [\cite{Einstein1935}]. Similarly, the detection of one photon of a two photon entangled number state, not only produces the reality of this photon but also that of a second one. 
A discussion of Einstein-Podolsky-Rosen entanglement can also be found in [\cite{Schnabel2015}]. 
Note that the EPR paradox can also be described as `quantum steering' [\cite{Schroedinger1935,Cavalcanti2009,Haendchen2012}]. 
It should also be mentioned that two-mode squeezing being detected with BHDs and not with photon counters cannot be used to violate a Bell inequality. The latter topic is outside the scope of this Review. 

Bi-partite squeezed states were first characterized with balanced homodyne detectors by the group of J. Kimble in 1992 [\cite{Ou1992}]. Generally, the EPR paradox becomes more pronounced the stronger the bi-partite squeezing is. A measure of the strength of EPR entanglement was introduced by M. Reid [\cite{Reid1985}]. According to this measure the result in Fig.\,\ref{fig:5} can be quantified to $\epsilon^2 = 0.0309$, where the critical value is one. It corresponds to the strongest Gaussian EPR entangled state generated so far.\\

For a long time it looked like that two-mode squeezed states are not useful for laser interferometers. The reason for that belief was that a laser interferometer, as any other measurement device too, is built to measure \emph{one} observable. It seems to be ideal already if the quantum noise in this single observable is squeezed. The increased quantum noise in the orthogonal observable is not harmful in this case, and squeezing in two different observables useless.
Only recently realistic scenarios were discussed in which two-mode squeezing in fact does improve the performance of a laser interferometer [\cite{Steinlechner2013b}]. The proof-of-principle experiment is reviewed in Sec.\,\ref{Sec:7}.\\

\subsection{Observations using photon counters}
Alternatively to field quadratures, an optical mode in a squeezed state can also be characterized, at least partly, by detecting its photon number distribution. 
For a pure squeezed vacuum state, such a measurement would reveal the existence of solely even photon numbers including a large probability for zero photons. The average photon numbers of squeezed vacuum states with feasible squeeze factors are very small, of the order of one per second and bandwidth in hertz, see Fig.\,\ref{fig:13}\,(a)\,--\,(c). A distribution with close to zero probability of odd photon numbers, however, has not been measured so far. The reason is the lack of ideal photon counters. 
First of all, the efficiency of these detectors, i.e.~their probability of converting one photon into one click and no photon into no click, must be almost perfect. `Lost' photons as well as dark counts wash out the odd/even oscillations. Furthermore, most detectors available can only distinguish between zero and one photon. This problem can be solved by distributing the squeezed mode onto a large number of single photon detectors using an array of beam splitters, such that all paths have a low probability of carrying more than one photon.
Photon number measurements on squeezed vacuum states, nevertheless, play an extremely important role in quantum optics. When the squeezing strength is very low, the probability of detecting more than 2 photons can be neglected, and the detection of a photon heralds the existence of a second one. \\

\begin{figure}[ht]
  \vspace{0mm}
  \includegraphics[width=13.6cm]{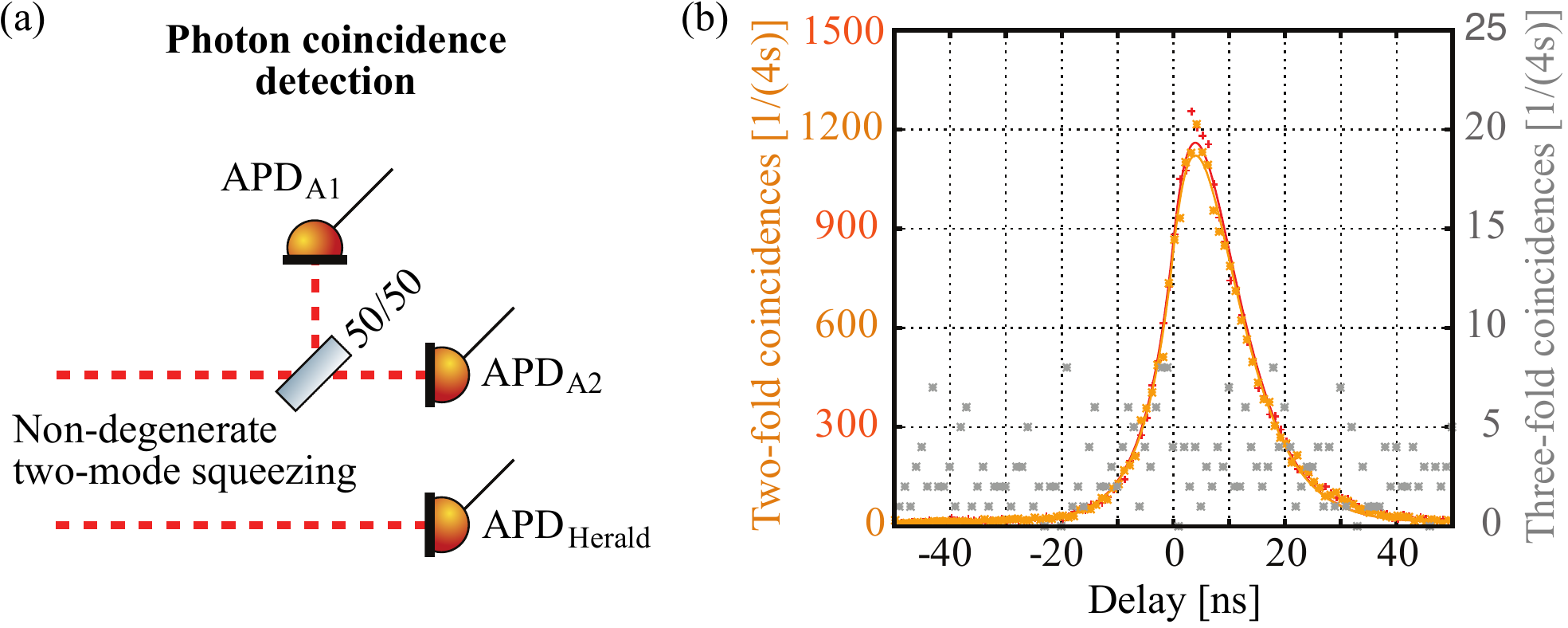} 
  \vspace{-1mm}
\caption{{\textbf{Coincidence clicks from non-degenerate photon pairs}} --  The first such experiment was reported in Ref.~[\cite{Hong1986}].
 (a) shows a setup with three avalanche photo-diodes (APDs) for proving the successful heralding of a single photon number state.  (b) Histograms of the two-fold coincidence detections at APD$_{\rm Herald}$ and APD$_{\rm A1}$ (red), and at APD$_{\rm Herald}$ and APD$_{\rm A2}$ (yellow) with theoretical models (solid lines). If the two-mode squeezing just carried one photon in each spatial subsystem, the three-fold coincidence detection should be zero. Indeed the according histogram (grey points, right y axis) shows only a few events. These are produced by false (dark) counts of the APDs.
The delay for the three-fold coincidences is defined as the time between counts at `A1' and `A2' given that the trigger APD$_{\rm Herald}$ detected a photon (within a 100\,ns time window).
The data was taken on photons that were up-converted from 1550\,nm to 532\,nm and it was first published in Ref.~[\cite{Baune2014}]. 
}
\label{fig:6}
\end{figure}

If a mode of light is always excited by either zero or two photons, `conditional' or `heralded' one-photon Fock states can be realized. (Measurements on an ensemble of the $n$-photon Fock state would always produce the measurement result $n$, i.e.~Fock states have a zero photon number uncertainty. They are also called `number states').   
The above concept of producing a one-photon Fock state obviously requires the deterministic and balanced distribution of the down-converted signal and idler fields into two different paths. In order to achieve this the signal and idler fields need to be non-degenerate. Usually, a mode in a squeezed state is composed of degenerate signal and idler fields, and this degeneracy thus needs to be removed. Possible ways are producing the down-converted fields at well separated wavelengths [\cite{Villar2005,Su2006,Li2010,Samblowski2011}], %
separating the upper and lower sidebands belonging to an ordinary squeezed mode by frequency filters [\cite{Schori2002,Hage2010}], and using spatial filters [\cite{Hong1987}]. A frequently used approach is, using type\,II parametric down-conversion where the photons within a pair are always orthogonally polarized [\cite{Ou1992,Kiess1993,Kwiat1995}]. \\
The list of experiments with conditional or heralded photon number states is long. They showed for instance nonclassical $g^{(2)}$-functions [\cite{Hong1987}] and violations of Bell inequalities [\cite{Weihs1998}].  
Fig.\,(\ref{fig:6}) shows a result from a more recent experiment, in which a bipartite-squeezed state with subsystems at 1550\,nm and 810\,nm was produced, the subsystem at 1550\,nm subsequently up-conver\-ted to 532\,nm, and the `quantum non-Gaussianity' of heralded up-converted single photons demonstrated [\cite{Baune2014}].\\
Squeezed states are also the resource for the conditional generation of superpositions of coherent states [\cite{Ourjoumtsev2006,Neergaard-Nielsen2006}] and so-called N00N-states [\cite{Afek2010}].\\ 

The generation of nonclassical states mentioned in the paragraph above is not stationary but relies on a probabilistic trigger event.
The production of squeezed states themselves usually happens in a stationary fashion.
This distinction has an important consequence for applications of nonclassical states in measurement devices.
Only (stationary) squeezed states allow for a continuous improvement of a measurement. Avoiding any loss of measuring time is generally of high relevance, for the detection of short-lived signals with unknown arrival time as well as for the detection of long-lived quasi-monochromatic signals since the signal-to-noise-ratio (S/N) improves with measuring time. \\

\subsection{Conclusions}

The detection of squeezed light produces measurement results that can be considered as remarkable. Let us focus on experiments where a mode in a bright coherent state is overlapped with a mode in a squeezed vacuum state, as shown in Figs.\,(\ref{fig:1}) and (\ref{fig:3}). In both setups, the squeezed vacuum field can easily be blocked, which allows us to compare the measurement results on a bright coherent state with and without the interference with the squeezed vacuum state. 
\emph{Without squeezing,} the photo-electric detectors measure a large number of photon events, with a large quantization noise (shot noise). The large noise reflects the fact that all photon events were independent from each other, as shown in Fig.\,\ref{fig:2}\,(b,i).
\emph{With squeezing,} the photo-electric detectors again measure a large number of photon events, with an expectation value that is even slightly higher, but nevertheless, the quantization noise of all detected photons is significantly reduced, Fig.\,\ref{fig:2}\,(b,ii). \\
Based on the discussion of EPR entanglement in Subsec.\,\ref{Ssec:2.4} the photo-electric detection of the output light of a squeezing-enhanced laser interferometer (with $\alpha^* \alpha \!\gg\! 1$) produces the reality of photons. This way we can keep the `wave picture', in which no photons exist, when light travels along the interferometer arms and when it interferes at the beam splitter. When the energy of the beam is elevating electrons to the conductance band of the photo-diode's semi-conductor, $n$ photon events simultaneously appear within the measuring interval with probability $P(n)$. What conclusion has to be drawn if the probabilities resemble a sub-poissonian statistic? -- The occurrence of photon events is still truly random but in this case not for individual photons. The occurrence of photons is correlated in such a way that the probability of detecting an additional photon in the same time interval reduces the larger the number of already detected photons is. 
What follows from the discussion of EPR entanglement for a photon counting experiment with pure squeezed vacuum and ideal photon counters? Here the probabilistic detection of one photon entails the detection of a second one with certainty. With some smaller probability a third photon is detected, which entails the detection of a fourth photon with certainty, and so on.\\ [-3mm]
 
\emph{If a photon of a mode that was not interrogated by the environment before is absorbed, its reality is created in this very moment. If the photon belongs to a squeezed state, this process instantaneously influences the probability of other photons becoming reality.} \\[-3mm]

Of course, a more general statement can be made, based on the insight that interaction with the environment creates the reality of any kind of quanta, including electrons, atoms, and molecules. \\

\section{Theoretical description of squeezed states} \label{Sec:3} 

\subsection{The quadrature amplitude operators}

Consider a single mode of light  at optical frequency $\omega$. Its Hamilton operator reads
\begin{equation}
\hat H_\omega = \hbar \omega \left(\hat n + \frac{1}{2} \right) =  \hbar \omega \left(\hat a^\dagger_\omega \hat a_\omega + \frac{1}{2} \right) = \hbar \omega \left( \hat X^2_\omega + \hat Y^2_\omega \right) \, ,
\label{eq:Homega}
\end{equation}
where $\hat n$ is the photon number operator, and $\hat a_\omega$ and $\hat a^\dagger_\omega$ are the annihilation and creation operators, which obey the commutation rule $\left [ \hat a_\omega, \hat a_\omega^\dagger \right] = 1$. 
The operator $\hat a_\omega$ 
has a complex-valued dimensionless eigenvalue spectrum and corresponds to the complex amplitude $\alpha_\omega$ 
in classical optics. $\hat X_\omega$ and $\hat Y_\omega$ are the hermitian amplitude and phase quadrature operators. 
The eigenvalues of the quadrature operators are also dimensionless and proportional to the electric fields at the oscillation's antinode and at the oscillation's node. In the above equation, they are defined such that their variances are $\Delta^2 \hat X_\omega = \Delta^2 \hat Y_\omega = 1/4$ if the oscillator is in its ground state, i.e.~if $ \left< \hat n \right> = 0$. 

Although Eq.\,(\ref{eq:Homega}) simply describes the energy of an harmonic oscillator it is the essence of quantum theory, since it mathematically describes the wave-particle dualism. Whereas the eigenvalues of $\hat n$ have a discrete spectrum, the eigenvalues of $\hat X_\omega$ and  $\hat Y_\omega$ have a continuous spectrum. In classical optics the phase quadrature is zero. In quantum optics its expectation value is also zero, but its uncertainty contributes to the overall energy. 

Eq.\,(\ref{eq:Homega}) describes a cavity mode as well as a section that is cut from a propagating quasi-monochromatic light beam. The latter example is of high relevance in actual experiments. \emph{By setting the section's time window, i.e.~the measuring time interval, the time-frequency (`modulation') mode is defined.} 

\begin{figure}[ht!!!!!!!!!!!!!!!!!!!!!]
  \vspace{-8mm}
  \includegraphics[width=13.6cm]{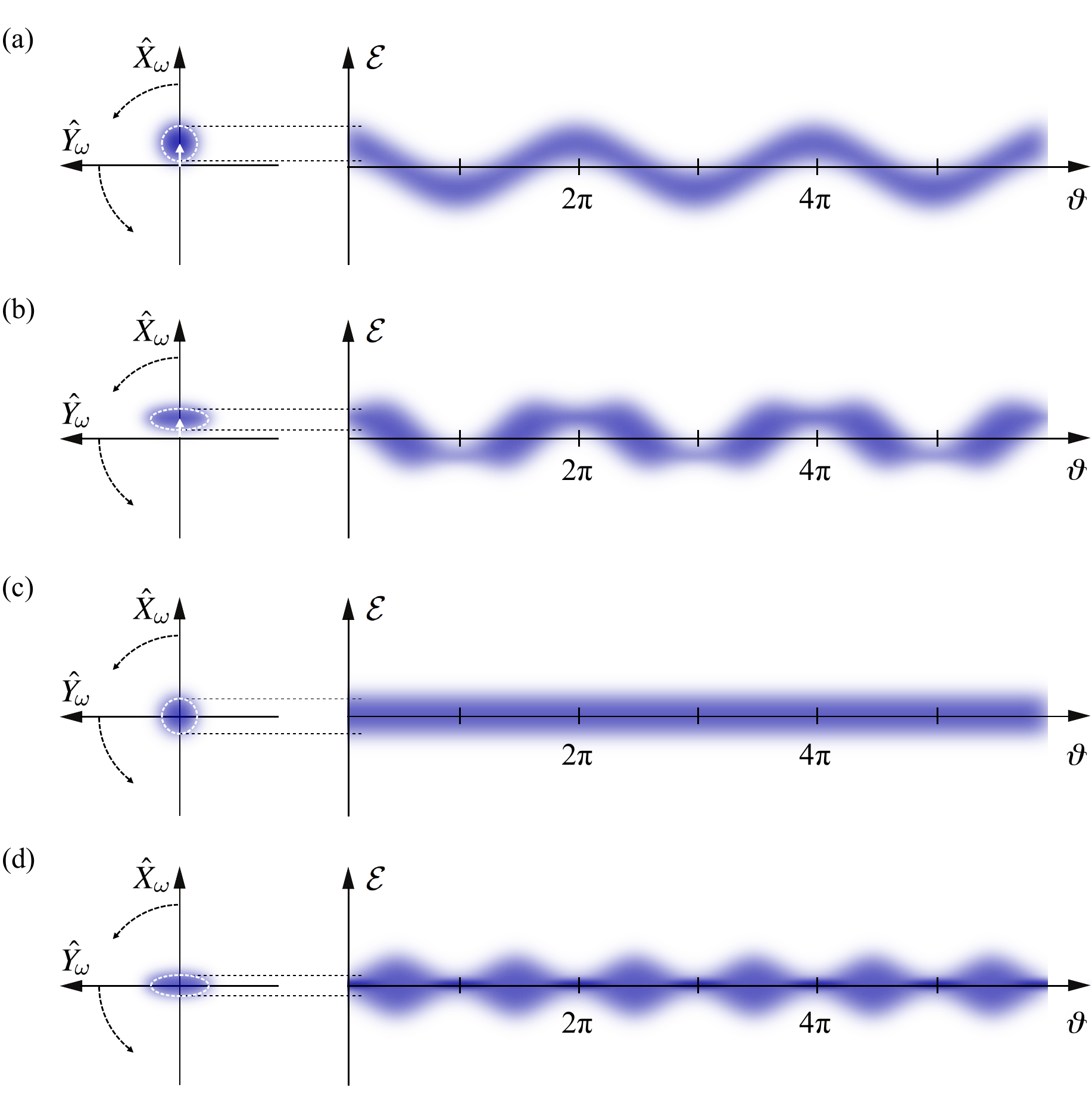}
  \vspace{-0mm}
\caption{\textbf{Phase spaces and electric field oscillations of monochromatic light} -- 
Top Left: Monochromatic light in a coherent state is represented by a phasor (white arrow) including its quantum uncertainty (white dashed circle and fuzzy area) located in the phase-space spanned by the quadratures $\hat X_\omega$ and $\hat Y_\omega$. 
When the phase space rotates with optical frequency $\omega/2\pi$, the projection of the quantum phasor onto a fixed (vertical) axis corresponds to the electric field ${\cal E} (t)$, as shown on the right side.
(a) Weakly displaced coherent state. (b) Corresponding amplitude squeezed state. The electric field uncertainty around the zero average field region is anti-squeezed. (c) Vacuum state at the same optical frequency. (d) Corresponding squeezed vacuum state. 
The meaning of the uncertainty could be carved out by supplementing them with monochromatic waves all having the optical frequency $\omega/2\pi$. Changing amplitudes then display amplitude quadrature noise. Changing shifts along the time axis model the electric field uncertainty at the expected zero crossing. They are not implemented in the graphics here, however, since any of those waves does not exist due to Heisenberg's uncertainty relation. 
}
\label{fig:7}
\end{figure}

The quadrature operators introduced in Eq.\,(\ref{eq:Homega}) and displayed in Fig.~\ref{fig:7} do not correspond to `$\hat X$'  and `$\hat Y$' that are of relevance in laser interferometry and in optical communication, and which were already discussed in Subsec.~\ref{Ssec:2.2} and \ref{Ssec:2.3}. 
The optical frequency of visible and near-infrared light is far too high to be transferred to an oscillation of photoelectric voltage. 
Quite general, a laser interferometer targets signals at audio or radio band frequencies  $\Omega_i \ll \omega$. 
Such a measurement is achieved, as stated before, by decomposing the photo-electric voltage from the photo diode at the interferometer output into a single-sided spectrum (positive frequencies only) of intervals of $\Omega \pm \Delta \Omega/2$. 
\FloatBarrier 
The signals as well as the quantum uncertainties carried by a beam of light are thus described by a spectrum of pairs of non-commuting quadrature operators. Mathematically, every such operator is defined by an integral over the Fourier components within the bandwidth. The spectral weighting of the Fourier components is called the `window function'.   
\emph{By going to sideband intervals, a spectrum of a new type of optical mode is defined, which describes the modulation of the electric field in the respective frequency interval $\Omega \pm \Delta \Omega/2$.} In this Review we call it a `modulation mode'. 

The quadrature operators that are defined around a modulation frequency $\Omega$ with a bandwidth of $\Delta \Omega$ are the 
quadrature amplitude operators that are relevant in laser interferometry. Whenever they are not related to a specific band we use the short form  $\hat X_{\rm \Omega,\Delta \Omega}(t) \equiv \hat X$ and $\hat Y_{\rm \Omega,\Delta \Omega}(t) \equiv \hat Y$, cf. Eqs.~(\ref{eq:X}) and (\ref{eq:Y}). 
These operators can slowly vary with time, where the time dependence is limited by $\Delta \Omega$.
(The time dependence is not due to quantum uncertainty, which usually is time independent, but for instance due to the time dependence of the signal, e.g.~a passing gravitational wave.) 
Let us consider now a pair of quadrature operators for a particular sideband $\Omega \pm \Delta \Omega /2$. The Hamilton operator of the corresponding modulation mode is found by switching to the  frame rotating at optical frequency $\omega$. The transition is done by applying the unitary transformation $\hat U = {\rm exp}(i\omega \hat a^\dagger \hat a t)$ generating a new Hamiltonian $\hat H = \hat U^\dagger \hat H_\omega \hat U - i \hbar \hat U \partial \hat U^\dagger / \partial t$. The Hamiltonian of the modulation mode reads
\begin{equation}
\hat H = \hbar \Omega \left(\hat n_\Omega + \frac{1}{2} \right) =  \hbar \Omega \left(\hat a^\dagger \hat a + \frac{1}{2} \right) = \hbar \Omega \left( \hat X^2 + \hat Y^2 \right) \, ,
\label{eq:H}
\end{equation}
where $\hat n_\Omega$ is the (occupation) number operator for the modulation mode, and $\hat a$ and $\hat a^\dagger$ its annihilation and creation operators. The commutation rule $\left [ \hat a, \hat a^\dagger \right] = 1$ is unchanged. $\hat X$ and $\hat Y$ are the \emph{amplitude and phase quadrature amplitude operators}, respectively. They correspond to the \emph{depth} of the amplitude modulation and, for weak excitations, to the \emph{depth} of  the phase modulation, respectively. They are the conventional hermitian field operators in experimental quantum optics. 
Note, that modulation modes at angular frequency $\Omega$ can be described by a superposition of three \emph{optical} frequencies, a carrier at $\omega$, an upper sideband at $\omega+\Omega$ and a lower sideband at $\omega - \Omega$. The quantum mechanical description of modulation states in connection to optical carrier and upper and lower sidebands is known as the `Two-Photon Formalism' [\cite{Caves1985a,Schumaker1985}].

The quadrature amplitude operators in Eq.\,(\ref{eq:H}) are again defined such that the variances of the uncertainty of a modulation field in its ground state or in a coherent state are 
\begin{equation}
\Delta^2 \hat X_{\rm vac} = \Delta^2 \hat Y_{\rm vac} = 1/4\, .
\label{eq:zp}
\end{equation}
Generally, quadrature operators $\hat X$ and $\hat Y$ as defined in Eqs.\,(\ref{eq:Homega}) and (\ref{eq:H}) are the real and imaginary parts of the annihilation operator
\begin{equation}
\hat a = \hat X + i \hat Y \;\; \Leftrightarrow  \;\; \hat a^\dagger = \hat X - i \hat Y 
\end{equation}
\begin{equation}
\Leftrightarrow \;\;\; \hat X = \frac{1}{2} \left( \hat a + \hat a^\dagger\right) \, , \; \hat Y = \frac{1}{2i} \left( \hat a - \hat a^\dagger\right) \, .
\end{equation}
They satisfy the commutation relation
\begin{equation}
\left [ \hat X, \hat Y \right] = \frac{i}{2} \, ,
\label{eq:XYcom}
\end{equation}
and their variances are limited by a Heisenberg uncertainty relation of the following form
 \begin{equation}
\Delta^2 \hat X \Delta^2 \hat Y  \geq \frac{1}{16} \, .
\label{eq:XYhup}
\end{equation}

A quantum state is called a ``squeezed state''~[\cite{Bachor2004}] if $\Delta^2 \hat X_\vartheta  <1/4$ for an arbitrary field quadrature $\hat X_\vartheta= \hat X \cos \vartheta + \hat Y \sin \vartheta$, see Eq.\,(\ref{eq:quadrature}). The angle of the lowest variance below 1/4 is called the \textit{squeeze angle} $\theta$. The largest factor by which the variance is below 1/4 is called the \textit{squeeze factor}, often given on a decibel (dB) scale using the following transformation
\begin{equation}
\label{eq:dB}
-10\cdot {\rm log}_{10} \left ( \frac{\Delta^2 \hat X_\theta}{\Delta^2 \hat X_{\rm vac}}  \right )\, .
\end{equation}
The squeeze factor can also be described by the  \textit{squeeze parameter} $r$
\begin{equation}
\label{eq:r}
e^{-2r}= \frac{\Delta^2 \hat X_\theta}{\Delta^2 \hat X_{\rm vac}} \, .
\end{equation}

When a squeezed state experiences optical loss, it remains squeezed but the squeeze factor is reduced. Also the state's purity is reduced, i.e.~the product of the quadrature uncertainties increases above the minimum value. Optical loss corresponds to mixing the state with the vacuum state. Let $\Delta^2 \hat X_\vartheta$ be the variance of a quadrature amplitude, $\Delta^2 \hat X_{\rm vac}$ the variance of the (quadrature angle independent) ground state uncertainty, and $(1 - \eta^2)$ the relative energy loss. Then the resulting quadrature variance reads 
\begin{equation}
\Delta^2 \hat X_\vartheta^{'} = \eta^2 \Delta^2 \hat X_\vartheta + (1 - \eta^2) \Delta^2 \hat X_{\rm vac} \, .
\label{eq:loss}
\end{equation}

To maximize the benefit from squeezed states in applications, strongly squeezed states need to be generated and optical loss minimized. Optical loss occurs due to absorption and scattering in the optical components in the path of the squeezed beam, including the squeezing resonator itself, and due to non-perfect matching to the interferometer mode, non-perfect interference contrast of the interferometer and non-perfect quantum efficiency of the photo diodes. The sum of all losses, including those outside the interferometer, need to be less then 10\% to allow a nonclassical quantum noise suppression of a factor of 10 in power, i.e.~10\,dB.

\subsection{Phase space representations of squeezed states}

\begin{figure}[h!!!!!!!!!]
  \vspace{-4mm}
  \includegraphics[width=12.6cm]{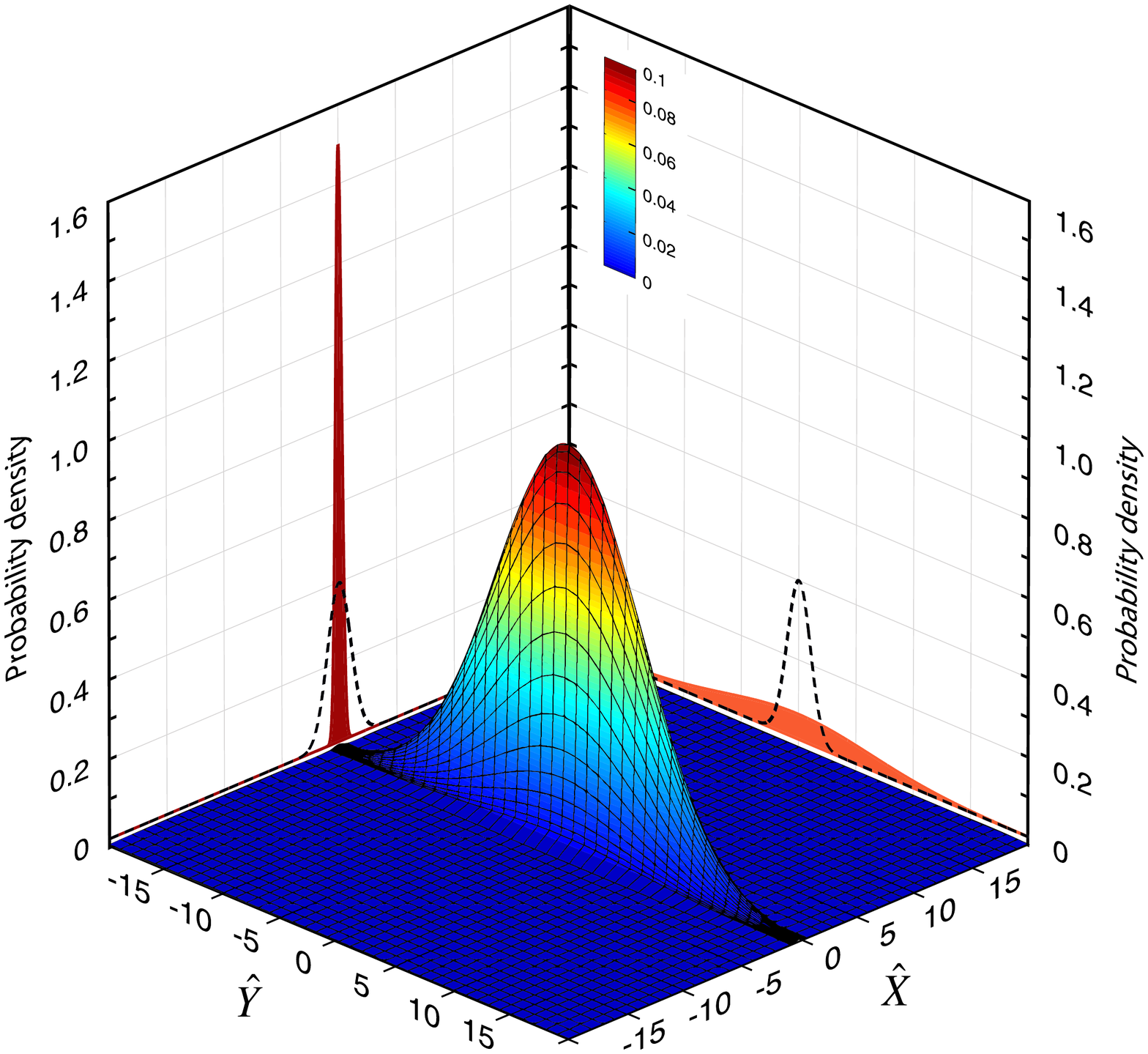}
  \vspace{0mm}
\caption{\textbf{Wigner function and its projections} -- Displayed is the full information of a squeezed vacuum state. The quasi probability density of the Wigner function (centre) is plotted along a third dimension and is color-coded. Also shown are the actual measurement results from which the Wigner function is reconstructed. They are represented by the squeezed and anti-squeezed Gaussian projections onto the $\hat X$ and $\hat Y$ axes. Their probability densities $p(X)$ and $p(Y)$ are given on the vertical axes. The Gaussian measurement statistic of the first clearly shows squeezing compared to the ground state statistic (dashed). The squeeze factor is 11.6\,dB and the squeeze parameter $r = 1.335$ [\cite{Mehmet2010}].
} 
\label{fig:8}
\end{figure}

\emph{The Wigner function} -- The properties of squeezed states are nicely displayed by the Wigner function $W(X,Y)$ [\cite{Wigner1932}]. An example in terms of a squeezed vacuum state is shown in Fig.\,\ref{fig:8}. It is a quasi-probability distribution, which contains the state's full information, including its quantum statistic. 
There are two ways how a Wigner function provides a sufficient criterion for nonclassicality. First, by containing negative values; second by features that have a smaller (squeezed) width compared with the Wigner function of the ground state.
Integrating the Wigner function over $Y$ provides the probability density of measurement results, i.e.~of the eigenvalues of the observable $X$ and vice versa

\begin{equation}
\int\limits_{-\infty}^\infty W(X,Y) {\rm d} Y = p(X) \, , \;\; \int\limits_{-\infty}^\infty W(X,Y) {\rm d} X = p(Y) \, ,
\label{eq:Wigner1}
\end{equation}
where $p(X)$ and $p(Y)$ are the observed probability distributions, also exemplarily shown in Fig.\,\ref{fig:8}. 

The ground state, coherent states as well as (quadrature) squeezed states have quadrature eigenvalue probability densities that are Gaussian. Their Wigner functions are also Gaussian and thus entirely positive. Wigner functions of other nonclassical states, for instance Fock states,  exhibit negative values. For this reason the Wigner function is called a \emph{quasi}-probability function.
\FloatBarrier
\begin{figure}[ht!!!!!!!!!!!!!!!]
  \vspace{-6mm}
  \includegraphics[width=9.6cm]{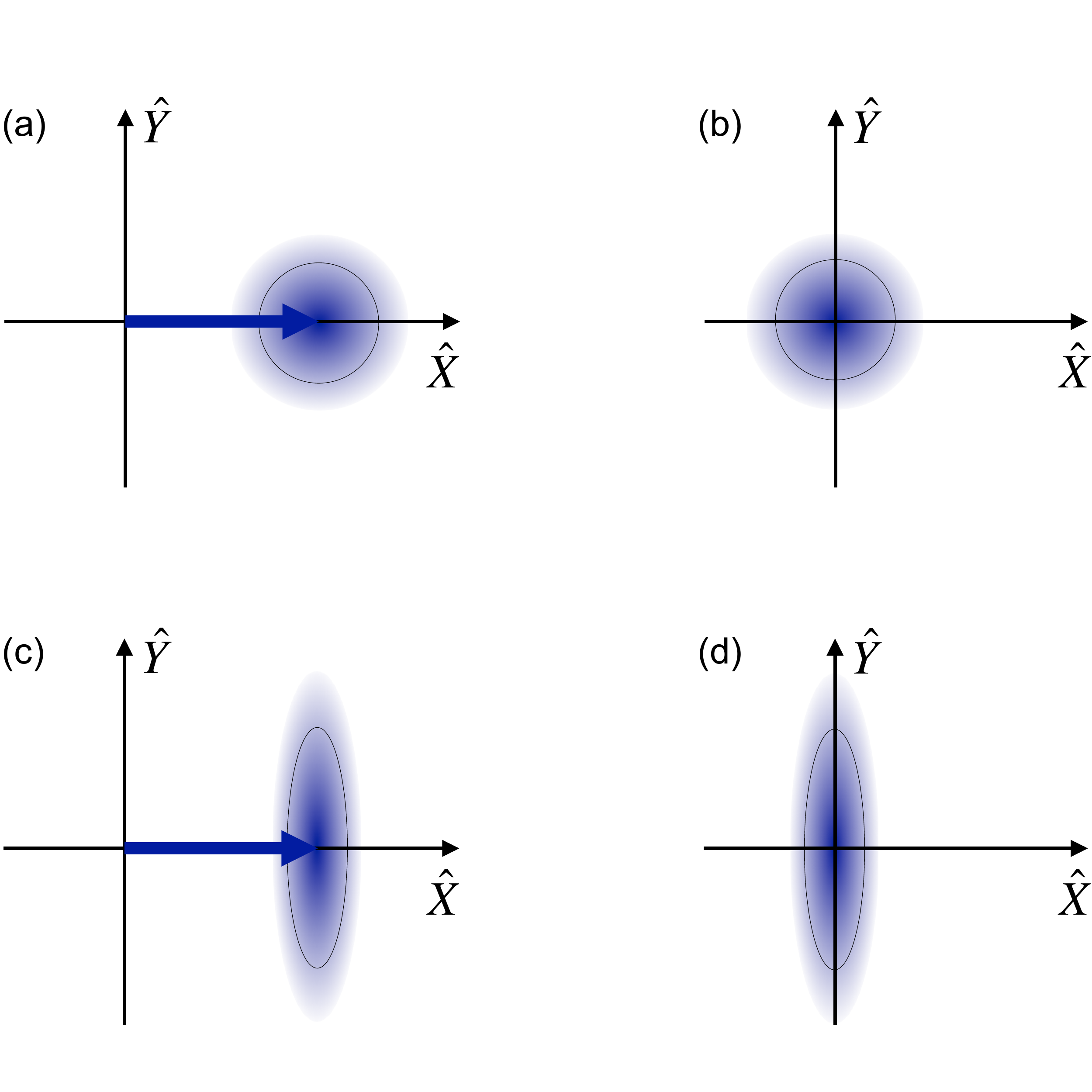}
  \vspace{-2mm}
\caption{\textbf{Simplified representation of Wigner functions} -- The darker the shaded areas, the larger is the phase-space quasi-probability. Shown are four different (time-independent) states of a modulation mode at frequency $\Omega$ for a specific resolution band width $\Delta \Omega$. Panel (a) represents a coherent state; the displacement ($\alpha$) corresponds to a classical amplitude modulation. Panel (b) represents the ground (vacuum) state, (c) a displaced squeezed state, and (d) a squeezed vacuum state, both with squeeze angle $\theta = 0$. The latter is in analogy to Fig.\,\ref{fig:8}. Again, the light field that carries the modulation is not part of the pictures.}
\label{fig:9}
\end{figure}

\begin{figure}[ht]
  \vspace{-2mm}
  \includegraphics[width=10.2cm]{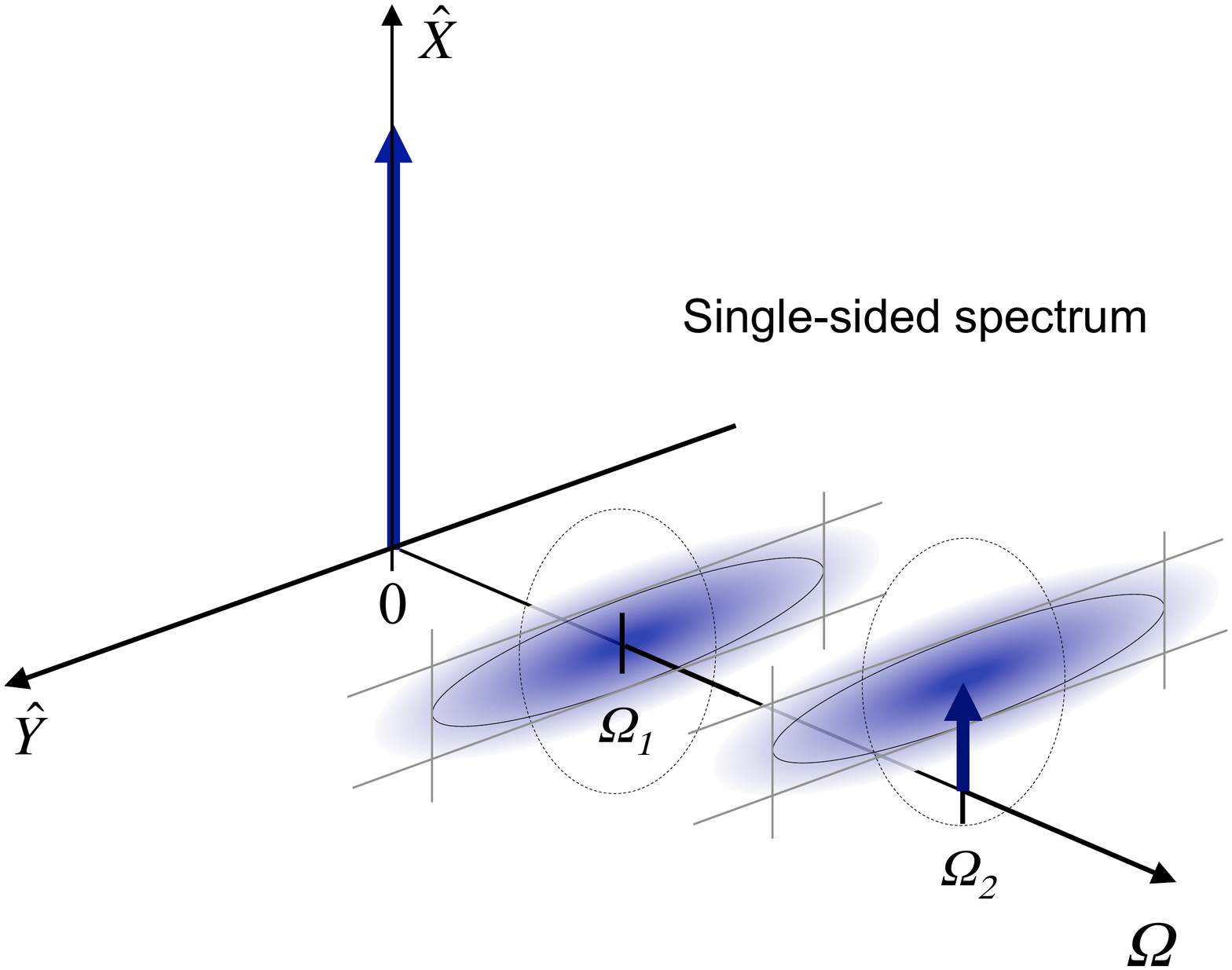}
  \vspace{-2mm}
\caption{\textbf{Hint of a Wigner function \emph{spectrum}} -- A single-sided spectrum (positive frequencies only) with respect to the carrier field can be used to visualize a broadband squeezed field. Shown are two examples displaying a squeezed vacuum state at $\Omega_1$ and a displaced amplitude squeezed state at $\Omega_2$.  The individual Wigner functions cover the resolution bandwidth $\Delta \Omega >0$ (not shown). In general, the squeezing strength as well as the squeeze angle and the displacement are a function of sideband frequency.   
}
\label{fig:10}
\end{figure}

Fig.~\ref{fig:9} shows the Wigner functions for (a) a coherent state, (b) the ground (vacuum) state, (c) a displaced squeezed state, and (d) a squeezed vacuum state. All Wigner functions describe a modulation of the carrier light, at sideband frequency $\Omega$ integrated over the frequency interval $\Delta \Omega$. The carrier light is not part of these Wigner functions. The displacement in (a) represents a classical amplitude modulation. (b) corresponds to the absence of any photons with a frequency offset of $\pm \Omega$ from the local oscillator field. (c) and (d) represent states whose amplitude modulation depth is more precisely defined than that of the ground state.   
Fig.~\ref{fig:10} shows Wigner function spectrum for a broadband squeezed vacuum field. Every Wigner function describes the modulation field at some modulation frequency $\Omega_i$ integrated over the resolution bandwidth (RBW) of $\Delta \Omega$.\\ 

\emph{The Glauber-Sudarshan $P$-function} -- The $P$-function [\cite{Glauber1963,Sudarshan1963}] is calculated by de-convoluting the Wigner function from the ground state uncertainty [\cite{Gerry2005}]. For displaced vacuum states (coherent states) the $P$-function corresponds to a displaced $\delta$-function. The mathematical expression of the $P$-function of a squeezed state contains infinitely high orders of derivatives of the $\delta$-function [\cite{Vogel2006}]. Such a function contains negativities but cannot be displayed. 
It is possible, however, to define a phase-space quasi probability function for squeezed states that can be displayed and that does show negativities as a sufficient and necessary condition for certifying the squeezing effect. This `\emph{nonclassicality function}' is calculated by de-convoluting the Wigner function from an uncertainty distribution that is steeper than the Gaussian distribution. A pronounced negativity of a squeezed vacuum state of up to 69 standard deviations was found [\cite{Kiesel2011a}].\\
\begin{figure}[h!!!!!!!!!!]
  \vspace{-11mm} 
  \includegraphics[width=11.1cm]{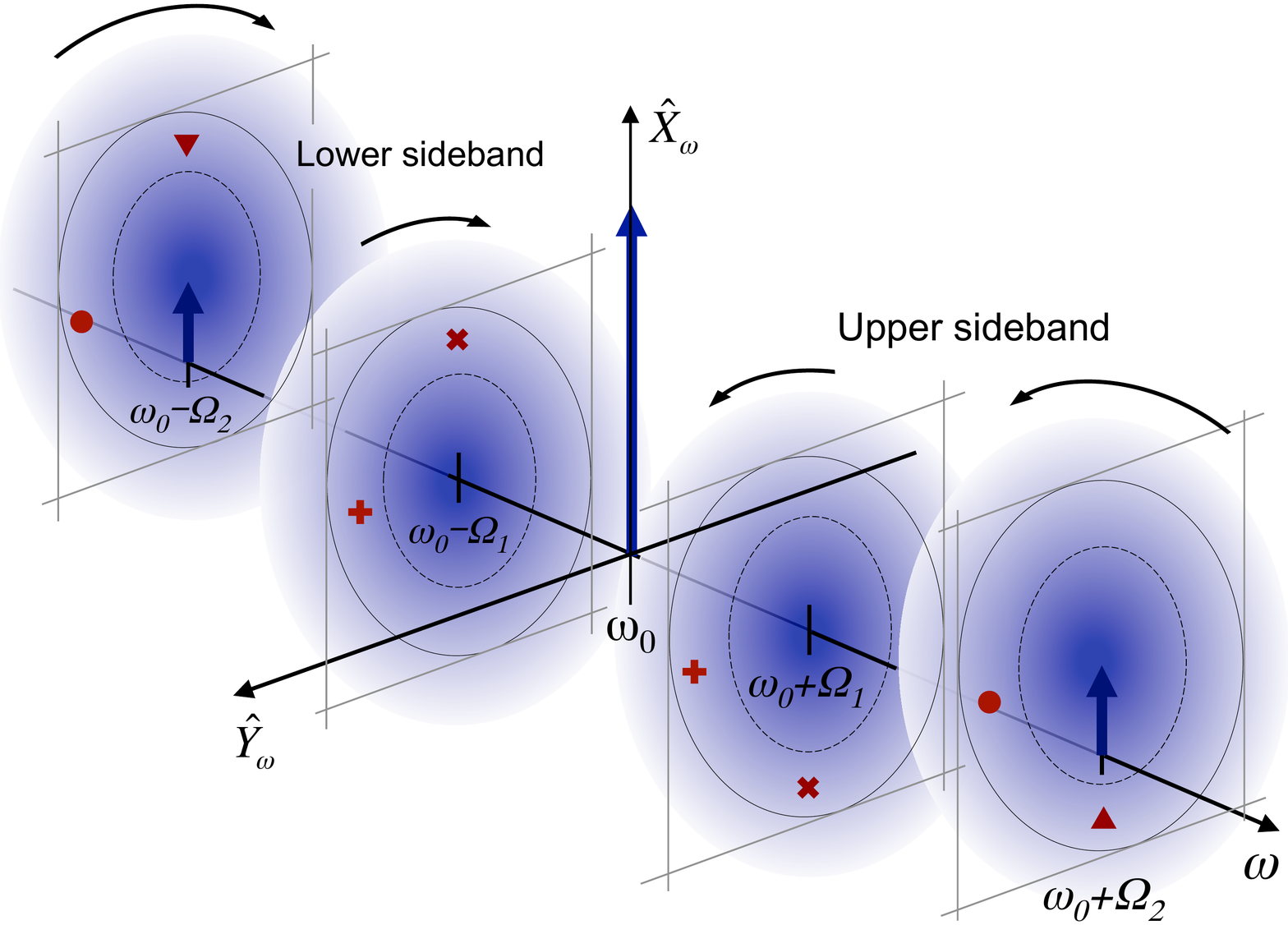}\\ [-1mm]
  \includegraphics[width=11.1cm]{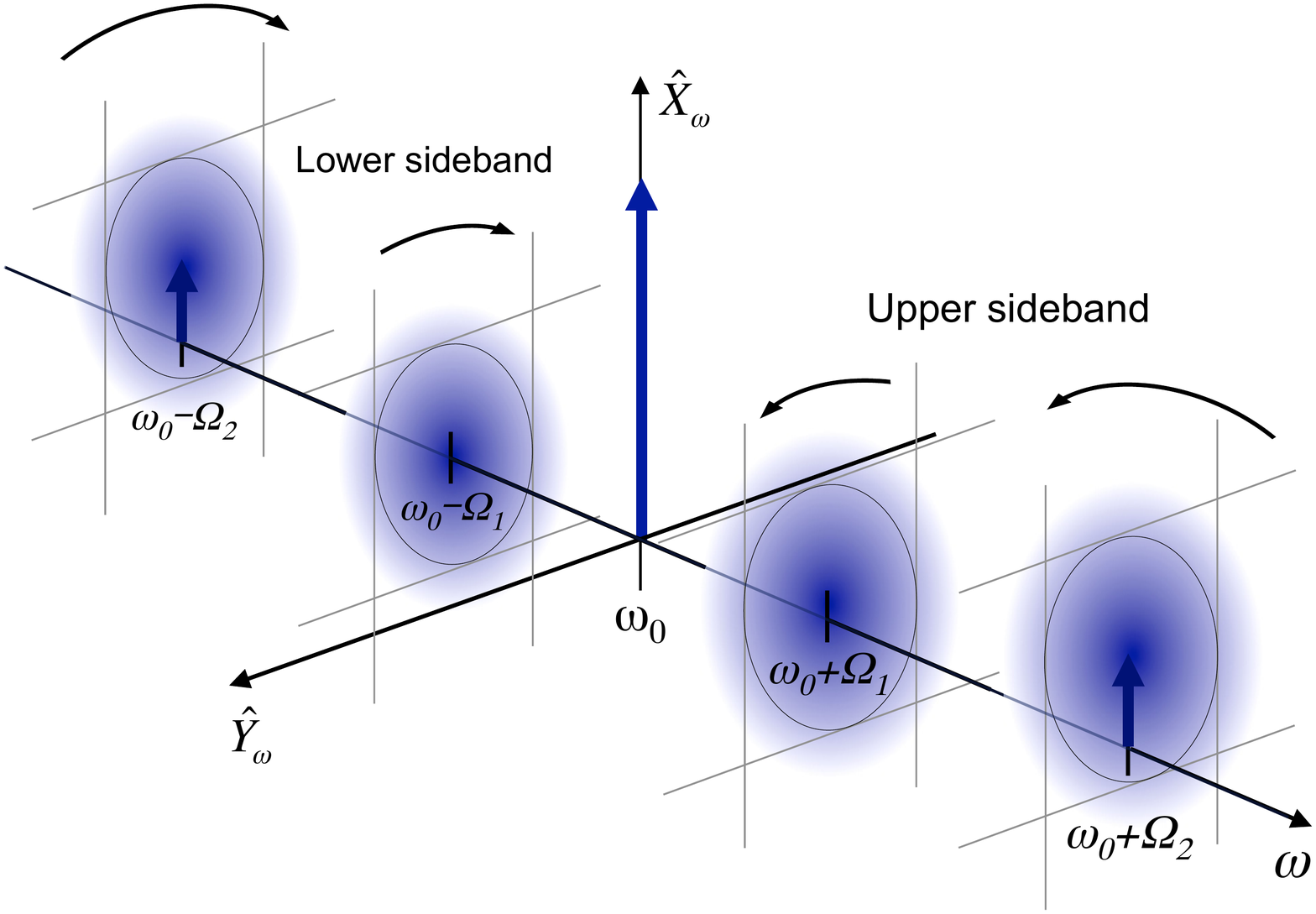}
    \vspace{-5mm}
\caption{\textbf{Double-sided phasor pictures} -- Phase spaces at optical frequency $\omega_0 \pm \Omega_i$ rotate around the frequency axis with sideband frequency $ \pm \Omega_i$. Its sign determines the direction of rotation with respect to the local oscillator in the rotating frame at $\omega_0$. A pair of phase spaces need to be superposed to provide a description of a modulation field at $\left | \Omega_i \right |$. 
Top: Amplitude quadrature squeezed field with a coherent displacement at $\left | \Omega_2 \right |$. The displacement corresponds to a classical amplitude modulation. The uncertainties of all optical frequencies are circular and larger than that of the ground state (dashed). Quantum correlations are indicated by additional symbols. More details are given in the main text.
Bottom: Corresponding spectrum of (displaced) vacuum states, which do not have any quantum correlations.}
\label{fig:11}
\end{figure}

\emph{The double-sided phasor picture} -- This phasor picture links quantum states of modulations with the quantum states of the contributing optical fields [\cite{Bachor2004}] and is mathematically described by the two-photon-formalism [\cite{Caves1985a,Schumaker1985}]. 
Generally, a weak amplitude or phase modulation at frequency $\Omega$ of a carrier field at optical frequency $\omega$ can be understood as the carrier's beat with two optical frequencies at $\omega \pm \Omega$. 
The double-sided phasor picture is able to display a spectrum of different and independent modulation frequencies in the rotating frame of the carrier field. The carrier light field is time-independent but the upper and lower sidebands are not. They rotate with $\pm \Omega_i / (2\pi)$, respectively, around the frequency axis. 

\FloatBarrier

Fig.~\ref{fig:11} shows such a double-sided phase space picture where the carrier's modulation at $\Omega_1$ is in a squeezed vacuum state and where the modulation at $\Omega_2$ is in a displaced squeezed state. The picture shows how a classical amplitude modulation as well as the quantum statistic of a modulation field is decomposed into contributions from upper and lower sidebands. For a squeezed modulation field the upper and lower sidebands show no squeezed but circular, thermally excited quantum uncertainties. The uncertainties of a pair of sidebands, however, show correlations as well as anti-correlations. In Fig.~\ref{fig:11} these (anti-) correlations are marked with $\boldsymbol \times$ and $\boldsymbol +$ for the modulation frequency $\Omega_1$ and with $\blacktriangle$ and \raisebox{-.3ex}{{\Large{$\bullet$}}} for the modulation frequency $\Omega_2$.

~\\
\subsection{Covariance matrix representation of (single-party) squeezed states}

Since squeezed states have a \emph{Gaussian} quantum statistic, four numbers are sufficient for their full description. These numbers are the second moment of the quadrature amplitude showing the strongest squeezing, and the second moment of its orthogonal quadrature amplitude, as well as their first moments describing the displacement. These four numbers are sufficient to calculate the Wigner function shown in Fig.~\ref{fig:8}. In general the quadrature of strongest squeezing is not perfectly aligned with one of the axes of the measurement's coordinate system. The so-called covariance matrix 
$ (V_{XX} V_{XY}; V_{YX} V_{YY})$ [\cite{Simon1994}] accounts for phase space rotations and enables the calculation of how these states evolve within an interferometric arrangement. Their components are normalized to the vacuum noise variance $\Delta^2 \hat X_{\rm vac} = \Delta^2 \hat Y_{\rm vac}$ and read
\begin{equation}
V_{XY} = \frac{\left<\hat X \hat Y + \hat Y \hat X \right> - 2 \left<\hat X\right>\left<\hat Y\right>}{ 2\Delta^2 \hat X_{\rm vac}  }\, .
\label{eq:CVM2}
\end{equation}

The following examples represent the ground state, a pure 10\,dB amplitude quadrature squeezed state and a pure 10\,dB squeezed state with a squeeze angle of 45$^\circ$,

\begin{center}
\vspace{-7mm}
\begin{equation}\label{eq:r1}
\hspace{-11mm}
{\bf V}_{\rm{vac}} \!=\!
\parenth{\!\begin{array}{cc}
1	&0	\\
0	&1	
\end{array}\!}, \hspace{4mm}
{\bf V}^{0^\circ}_{\rm{0.1}} \!=\!
\parenth{\!\begin{array}{cc}
0.1	&0	\\
0	&10	
\end{array}\!}, \hspace{4mm}
{\bf V}^{45^\circ}_{\rm{0.1}} \!=\!
\parenth{\!\begin{array}{cc}
5.05	& 4.95	\\
4.95	& 5.05	
\end{array}\!} \, ,
\end{equation}
\end{center}
with {\bf V}$^{45^\circ}_{\rm{0.1}}=${\bf R}$^{\rm T}_{45^\circ}${\bf V}$^{0^\circ}_{\rm{0.1}}${\bf R}$_{45^\circ}$, where {\bf R}$_\alpha = ({\rm cos}\alpha \, -{\!\rm sin}\alpha; \,{\rm sin}\alpha \; {\rm cos}\alpha)$ is the rotation matrix.
%
%
~\\
\subsection{Phase space representation of two-mode (bi-partite) squeezed states} \label{ssec:EPR}
\begin{figure}[h!]
  \vspace{-3mm}
  \includegraphics[width=9.6cm]{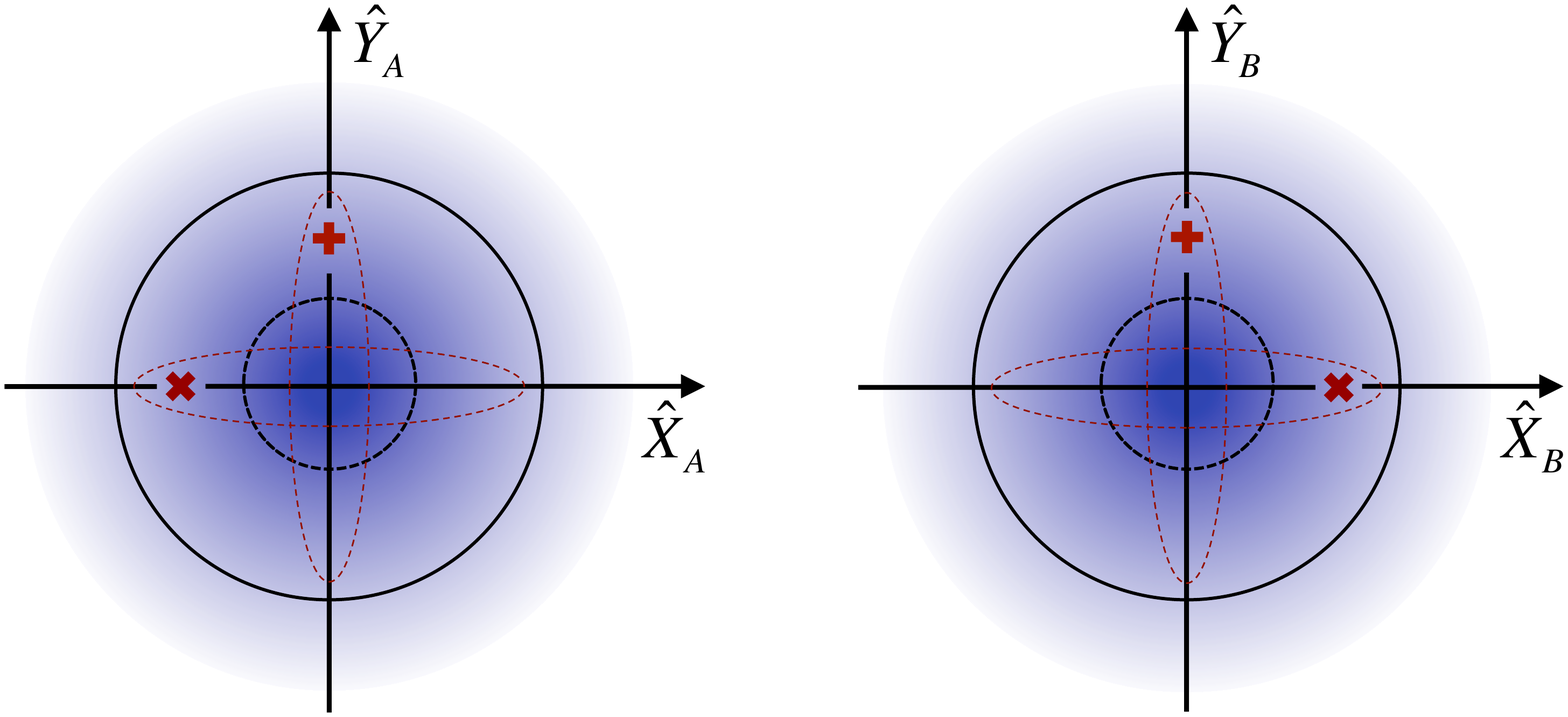}
  \vspace{-14mm}
\caption{\textbf{Bi-partite squeezed vacuum state} -- Shown is a Wigner-function-based phase space representation, in close analogy to the single party version in Fig.~\ref{fig:9}\,(d). The picture describes a single modulation at frequency $\Omega$ with bandwidth $\Delta \Omega$. Measurements at party $A$ and $B$ reveal local Wigner functions that correspond to thermal states, since the uncertainties (indicated by the color and by the large circles) are symmetric and larger than that of the ground state (indicated by the small dashed circles). The uncertainties, however, show correlations and anti-correlations, here indicated by $+$ and $\times$, respectively. The strength of these (anti-)correlations are indicated by ellipses. Bi-partite squeezing, i.e.~entanglement, is present, if the short axes of the ellipses are shorter than the diameter of the ground state uncertainty.
The picture, in fact, represents Einstein-Podolsky-Rosen entanglement [\cite{Einstein1935}]. From a measurement of $\hat X_A$ or $\hat Y_A$, the corresponding measurement result at party $B$ can always be inferred with an uncertainty that is smaller than the ground state uncertainty.
}
\label{fig:12}
\end{figure}
\FloatBarrier
A bi-partite state enables a measurement on subsystem $A$ and simultaneous a measurement on subsystem $B$. For a large number of simultaneous ensemble measurements of the same quadrature amplitude $\hat X_\vartheta$ the following two joint quadrature variance can be calculated
\begin{equation} \label{eq:Duan}
\Delta^2 (\hat X_{\vartheta}^A \pm \hat X_{\vartheta}^B).
\end{equation}
A state that is symmetrically shared between two parties ($A$ and $B$) is called a \emph{two-mode squeezed state} if the variances of joint quadrature measurements fulfill the following inequality [\cite{Duan2000}], i.e.
\begin{equation} \label{eq:Duan}
\frac{\Delta^2 (\hat X^A - \hat X^B)}{\Delta^2 (\hat X^A_{\rm vac} - \hat X^B_{\rm vac})} + 
\frac{\Delta^2 (\hat Y^A + \hat Y^B)}{\Delta^2 (\hat Y^A_{\rm vac} + \hat Y^B_{\rm vac})} < 2 \, ,
\end{equation}
with $\Delta^2 (\hat X^A_{\rm vac} \pm \hat X^B_{\rm vac}) = \Delta^2 (\hat Y^A_{\rm vac} \pm \hat Y^B_{\rm vac}) = 2 \Delta^2 \hat X_{\rm vac}$. 
A `two-mode squeezed state' reveals entanglement in the second moments of the measurement statistics. It is thus a `bi-partite Gaussian entangled state'. 

Fig.~\ref{fig:12} displays a (pure) bi-partite  squeezed vacuum state distributed between $A$ and $B$. The state shows full symmetry regarding its subsystems at the two sites. The large circles and the colored area represent Wigner functions of the subsystems. Measurements of the quadrature amplitudes $\hat X_A$, $\hat Y_A$, $\hat X_B$, and $\hat Y_B$ show identical variances and the correlations and anti-correlations have identical strength since $\Delta^2 (\hat X_A - \hat X_B) = \Delta^2 (\hat Y_A + \hat Y_B) < 1/2$ for our normalization of quadrature amplitudes having a ground state variance of $1/4$.

Generally, a symmetric bi-partite squeezed state fulfills another quantitative (Gaussian) entanglement criterion if less than 50\% of the vacuum state is symmetrically mixed into the initially pure state. Bi-partite squeezed states are always entangled but in this case they are even Einstein-Podolsky-Rosen (EPR) entangled [\cite{Reid1989}], allowing the demonstration of the quantum steering effect [\cite{Einstein1935,Schroedinger1935,Reid1989,Cavalcanti2009}]. The first such experiment was performed by Ou et al. [\cite{Ou1992}] using type\,II parametric down-conversion (PDC). Later experiments produced bi-partite squeezed vacuum states by overlapping two squeezed vacuum states, each produced with type\,I PDC, on a balanced beam splitter and used the entangled output for the demonstration of quantum teleportation [\cite{Furusawa1998,Bowen2003,Bowen2003b}]. The criterion in Eq.~(\ref{eq:Duan}) and the EPR criterion from [\cite{Reid1989}] was experimentally compared in Ref.~[\cite{Bowen2003a}]. The steering effect in asymmetric bi-partite squeezed states were recently experimentally characterized in Ref.~[\cite{Haendchen2012}].

Fig.~\ref{fig:12} shows features similar to those in the top part of Fig.~\ref{fig:11}. This is not a coincidence and shows that a bi-partite squeezed state can also be generated by spatially splitting the upper and lower sideband of a (single-party) squeezed state. This was first experimentally demonstrated by the group of E. Polzik [\cite{Schori2002}] and later used for EPR multiplexing of a single longitudinal mode of a squeezing resonator [\cite{Hage2010}].\\

\FloatBarrier
\subsection{Covariance matrix representation of bi-partite squeezed states}

Also the full information of bi-partite states, including the entanglement, can be cast by the covariance matrix [\cite{Simon1994}], which can be used to calculate the propagation of these states in laser interferometers.
Again all variances are normalized to the vacuum noise variance, in full analogy to Eq.~(\ref{eq:CVM2}). The generic bi-partite covariance matrix has dimension 4$\times$4 and reads
\begin{center}
\vspace{2mm}
{\bf V}$^{\rm{bp}} \!=\!$
\parenth{\!\begin{array}{cccc}
V_{X_A X_A}		&V_{X_A Y_A}		&V_{X_A X_B}		&V_{X_A Y_B}\\
V_{Y_A X_A}		&V_{Y_A Y_A}		&V_{Y_A X_B}		&V_{Y_A Y_B}\\
V_{X_B X_A}		&V_{X_B Y_A}		&V_{X_B X_B}		&V_{X_B Y_B}\\
V_{Y_B X_A}		&V_{Y_B Y_A}		&V_{Y_B X_B}		&V_{Y_B Y_B}
\end{array}\!},\hspace{1mm} with \hspace{15mm} \vspace{3mm}
\begin{equation}
V_{X_AY_B} = \frac{\left<\hat X_A \hat Y_B + \hat Y_B \hat X_A \right> - 2 \left<\hat X_A\right>\left<\hat Y_B\right>}{ 2\Delta^2 \hat X_\vartheta^{\rm vac}  }.
\label{eq:CVM4}
\end{equation}
\vspace{3mm}
\end{center}
Due to the symmetry in Eq.~(\ref{eq:CVM4}), the 4$\times$4 covariance matrix is fully specified by just ten independent coefficients. If the phase spaces at $A$ and $B$ are aligned along the strongest correlations and anti-correlations, the matrix components referring to different quadrature amplitudes, e.g.~$V_{X_A Y_A}$, are zero. Such entangled states can be produced by overlapping two squeezed fields with a squeeze angle difference of 90$^\circ$ on a balanced beam splitter.

A symmetric bi-partite squeezed vacuum state, which is also called an `S-class' [\cite{DiGuglielmo2007}] bi-partite squeezed vacuum state, shows (anti-)correlations in two joint quadratures as defined in Eq.~(\ref{eq:Duan}). For a pure such state of 10\,dB squeezing, the covariance matrix reads 
\begin{center}
\vspace{2mm}\hspace{0mm}
{\bf V}$_{\rm{10|10}}^{\rm{bp}} = $
\parenth{\!\begin{array}{cccc}
5.05		\!&0		\!&4.95	\!&0\!\\
0		\!&5.05	\!&0		\!&-4.95\!\\
4.95		\!&0		\!&5.05	\!&0\!\\
0		\!&-4.95	\!&0		\!&5.05\!
\end{array}\!}.\\
\vspace{3mm}
\end{center}
The following covariance matrix describes a so-called `V-class' 10\,dB bi-partite squeezed vacuum state. Here, only one joint quadrature shows 10\,dB squeezing whereas the orthogonal joint quadrature shows vacuum noise. The state is obtained by overlapping one 10\,dB squeezed state with a vacuum state on a balanced beam splitter.
\begin{center}
\vspace{2mm}\hspace{0mm}
{\bf V}$_{\rm{0|10}}^{\rm{bp}} \!=\!\!$
\parenth{\!\begin{array}{cccc}
0.55		\!&0		\!&0.45	\!&0\!\\
0		\!&5.5	\!&0		\!&-4.5\!\\
0.45		\!&0		\!&0.55	\!&0\!\\
0		\!&-4.5	\!&0		\!&5.5\!
\end{array}\!}.\\
\vspace{3mm}
\end{center}
The first measurement of all elements of such a covariance matrix was achieved in [\cite{DiGuglielmo2007}].\\

\FloatBarrier
\subsection{Photon numbers of squeezed states} \label{ssec:N}

In contrast to the ground state, squeezed vacuum states do have photon excitations. As said earlier, quantum theory links the wave and the particle pictures. Indeed, the squeeze factor of a modulation mode is directly connected to a certain photon number excitation. Squeezed states of light are produced via spontaneous photon pair generation, e.g. by parametric down-conversion. The following operator $\hat S$ is called the `squeeze operator' [\cite{Gerry2005}]. It creates and annihilates photon pairs,
\begin{equation}
\left |r,\theta\right> = \hat S(r,\theta) \left |0 \right> \, ,
\end{equation}
where $\left |r,\theta\right>$ is a squeezed vacuum state with squeeze parameter $r$ and squeeze angle $\theta$, and $\left |0 \right>$ is the vacuum state. The definition of the squeeze operator is
\begin{equation}
\hat S(r,\theta) = {\rm exp} \left[ \frac{1}{2} \left(r {\rm e}^{-i \theta} \hat a^2 -  r {e}^{i \theta} \hat a^{\dagger\,2}  \right ) \right] \, .
\end{equation}
The following shows that this definition indeed results in a state with squeezed quadrature amplitude variances. Lets set $\theta = 0$
\begin{equation}
\left <0\right| \hat S^\dagger(r,0) \,\hat X\, \hat S(r,0) \left | 0 \right> = \frac{1}{2} \left <0\right| \hat S^\dagger(r,0) \left(\hat a + \hat a^\dagger \right)  \hat S(r,0) \left | 0 \right> \, ,
\label{eq:SX}
\end{equation}
\begin{equation}
\left <0\right| \hat S^\dagger(r,0)\, \hat Y\, \hat S(r,0) \left | 0 \right> = \frac{i}{2} \left <0\right| \hat S^\dagger(r,0) \left(\hat a - \hat a^\dagger \right)  \hat S(r,0) \left | 0 \right> \, .
\label{eq:SY}
\end{equation}
Using the Baker-Hausdorff formula we get
\begin{equation}
\hat S^\dagger(r,\theta) \,\hat a \,\hat S(r,\theta) = \hat a\, {\rm cosh}\,r - \hat a^\dagger {e}^{i \theta} {\rm sinh}\,r \, ,
\label{eq:BH1}
\end{equation}
\begin{equation}
\hat S^\dagger(r,\theta)\, \hat a^\dagger\, \hat S(r,\theta) = \hat a^\dagger {\rm cosh}\,r - \hat a {e}^{-i \theta} {\rm sinh}\,r \, .
\label{eq:BH2}
\end{equation}
Since $\left< 0 \right| \hat X \left| 0 \right> = \left< 0 \right| \hat Y \left| 0 \right> = 0$ , also Eqs.~(\ref{eq:SX}) and (\ref{eq:SY}) are zero. To finally calculate the variances we need
\begin{equation*}
\left <0\right| \hat S^\dagger(r,0) \,\hat X^2\, \hat S(r,0) \left | 0 \right> = \frac{1}{4} \left <0\right| \hat S^\dagger(r,0) \left(\hat a^2 + \hat a^\dagger \hat a + \hat a \hat a^\dagger + \hat a^{\dagger 2} \right)  \hat S(r,0) \left | 0 \right> \, ,
\label{eq:SX2}
\end{equation*}
\begin{equation*}
\left <0\right| \hat S^\dagger(r,0) \,\hat Y^2\, \hat S(r,0) \left | 0 \right> = -\frac{1}{4} \left <0\right| \hat S^\dagger(r,0)\left(\hat a^2 - \hat a^\dagger \hat a - \hat a \hat a^\dagger + \hat a^{\dagger 2} \right)  \hat S(r,0) \left | 0 \right> \, .
\label{eq:SY2}
\end{equation*}
Given that $\hat S \hat S^\dagger$ is the identity, and using again Eqs.~(\ref{eq:BH1}) and (\ref{eq:BH2}) we get the expected variances 
\begin{equation*}
\Delta^2 \hat X = \frac{1}{4} \left [ {\rm cosh}^2\,r - 2 {\rm cosh}\,r \; {\rm sinh}\,r  + {\rm sinh}^2\,r  \right] = \frac{1}{4}  {e}^{-2r} \, ,
\end{equation*}
\begin{equation*}
\Delta^2 \hat Y = \frac{1}{4} \left [ {\rm cosh}^2\,r + 2 {\rm cosh}\,r \; {\rm sinh}\,r  + {\rm sinh}^2\,r  \right] = \frac{1}{4} {e}^{2r} \, .
\end{equation*}

Since the squeeze operator can only create and annihilate photon \emph{pairs}, a squeezed vacuum state without photon loss must correspond to an even number of photons. But not only photon loss, also a coherent displacement leads to flattening out the odd-even oscillations.
The probability of detecting $N$ photons in a pure displaced squeezed state are derived for instance in [\cite{Gerry2005}] and read

\begin{figure}[ht!!!!!!!!!!!!]
  \vspace{0mm}
  \includegraphics[width=13.4cm]{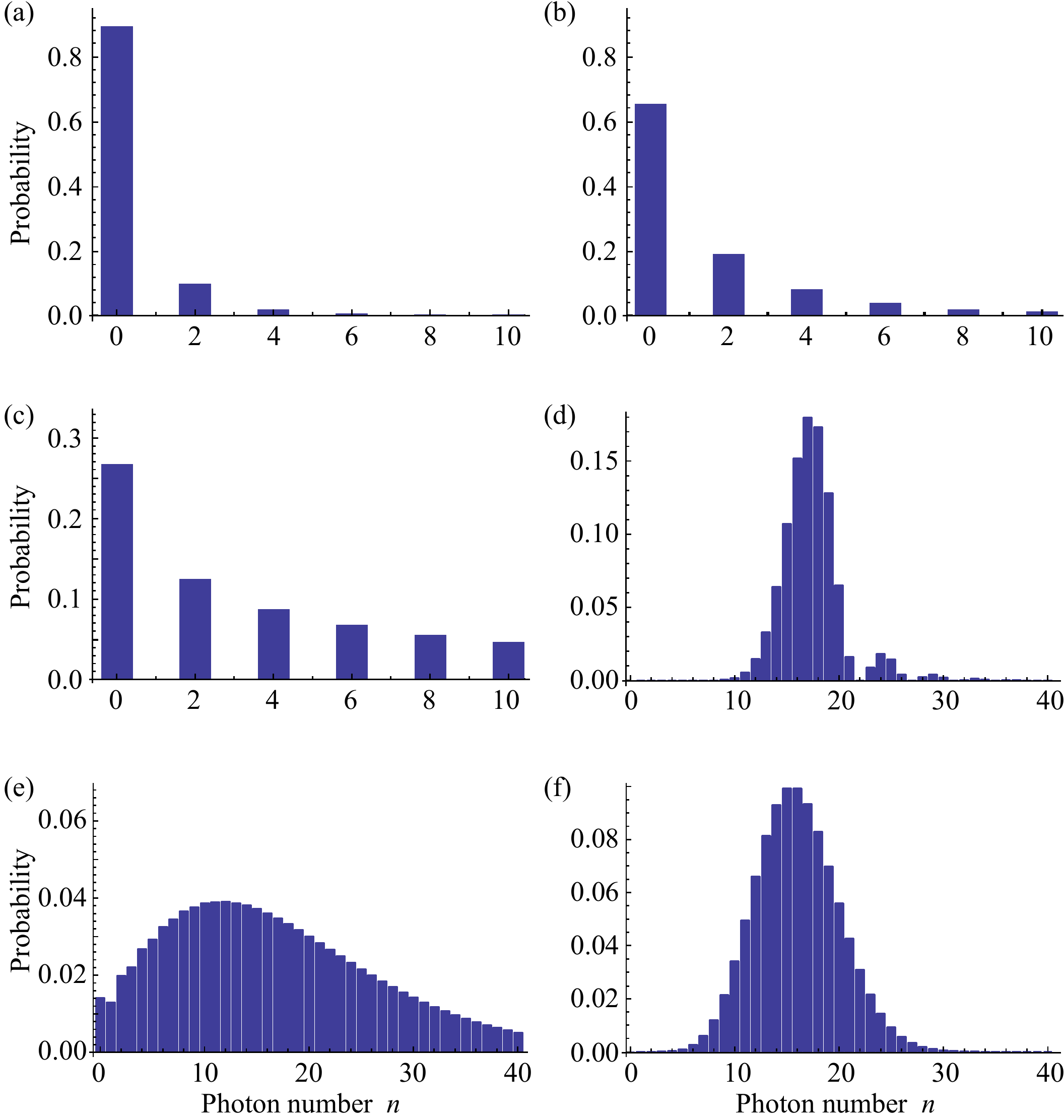}
  \vspace{1mm}
\caption{\textbf{Photon number distributions} -- All panels represent pure states. 
(a) 4.3\,dB squeezed vacuum state ($r=0.5, \,\alpha = 0$).
(b) 8.6\,dB squeezed vacuum state ($r=1, \,\alpha = 0$).
(c) 17.2\,dB squeezed vacuum state ($r=2, \,\alpha = 0$).
(d) displaced 8.6\,dB squeezed state ($r=1, \theta = 0, \,\alpha = 4$).
(e) displaced 8.6\,dB squeezed state ($r=1, \theta = \pi/2, \,\alpha = 4$).
(f) Coherent state ($r=0, \,\alpha = 4$).
The average photon numbers are  about 0.27, 1.4, 13.1, 17.4, 17.4, and exactly 16, see Eq.~(\ref{eq:nsqz}).
}
\label{fig:13}
\end{figure}

\begin{eqnarray*}
P(N) = \left| \left< N | \alpha,r,\theta \right> \right|^2 
        = \frac{\left ( 0.5 \,{\rm tanh}\, r \right)^N}{N! \,{\rm cosh} \, r} {\rm exp} \left[ - |\alpha|^2 -\frac{1}{2}(\alpha^{*2} e^{i\theta} + \alpha^{2} e^{-i\theta}) {\rm tanh}\, r \right] 
\end{eqnarray*}
\vspace{-5mm}
\begin{eqnarray} \label{eq:P}        
        \hspace{35mm} \times \left| H_N \left[ (\alpha {\rm cosh}\, r + \alpha^* e^{i \theta} {\rm sinh}\, r) \sqrt{e^{i \theta} {\rm sinh}(2r)}  \right] \right|^2 \, ,
\end{eqnarray}
where $H_N$ is the $N^{\rm th}$ Hermite polynomial.

Fig.~\ref{fig:13} shows the photon number distributions for 5 different pure squeezed states according to Eq.~(\ref{eq:P}). Panels (a) to (c) show squeezed vacuum states with 4.3\,dB, 8.6\,dB, and 17.2\,dB of squeezing. Panel (d) shows the more general case of a squeezed state with a coherent displacement $\alpha \ne 0$. Due to $\theta = 0$ the state is amplitude quadrature squeezed. Panel (e) refers to the corresponding phase quadrature squeezed state.
For comparison, panel (f) shows the photon number distribution of the coherent state with the same displacement. 

The panels in Fig.~\ref{fig:13} represent the diagonal elements of the state's density matrix in number basis. Only the latter also contains the coherences between photon numbers [\cite{Gerry2005}]. Figures as shown here generally do not give full descriptions of the states.\\

A squeezed vacuum state ($r\neq0$) always has a non-zero photon number and can not be the ground state.
The average photon number of a pure squeezed vacuum state can be calculated using Eq.~(\ref{eq:H}). With the maximally squeezed quadrature variance $\Delta^2 \hat X_{\rm sqz}$ the average photon number is given by
\begin{equation}
\label{eq:nsqz}
\overline{n} = \langle n \rangle_{\ket{\alpha=0,\theta,r}} =  \Delta^2 \hat X_{\rm sqz} + \frac{(\Delta^2 \hat X_{\rm sqz})^{-1}}{16} - \frac{1}{2}
=  \frac{e^{-2r}}{4} + \frac{e^{2r}}{4} - \frac{1}{2}
\, ,
\end{equation}
with the vacuum noise variance normalized to one quarter. A coherent displacement further adds $| \alpha |^2$ photons on average.

~\\
\FloatBarrier

\section{Squeezed-light generation} \label{Sec:4} 

\subsection{Overview}
Squeezed light was first produced in 1985 by Slusher \emph{et al.}\ using four-wave-mixing in sodium atoms in an optical cavity [\cite{Slusher1985}]. Shortly after, squeezed light also was generated by four-wave-mixing in an optical fibre [\cite{Shelby1986}] and by degenerate parametric down-conversion (PDC) in a 2$^{\rm nd}$-order nonlinear crystal placed in an optical cavity [\cite{Wu1986}]. The pumped cavity was operated below its oscillation threshold, i.e.~the parametric gain did not fully compensate the round trip losses, which is also called `cavity-enhanced optical-parametric amplification (OPA)'. 

The early day experiments achieved squeeze factors of a few percent up to about 3\,dB. Today, squeeze factors of more than 10\,dB are directly observed in several experiments  [\cite{Vahlbruch2008,Eberle2010,Stefszky2012,Vahlbruch2016}]. All of them are based on cavity-enhanced OPA (below threshold). The parametrically amplified mode is \emph{degenerate}, i.e.~signal and idler modes are identical. In particular, the down-conversion process is of `type\,I', which means that the amplified mode has a well-defined polarization. 
Squeezed states can also be generated \emph{above} oscillation threshold. In Refs.~[\cite{Villar2006,Jing2006}], bi-partite squeezing was generated with above-threshold PDC. Both experiments used type\,II PDC, which provides orthogonally polarized signal and idler fields. 
Type\,II PDC below threshold was also used to generate squeezed and bi-partite squeezed fields [\cite{Grangier1987,Ou1992}]. 
All these experiments were performed in the continuous-wave regime, which is also the focus of this Review.
Squeezed states of modulations of  trains of laser \emph{pulses}, however, have been also generated since the 1980s using either PDC or the optical Kerr effect [\cite{Slusher1987,Bergman1991,Ourjoumtsev2006,Dong2008}]. 
For an overview of the developments in squeezed-light generation in the continuous-wave as well as pulsed regime, see Ref.~[\cite{Bachor2004}].
Squeezed-light generation in opto-mechanical setups [\cite{Aspelmeyer2014}], which use the intensity dependent phase shift from radiation pressure, was   discussed in Refs.~[\cite{Pace1993,Rehbein2005,Corbitt2006}] and recently experimentally achieved by several groups [\cite{Brooks2012,Safavi-Naeini2013,Purdy2013b}].  

\begin{figure}[th!!!!!!!!!!!!!!!!]
  \vspace{-3mm}
  \includegraphics[width=12.8cm]{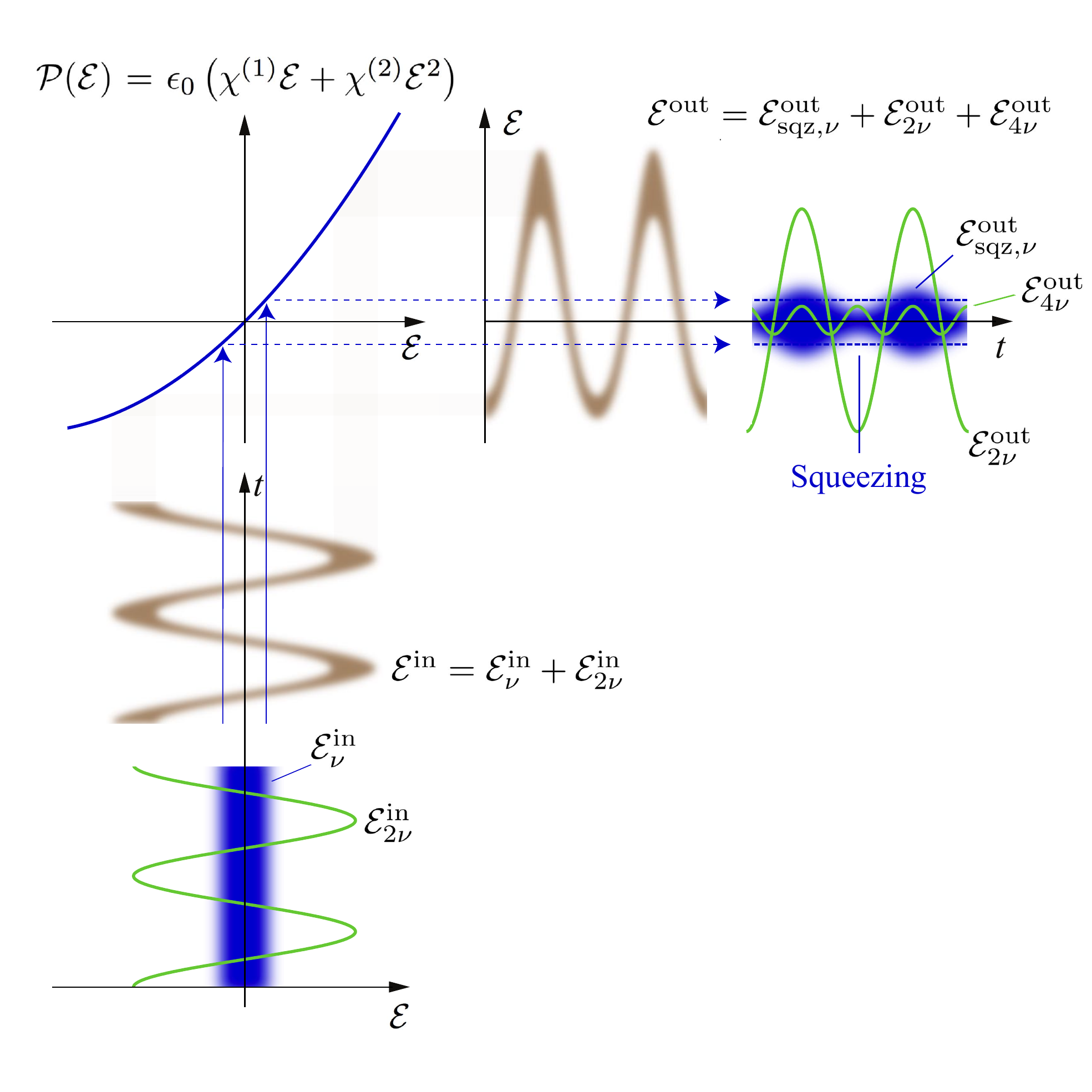}
  \vspace{-4mm}
\caption{\textbf{Optical parametric amplification of a vacuum state} -- The upper left corner shows the crystal polarization ${\cal P}({\cal E}) =  \epsilon_0 \left(\chi^{(1)}{\cal E}  +  \chi^{(2)}{\cal E}^2\right)$, i.e.~the separation of charge carriers by the electric component of an optical field ${\cal E}$. The graph illustrates how an input quantum field (from below) is projected into an output quantum field (towards the right). In the example shown here, the input field is composed of a classical pump field ${\cal E}^\text{in}_{2\nu}$ at frequency $2\nu$ and zero-point fluctuations ${\cal E}^\text{in}_{\nu}$ of a field at frequency $\nu$, cf. Fig.\,\ref{fig:7}(c). The superposition ${\cal E}^\text{in}$ of these two fields is transferred into a time-dependent dielectric polarization that is the source of (and thus directly proportional to) the electric component of the output field ${\cal E}^\text{out}$. The quantum uncertainty of the output field shows a phase-dependent (parametric) amplification at frequency $2\nu$. Spectral decomposition of the output field ${\cal E}^\text{out}$ reveals coherent amplitudes at frequencies $2\nu$ and $4\nu$ and a squeezed vacuum state ${\cal E}^\text{out}_{\text{sqz,}\nu}$, cf. to Fig.\,\ref{fig:7}(d). The concept of this figure was published in Ref.~[\cite{Bauchrowitz2013}].
} 
\label{fig:14}
\end{figure}

\subsection{Degenerate type\,I optical-parametric amplification (OPA)} 

This section provides a graphical description of how degenerate type\,I OPA/PDC turns a vacuum state into a squeezed vacuum state, and a coherent state into a displaced squeezed state. The process requires a bright pump field and a 2$^{\rm nd}$-order nonlinear crystal. For simplicity we set all nonlinearities above 2$^{\rm nd}$-order to zero.
\begin{figure}[h!!!!!]
  \vspace{-3mm}
  \includegraphics[width=12.8cm]{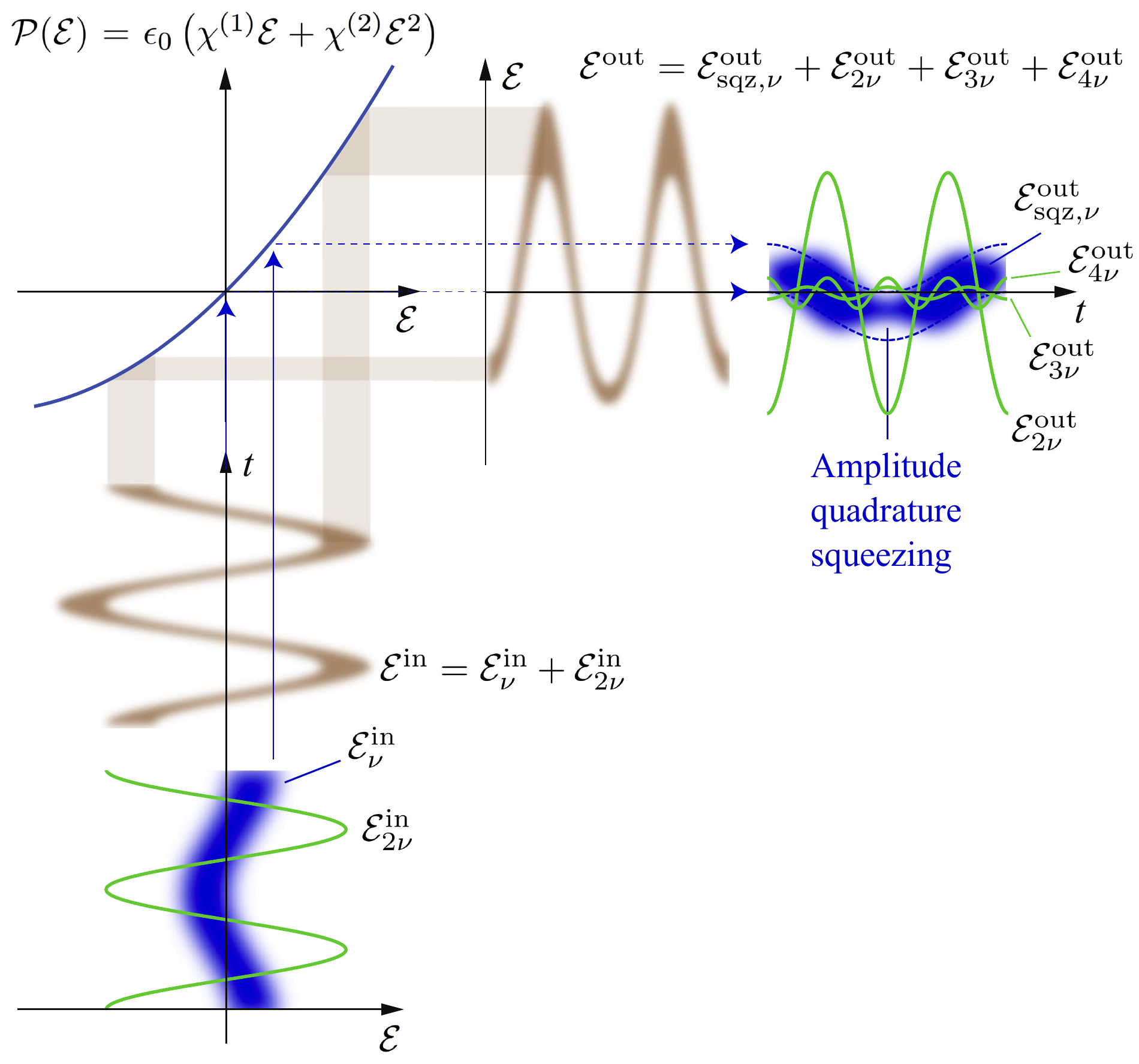}
  \vspace{-4mm}
\caption{\textbf{Optical parametric amplification of a coherent state} -- The picture shows how a displaced vacuum state is transformed into a displaced squeezed state. The pump's electric field is maximal when the input field at fundamental frequency $\nu$ shows a zero crossing. The phase relation described results in an output state that is amplitude quadrature squeezed. If the pump field at the input in phase was shifted by half of its wavelength, the squeezed output field were a phase quadrature squeezed. The squeezing generation displayed here corresponds to the transition from Fig.\,\ref{fig:7}(a) to Fig.\,\ref{fig:7}(b), but with an additional reduction of the displacement. The concept of this figure was published in Ref.~[\cite{Bauchrowitz2013}].
} 
\label{fig:15}
\end{figure}

Let us  consider a short segment of the second-order nonlinear crystal, pumped with light of optical frequency $2\nu$. All other modes that enter the crystal shall not contain any photons, i.e.~are in their vacuum states. Of these, the only mode of interest is that at optical frequency $\nu$, which spatially overlaps with the pump mode.
Fig.\,\ref{fig:14} shows the total electric field of the optical input $\cal E^{\rm in}$ and the 2$^{\rm nd}$-order nonlinear dielectric polarisation of the crystal $\cal P$. The latter is proportional to the total electric field of the output $\cal E^{\rm out}$. The pump field at $2\nu$ periodically drives the vacuum field at $\nu$ between regions of low and high polarisation. This process transforms the vacuum state into a squeezed vacuum state in the output [\cite{Bauchrowitz2013}].
The output further contains the hardly depleted pump field and frequency doubled parts of the pump field at $4\nu$. 
It is again emphasized that Fig.\,\ref{fig:14} displays OPA in a small segment of the crystal. In reality the nonlinear effect accumulates over the crystal length, or even over several passages, since the crystal is usually put into an optical resonator. A noticeable effect is achieved if all infinitesimal contributions constructively interfere. This is achieved in case of \emph{phase matching}, i.e.~if the wave fronts of the modes at $2\nu$ and $\nu$ propagate with the same speed and thus do not run out of phase. 
Note that in actual squeezing experiments the $4\nu$ component is usually suppressed by phase miss-matching. 

\begin{figure}[h!!!!!]
  \vspace{0mm}
  \includegraphics[width=12.6cm]{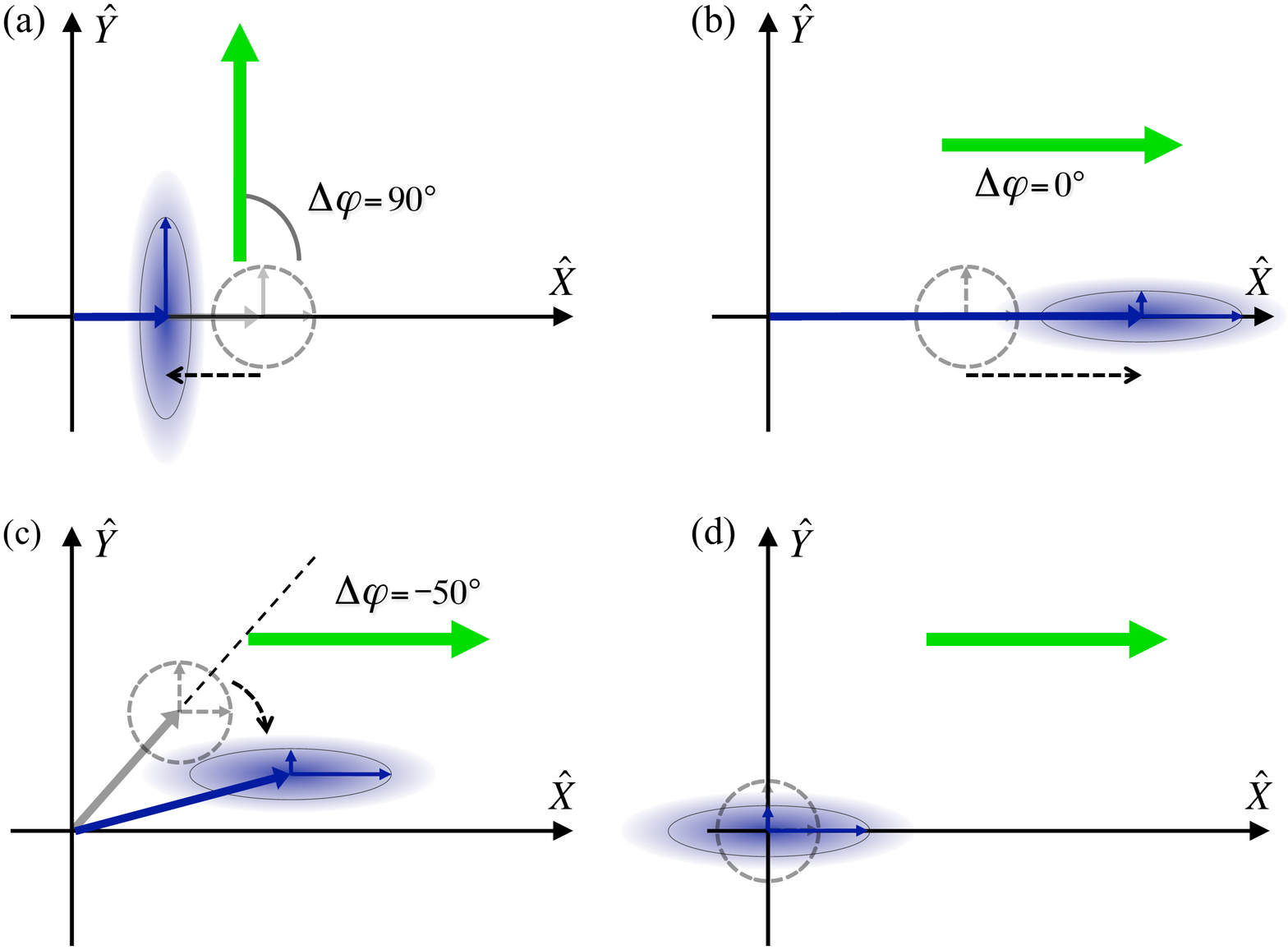}
  \vspace{-1mm}
\caption{\textbf{Phase-space illustration of degenerate OPA} --  The (displaced) dashed circle in each diagram represents the uncertainty of the initial state at optical frequency $\nu$. The (displaced) shaded area represents the state after degenerate optical parametric amplification. All quantum uncertainties shown correspond to pure states. The bold green arrow describes the bright second-harmonic pump field, whose uncertainty can be neglected. The phase between the 2$^{\rm nd}$ harmonic pump and the initial state ($\Delta \varphi = \varphi_{2\nu} - 2 \varphi_{\nu}$) determines the result of the parametric amplification.
}
\label{fig:16}
\end{figure}

Fig.\,\ref{fig:15} shows the same process, but now for an input field at frequency $\nu$ in a coherent state. In this case the relative phase between the two input states $\Delta \varphi = \varphi_{2\nu} - 2 \varphi_{\nu} $ is relevant. In  Fig.\,\ref{fig:15} the relative phase is set such that the expectation value of the field at frequency $\nu$ is zero when the pump field reaches its maximum ($\Delta \varphi = 90^\circ$). The output at the fundamental frequency is then an amplitude squeezed state, with a deamplified coherent amplitude.

Fig.\,\ref{fig:16} summarizes the squeezing operation on the vacuum state as well as on displaced vacuum states for different phase relations $\Delta \varphi$ between the two input fields.\\

\subsection{Cavity-enhanced OPA} 

Placing the nonlinear crystal inside a cavity can greatly enhance the down-conversion efficiency, but not only that. A cavity introduces a threshold for the pump power above which the parametric gain is infinite, just limited by the finite pump power. 
In this case, the vacuum uncertainty of the input field at frequency $\nu$ is amplified to a bright laser field at frequency $\nu$. The device is then called an optical-parametric oscillator (OPO). 
For the generation of squeezed states, however, the pump power is usually kept (slightly) below threshold. Due to nonzero optical loss, there exists a pump power smaller than the threshold above which the tiny improvement of squeezing is not noticeable anymore. Getting the pump power closer to the threshold could even reduce the observed squeeze factor if a fluctuating squeeze angle projects anti-squeezing into the observed quadrature amplitude [\cite{Franzen2006,Suzuki2006,Dwyer2013}].
The cavity has another important purpose. It confines the transverse spatial mode, usually to TEM00. This mode confinement is crucial for any efficient application of the squeezed state in laser interferometry since it allows the suppression of anti-squeezing from other transversal modes. 
The squeezing process requires a nonlinear material that should show negligible absorption at both optical frequencies involved, in particular at the wavelength of the squeezed mode. In Refs.~[\cite{Vahlbruch2008,Mehmet2009}] 10\,dB and 11.6\,dB of squeezing were achieved using MgO:LiNbO$_3$. The highest squeeze factors today are produced in (quasi phase matched) periodically poled KTP [\cite{Eberle2010,Mehmet2011,Stefszky2012,Vahlbruch2016}]. 

\FloatBarrier

The optical cavity that is built around the nonlinear crystal is vital for squeezed-light generation, and it deserves a detailed consideration. 
Generally, the mode propagating away from a cavity is the result of interference at the cavity coupling mirror. 
One contribution is given by the intra-cavity field attenuated by the amplitude transmission coefficient $t$ of the outcoupling mirror. The second contribution is given by the outside field that is reflected by the same mirror with amplitude reflectivity $r = \sqrt{1-t^2}$ and spatially overlapped with the first. 
Also the mode from a squeezing resonator is such an interference product. \\

\begin{figure}[h!!!]
  \vspace{-19mm} \hspace{0mm}
  \includegraphics[width=10.6cm]{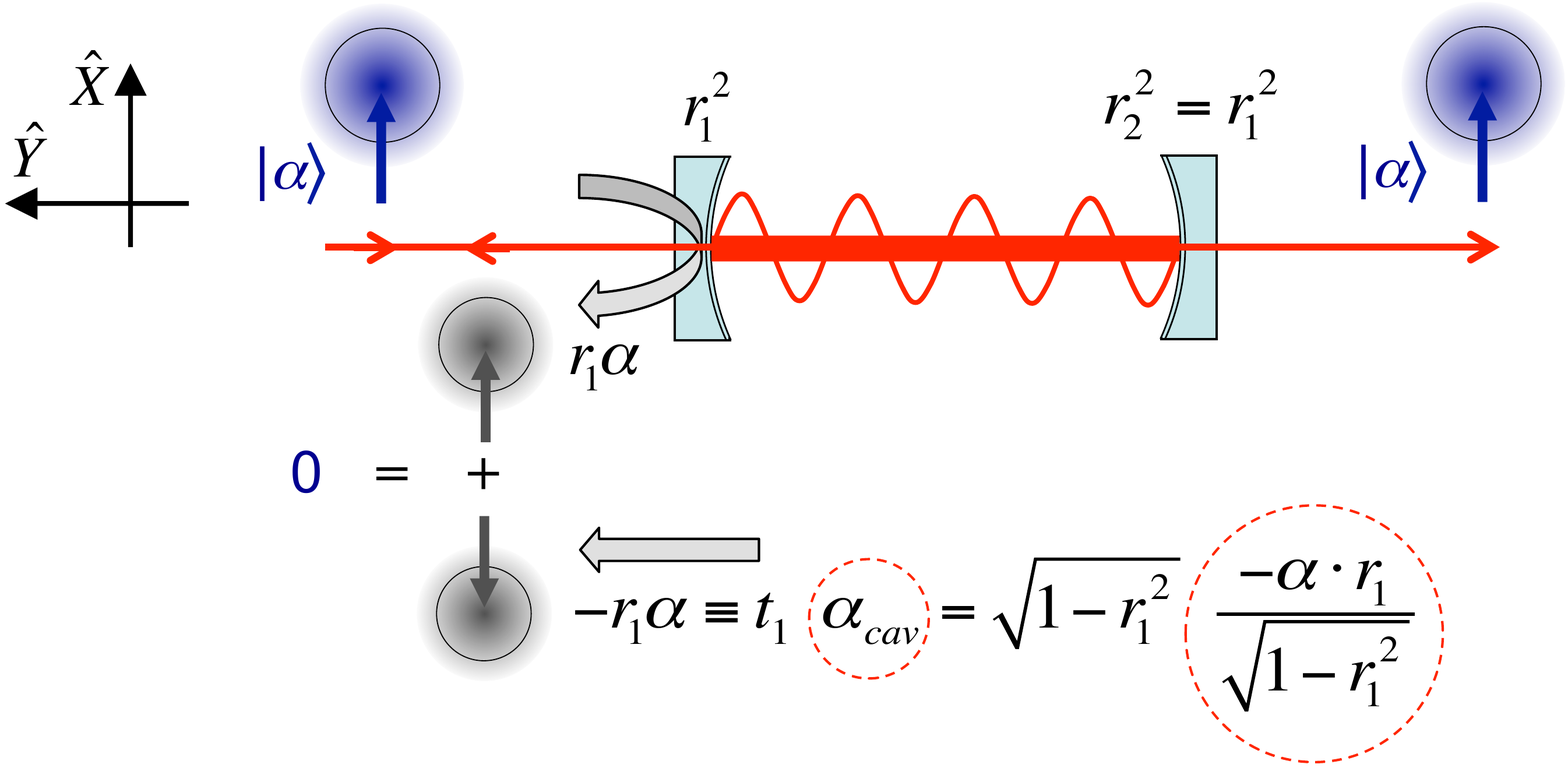}
  \vspace{-14mm}
\caption{\textbf{Empty, impedance-matched resonator} -- Mode-matched and resonant light that enters the cavity from the left, here displayed by a quantum phasor for a coherent state, is fully transmitted, including its quantum uncertainty. The back-reflected state destructively interferes with itself for all frequencies well within the cavity linewidth. The 180$^\circ$ phase shift of the transmitted cavity field amplitude $-  r_1 \alpha$ is a necessary condition in order to fulfill energy conservation on cavity resonance. Impedance matching is for instance achieved for a resonant cavity with matched mirror reflectivities ($r^2_2=r^2_1$) and zero optical loss. The complex amplitude of the field traveling towards left \emph{inside} the cavity is readily deduced from the figure and reads $\alpha_{\rm{cav}} = - \alpha r_1 (1-r_1^2)^{-1/2}$ (highlighted by the dashed circle). In the displayed setup, zero field uncertainties are reflected, however, also the vacuum state that enters the cavity from the right is fully transmitted (not shown).
}
\label{fig:17}
\end{figure}
\begin{figure}[b]
  \vspace{0mm}\hspace{2mm}
  \includegraphics[width=10.6cm]{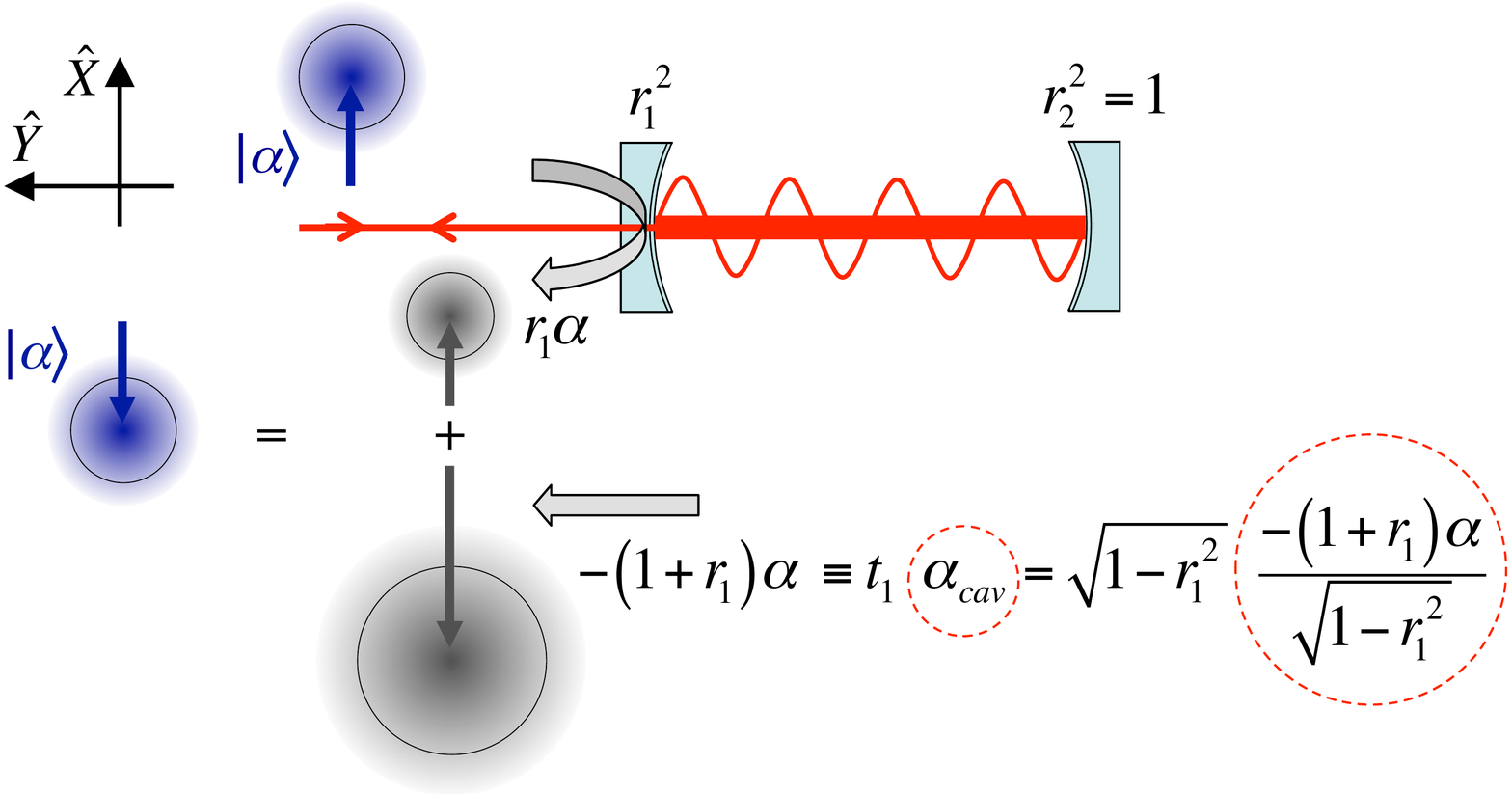}
  \vspace{-10mm}
\caption{\textbf{Empty, maximally overcoupled resonator} --  Maximal overcoupling is achieved for a resonant cavity with a perfect end-mirror reflectivity ($r^2_2=1$) and zero optical loss. For a given input-mirror reflectivity $r^2_1$ the intracavity light power is maximal. Mode-matched and resonating light entering the resonator from the left is fully reflected. The complex amplitude of the field traveling towards left \emph{inside} the cavity is readily deduced from energy conservation to $\alpha_{\rm{cav}} = - \alpha (1+r_1) (1-r_1^2)^{-1/2}$ (dashed circle). In this setup, no uncertainty from the right couples to the left side of the cavity.
}
\label{fig:18}
  \vspace{-4mm}
\end{figure}

\emph{\\[-3mm]The impedance matched resonator}\\[1mm]
Let us consider first an empty, optically stable and loss-less Fabry-Perot resonator built from two identical mirrors, each with amplitude reflectivity $r = r_1 = r_2 < 1$. A propagating field be perfectly mode-matched to one of the cavity resonances. In this setup, the resonator shows zero reflection, and the resonator is said to be \emph{impedance matched} (for all such input fields). 

\FloatBarrier

Obviously, the interference described in the previous paragraph is fully destructive. The same resonator also shows zero reflection of the input field's quantum uncertainty, since the interference happens between parts of the same quantum state. The mode propagating away from such a resonator, however, is not in a nonclassical but in a vacuum state, because the vacuum state that enters the cavity through the opposite site is also fully transmitted. The interference at the coupling mirror of an impedance matched resonator is displayed in Fig.\,\ref{fig:17}.

\begin{figure}[ht]
  \vspace{-2mm}\hspace{0mm}
  \includegraphics[width=10.6cm]{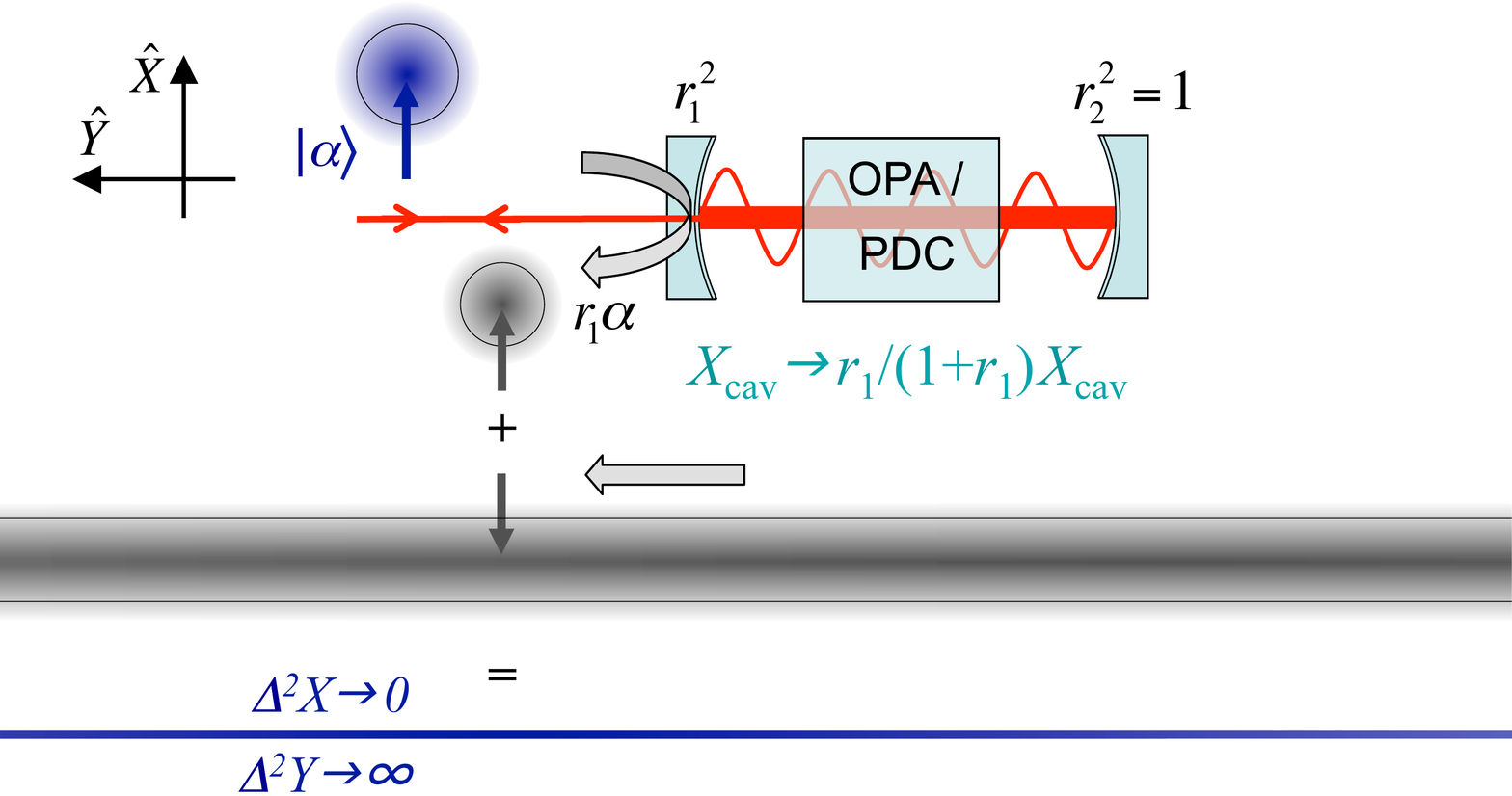}
  \vspace{-2mm}
\caption{\textbf{Squeezing resonator} -- Shown is the interference at the zero-loss squeezing resonator operated at threshold. The lower line represents the perfectly squeezed mode propagating away from the cavity towards the left. The parametric gain medium inside the cavity deamplifies the $X$ quadrature of the cavity mode ($X_{\rm{cav}}$) by the factor $r_1/(1+r_1)$, which is the ratio of the intra-cavity field amplitudes of the two previous figures. The $X$ quadrature of the field that is back-reflected towards the left destructively interferes with itself, similar to the situation of the impedance matched cavity in Fig.\,\ref{fig:17}. The parametric power gain per resonator round-trip ($G$) needs to mimic the effect of an end mirror with reflectivity $r^2_2 = r^2_1$. For this reason, the deamplification of $X_{\rm{cav}}$ corresponds to a \emph{round-trip} deamplification factor of $r_1 \equiv \sqrt{1/G}$. The round-trip amplification factor for $Y_{\rm{cav}}$ then is $1/r_1 \equiv \sqrt{G}$, which exactly compensates for the outcoupling and thus determines the parametric oscillation threshold (threshold for bright lasing). The variances of the quantum uncertainties $\Delta^2 Y_{\rm{cav}}$ and $\Delta^2 Y$ are thus infinite.
In this setup, no field uncertainty from the right couples to the left of the cavity and a perfectly $X$-quadrature-squeezed field \emph{outside} the squeezing resonator is produced. 
}
\label{fig:19}
  \vspace{6mm}
\end{figure}
\emph{\\[-3mm]The perfectly over-coupled, single-ended resonator}\\[1mm]
We now increase the reflectivity of the far mirror `2' to being perfect ($r_2 = 1$). This way the counter-propagating vacuum state can not enter the cavity. Again a propagating field be perfectly mode-matched through mirror `1'  to one of the cavity resonances. For frequencies well inside the cavity linewidth the situation is displayed in Fig.\,\ref{fig:18}. The setup protects the left side of the cavity against vacuum fluctuations entering through mirror `2', but of course does not squeeze quantum noise. The intra-cavity built-up factor is too high for achieving destructive interference below the vacuum uncertainty on the left side of the resonator.\\
\emph{\\[-3mm]
The impedanced-matched, single-ended squeezing resonator}\\[1mm]
Building on the two previous concepts, the straight forward approach now is to start from the perfectly over-coupled single-ended resonator and insert an attenuator into the cavity that does not couple the cavity mode to any bath, but still results in a roundtrip efficiency of precisely $r_1$($<1$) in amplitude. Optical loss is not appropriate since it increases the coupling of the cavity mode to a thermal bath, neither would any phase-insensitive atenuator be appropriate. It is easy to show that a phase-insensitive attenuator adds additional uncertainty, since otherwise the commutation relation $[\hat a,\hat a^{\dagger}]=1$ is violated. The amplification process that matches our requirement is OPA. To achieve infinite squeezing in $\hat X$ on cavity resonance, a second-order nonlinear crystal needs to be put into the cavity and pumped such that the intra-cavity amplitude quadrature is attenuated by the factor $(1+r_1)/r_1$ (on cavity resonance) with respect to the empty cavity. This factor is readily deduced from Figs.\,\ref{fig:17} and \ref{fig:18}. Due to the symmetry in parametric amplification, the intra-cavity phase quadrature is then amplified by $(1+r_1)/r_1$, and the round-trip gain has a value of $1/r_1$ in amplitude. In this situation not only infinite squeezing but also the (laser) threshold of the resonator is achieved, since the round-trip gain of the intra-cavity phase quadrature equals its roundtrip loss, here fully given by the incoupling mirror. \\

The physical descriptions in Figs.\,\ref{fig:17} to \ref{fig:19} are fully consistent with observations in squeezing experiments. The consideration above in particular shows that the intra-cavity field shows a finite squeezing strength while the external field shows infinite squeezing. 
The strongest intra-cavity squeeze factor possible is $(1+r_1)^2/r_1^2$. In the high reflectivity limit, this factor corresponds to 6\,dB.
Averaged over the full cavity mode, the squeeze factor of the cavity mode is, in this limit, even limited to 3\,dB [\cite{Walls2008}]. Higher intra-cavity squeeze factors are possible for lower mirror reflectivities.\\

\subsection{The generation of squeezed light for laser interferometry} %
\label{Ssec:sqz}
\begin{figure}[b!!!!!!!!!!!!!]
  \vspace{-0mm}
  \includegraphics[width=11.0cm]{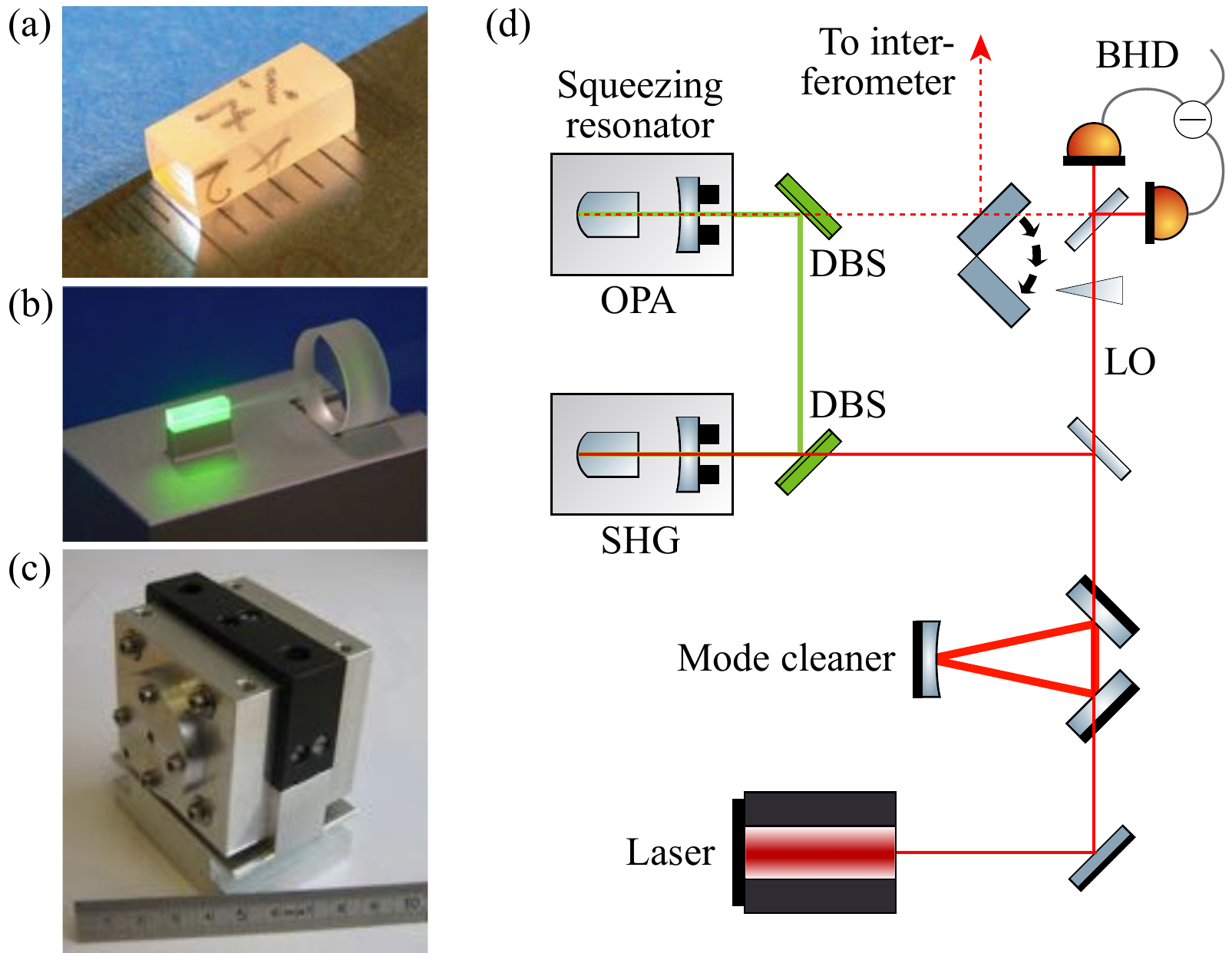}
  \vspace{0mm}
\caption{\textbf{Generation of squeezed light} --  
(a) Example of a 2$^{\rm nd}$-order nonlinear crystal for the squeezed-light generation at 1064\,nm. Shown is a bi-convex 6.5\,mm long 7\%MgO:LiNbO$_3$ crystal whose polished surfaces also carry the mirror coatings of the resonator. The crystal thus realizes a monolithic squeezing resonator, as it was used for the first demonstration of 10\,dB squeezing [\cite{Vahlbruch2008}]. (b) Optical configuration of a half-monolithic (hemilithic) standing-wave squeezing resonator. Here, the cavity length can be adjusted by displacing the coupling mirror. The crystal surface inside the cavity is anti-reflection coated. The photograph shows a 10\,mm long PPKTP crystal squeezing resonator as used for the GEO\,600 squeezed-light source [\cite{LSC2011}]. (c) Mechanically stable housing of a standing-wave squeezing resonator. The crystal's temperature is stabilized at its phase matching condition using Peltier elements. (d) Schematic for the squeezed-light generation. After spatial filtering of continuous-wave laser light two hemilithic standing-wave resonators are employed. The first generates second harmonic pump light (SHG). The second (OPA) generates a squeezed vacuum field at the initial wavelength. The squeezed states are observed by a balanced homodyne detector (BHD), or alternatively sent and mode-matched to the optical mode of an interferometer beforehand. LO: local oscillator; DBS: dichroic beam splitter.
}
\label{fig:20}
\end{figure}
With the insights gained in the previous subsection we now turn to actual experiments. The application of squeezed states in laser interferometry certainly requires large squeeze factors (idealy accompanied with the highest possible purity) to maximize the impact in terms of sensitivity improvement. 
In cavity-enhanced OPA, the highest parametric gain is achieved on cavity resonance, i.e.~at zero sideband frequency. But this is not the main reason why this Subsection focusses on the generation of squeezed states at low sideband frequencies. The application of squeezed states in a laser interferometer requires that their sideband frequencies cover the device's signal band. 
Ground-based gravitational wave (GW) detectors have a detection band from about 10\,Hz to 10\,kHz, frequencies which can be considered as `low' compared to typical frequencies in quantum optics experiments. 

Squeezing at MHz sideband frequencies is easier to observe than at acoustic frequencies because the latter are often polluted with excess noise from light beams that serve as control beams [\cite{Bowen2002c,McKenzie2004}] and parasitic interferences from back-scattered light [\cite{Vahlbruch2007}]. Furthermore, the observation of squeezing at low sideband frequencies requires a more stable setup since larger measuring times are necessary. The observation of strong squeezing at MHz frequencies, however, already sets an upper limit to the optical loss of the setup. At least the same squeeze factor can be observed at lower frequencies. 

There are two different main topologies for squeezing resonators. 
The Fabry-Perot-type standing-wave resonator consists of a minimum number of mirror surfaces and has the advantage of being compact and thus robust against mechanical vibrations. Usually one or even two mirror coatings are directly placed on the spherical and polished surfaces of the nonlinear crystal itself [\cite{Wu1986,Grangier1987,Breitenbach1998,Vahlbruch2008,Eberle2010,Vahlbruch2016}].
The Bowtie traveling-wave resonator has the advantage of providing a separately accessible counter propagating mode for cavity length control [\cite{Ou1992,Takeno2007}]. It shows no direct back-reflection of incoupled light, which helps reducing parasitic interferences [\cite{Stefszky2012}].

 \FloatBarrier
 
 Fig.\,\ref{fig:20} (a) and (b) show photographs of typical nonlinear crystals used for squeezed-light generation at near infra-red wavelengths. The crystals shown here form a monolithic standing-wave squeezing resonator (a) or are part of a half-monolithic standing-wave squeezing cavity. (c) shows a temperature stabilized and mechanically stable housing of the squeezing resonator.
(d) shows a schematic of a full setup for the generation of squeezed vacuum states of light for an application in a laser interferometer.
The only bright input required for the squeezing resonator (OPA) is the second-harmonic pump field. The resonator mode at fundamental frequency is thus initially not excited by photons, i.e.\,it is in its ground state, characterized by vacuum fluctuations due to the zero point energy, see Fig.\,\ref{fig:7}\,(c) [\cite{Gerry2005}]. 
The pump field spontaneously decays in the degenerate pair of signal and idler fields. The combined down-converted field leaving the resonator exhibits quantum correlations which give rise to a squeezed photon counting noise when overlapped with a bright coherent local oscillator beam. The detection is done either in a balanced homodyne detector (BHD) or with a single photo diode. 
The squeeze factor increases the closer the pump power of the squeezing resonator gets to the oscillation threshold, and the lower the optical loss on down-converted photon pairs is.\\[5mm] 
\subsubsection{High squeeze factors -- minimizing decoherence} 
Squeezed states of light have significant impact on the sensitivity of laser interferometers, if large squeeze factors can be produced. Squeezing of 3\,dB improves the signal-normalized quantum-noise spectral density by a factor of 2. 
This factor corresponds to doubling the (coherent state) light power circulating inside the interferometer.  
Squeezing of 10\,dB corresponds to a ten-fold power increase. The experimentally demonstrated squeeze factors were considerably improved in recent years
[\cite{Takeno2007,Vahlbruch2008,Polzik2008,Eberle2010,Stefszky2012}], culminating in a value of as large as 15.0\,dB [\cite{Vahlbruch2016}]. This  value corresponds to the same reduction of signal-normalized quantum noise that is achieved by increasing the light power by a factor of 32. 
(At this point it is already noted that squeezing the quantum noise can simultaneously reduce quantum measurement noise (shot noise) as well as quantum back action noise (radiation pressure noise). This is not possible with scaling the light power of coherent states, see Subsec.\,\ref{Ssec:5.5}.)

Ideally, a parametric squeezed-light source can produce an infinite squeezing level, see Fig.\,\ref{fig:19}, fundamentally just limited by the energy provided by the pump field. 
In practice, the limit is set by decoherence mechanisms. The by far most important one is optical loss.
Optical loss occurs during squeezed-light generation, its propagation through the interferometric setup including imperfect mode matchings, and finally the photo-electric detection. Also detector dark noise [\cite{Schneider1998}], phase noise [\cite{Takeno2007}], and excess noise [\cite{Bowen2002c}] impair the observable squeezing strength.  

Optical loss is usually understood as coupling the squeezed mode to a zero temperature bath, i.e.~overlapping it with a vacuum mode. 
For any amount of loss the resulting state is still squeezed. But to be able to directly observe, say, 10\,dB of squeezing, the total loss on the state needs to be less than 10\% in this example, cf. Eq.\,(\ref{eq:loss}). 
To minimize optical loss, the nonlinear crystal as well as lenses and beam splitters in the interferometric path need to show very low absorption and scattering at the wavelength of the squeezed light. PPKTP shows absorption of about 10$^{-4}/$cm and below at near-infrared wavelengths. Low OH content fused silica is a suitable material for all other optics. Absorptions of less than 10$^{-6}/$cm were measured [\cite{Hild2007Diss}]. Coatings on crystal surfaces and on all other optical components should also show lowest optical loss. Total loss of the 10$^{-6}$ level are available today. 
Superpolished surfaces, which show roughnesses with less than 1\,\AA~root mean square (integrated over spatial scales from approximately 1 micron to 100 microns) and thus very low scattering, are necessary to achieve these low numbers. Minimizing the total number of optical components is essential. From this perspective a monolithic squeezing resonator as shown in Fig.\,\ref{fig:20} (a) is the optimum choice. 
The squeezed mode needs to be matched to the mode of the laser interferometer or to the mode of the balanced homodyne detector. Visibilities of up to 99.8\% have been achieved [\cite{Eberle2010}], which corresponds to a loss of about 0.4\%. Of great importance also is the quantum efficiency of the photo-diodes used for detecting the squeezed field (together with the interferometric signal). Recently a quantum efficiency of photo-diodes in a squeezing experiment of $(99.5 \pm 0.5)$\% was measured [\cite{Vahlbruch2016}]. To minimize photon loss, the photo-diodes had no protection window, an anti-reflection coating on the semi-conductor material, and the remaining reflection was re-focussed with an external mirror.

Also the dark-noise spectral density of the detection electronics  reduces the observable squeezing and needs to be as low as possible. Similar to optical noise it also provides a contribution to the observed variance. The dark noise of the detection electronics needs to be much lower than the detected photon counting noise. In [\cite{Vahlbruch2016}] it was 28\,dB below shot noise but still reduced the observable squeeze factor from 15.3\,dB to 15.0\,dB.

Excess noise emerges if the squeezed mode couples to a nonzero temperature bath or to a mode whose excitation is strongly fluctuating.  
(The coupling process can always be understood as a beam splitter coupling and is physically described by overlapping electric fields. Coupling to a zero temperature bath leads to Eq.\,(\ref{eq:loss}).)
The captured excess noise variance then needs to be added to the initial squeezing variance which deteriorates the observed squeezing stronger than just mixing in the vacuum mode. Excess noise is less likely to occur at MHz frequencies, but can be significant at audio-band sideband frequencies and below, and is thus a serious issue in gravitational-wave detectors [\cite{Chua2014}]. The reason for that is that acoustically or thermally excited motions of surfaces produce frequency shifts of back-scattered light mainly at these low frequencies [\cite{Vahlbruch2007}].

Phase noise corresponds to stochastic phase fluctuations between the squeezed field and the local oscillator within the measuring time. It corresponds to mixing the squeezed mode with itself with a fluctuating squeeze angle [\cite{Suzuki2006,Franzen2006}]. Phase noise in squeezing experiments typically is less of an issue than optical loss [\cite{Dwyer2013,Oelker2016,Vahlbruch2016}]. 
The setup's phase noise can be reduced by making the squeezing resonator more compact and thus mechanically more stable against acoustic and thermal fluctuations of the environment, and by improving the quality of the servo loops for cavity length and propagation length controls. Operating a squeezed-light resonator in vacuum might also be beneficial. The ability to run a high performance squeezed-light generator in vacuum was demonstrated in [\cite{Wade2015}]. \\[3mm] 

\subsubsection{Squeezing in the gravitational-wave\,(GW) detection band} 
High squeeze factors have been first demonstrated at sideband frequencies of a few MHz and above, where excess noise is generally negligible when working with visible or near-infra-red light. Today we know that extending the squeezing spectrum towards the audio-band and even below is technically not always easy but straight forward, once a high squeeze factor is achieved at  MHz frequencies. 
\begin{figure}[ht!!!!!!!!!!!!!!!!]
  \vspace{-3mm}
  \includegraphics[width=13.6cm]{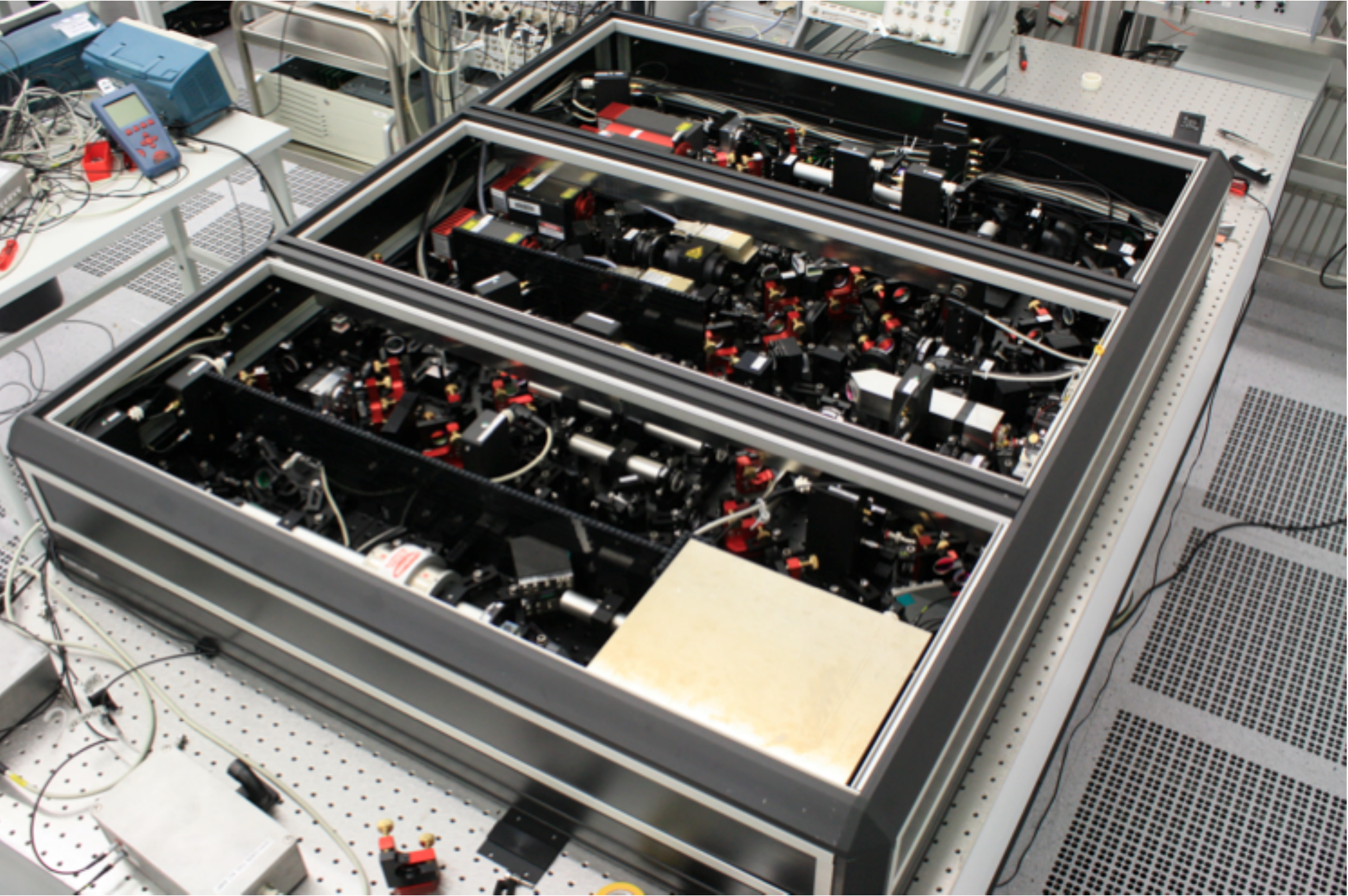}
  \vspace{-1mm}
\caption{\textbf{Photograph of the GEO\,600 squeezed-light source} -- The breadboard dimensions are 135\,cm $\times$ 113\,cm. The squeezing resonator is high-lighted by the white arrow and is set up as a standing-wave hemilithic cavity containing a plano-convex PPKTP crystal of about 10\,mm length (see also Fig.\,\ref{fig:20}\,b). It is pumped with continuous-wave 532\,nm light that is produced by frequency doubling of light (at angular frequency $\omega$) from a commercial Nd:YAG laser. Two more laser fields at about 1064 nm having frequency offsets of more than 10 MHz with respect to $\omega/(2\pi)$ serve as optical control fields. Both fields are mode-matched and injected into the squeezing resonator together with the second-harmonic pump field.                      
}
\label{fig:21}
\end{figure}
In most squeezing experiments the main laser light at the squeezing wavelength is accompanied by significant noise up to the laser relaxation oscillation. For this reason, laser control fields at the optical carrier-frequency in the optical path of the squeezed mode need to be avoided [\cite{Bowen2002c,Schnabel2004,McKenzie2004}], and the squeezing resonator length and the optical path stabilized by other means [\cite{McKenzie2005,Vahlbruch2006}]. 
Furthermore, and most importantly, excess noise due to back-scattering is an issue.  Back-scattering (also called `parasitic interferences') is produced if DC light scatters out of the optical path, hits a vibrating surface and re-scatters back into the optical path [\cite{Vahlbruch2007}].
Significant back-scattering can be produced in interferometers for the detection of gravitational waves since light powers of several hundreds of kilowatts are used. Even back-scattering from the milliwatt local oscillator of  balanced homodyne detectors is an issue at acoustic sideband frequencies and below.
The recipe for avoiding parasitic interferences turns out to be threefold: (i) avoiding scattering by using ultra-clean superpolished optics with close to perfect anti-reflex coatings, (ii) avoiding back-scattering by carefully blocking all residual (faint) light fields, and (iii) reduce the vibrationally and thermally excited motion of all mechanical and optical parts, that could potentially act as a re-scattering surface, with respect to the optical path [\cite{Vahlbruch2007,McKenzie2007}].\\ 
The insights described above led to the first demonstration of audio-band squeezing at frequencies down to 200\,Hz [\cite{McKenzie2004}], and later to the first demonstration of squeezing over the full gravitational-wave detection band, even from as low as 1\,Hz [\cite{Vahlbruch2007}]. 
While a standing-wave squeezing resonator [\cite{Ou1992}] can be built in a very compact way that is rather insensitive against mechanical vibrations [\cite{Chelkowski2007}], a traveling-wave bow-tie squeezing resonator [\cite{Wu1986}] is more tolerant against back-scattered light [\cite{Chua2011}].
The strongest squeezing in the audio-band of up to 11.6\,dB was reported in Ref.~[\cite{Stefszky2012}].
\\[0mm]
\FloatBarrier

\begin{figure}[ht]
  \vspace{-1mm}
  \includegraphics[width=11.6cm]{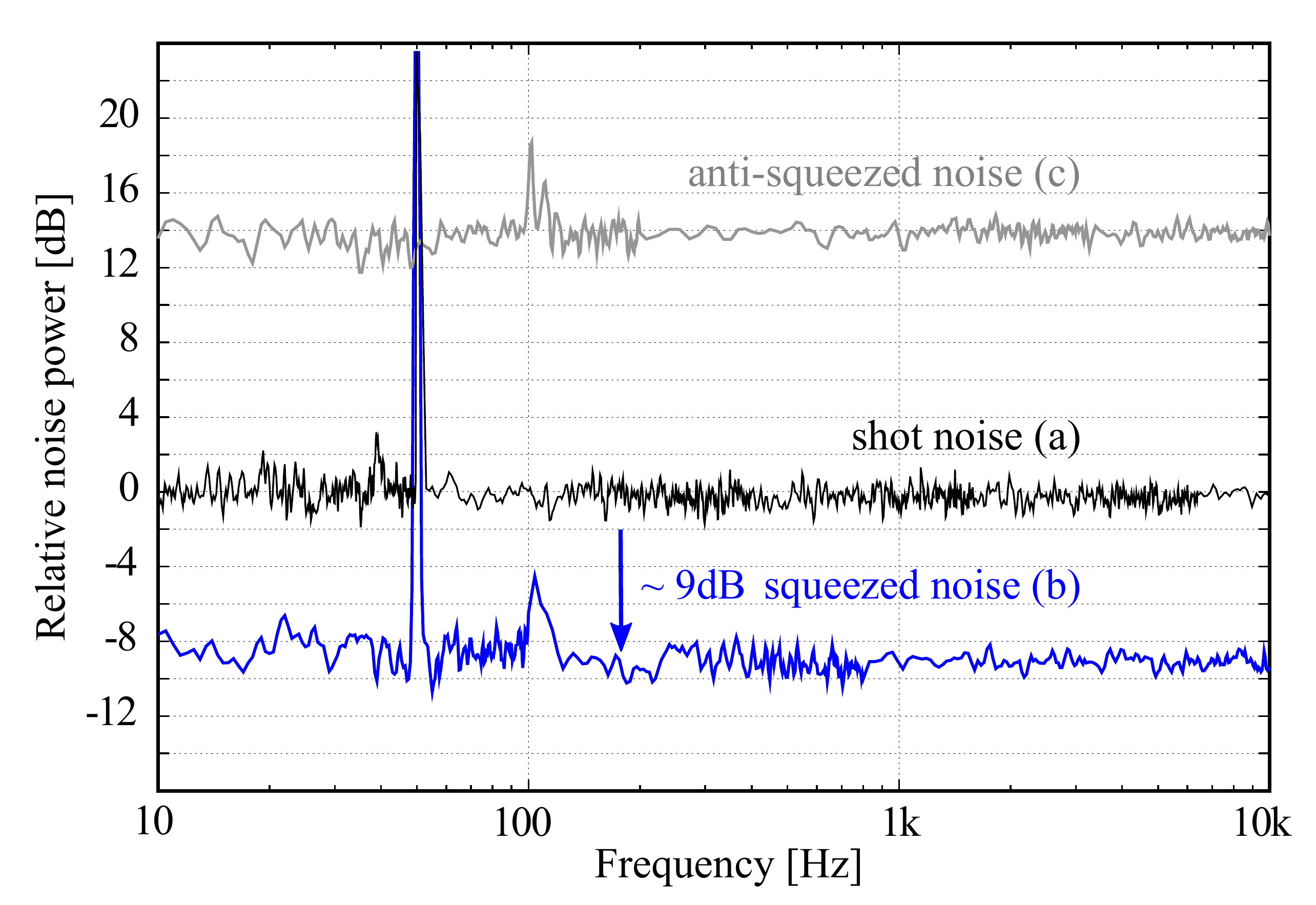}
  \vspace{-1mm}
\caption{\textbf{Broadband squeezing spectrum} -- 
Noise power spectra measured on the output of the GEO\,600 squeezed-light source shown in Fig.\,\ref{fig:21} with a balanced homodyne detector. The  traces correspond to the spectra of quadrature amplitude variances normalized to vacuum noise. The resolution bandwidth used increases towards higher frequencies to reduce the measurement time. (a) Shot noise, normalized to unity, which serves as the reference level (0\,dB). (b) Squeezed noise, covering the complete detection band of ground-based GW detectors. (c) Anti-squeezed noise. Peaks at 50\,Hz and 100\,Hz are the electric mains frequency and its first harmonic. The data was published in Ref.~[\cite{Vahlbruch2010}].
}
\label{fig:22}
\end{figure}

\subsubsection{The first squeezed-light source for GW detection} 
The first squeezed-light source for the continuous operation in GW detectors had been designed and completed between 2008 and 2010 [\cite{Vahlbruch2008Diss,Vahlbruch2010}]. Since then, this source has been producing squeezed vacuum states in a fully phase controlled way using co-propagating frequency-shifted bright control beams [\cite{Vahlbruch2006}] as an integral part of the GW detector GEO\,600. The source is a turn-key device with a fully automated re-lock system [\cite{Vahlbruch2010,Khalaidovski2012}]. Re-locking is required if the temperature of the environment changes significantly, which drives the actuators outside their dynamic ranges. 

Up to 9\,dB of squeezing over the entire GW detection band was observed using a balanced homodyne detector (BHD) located close to the squeezing resonator. The squeeze factor has been limited by optical loss due to absorption in the PPKTP crystal, transmission of the back-surface, and the non-perfect AR-coating of the crystals's intra-cavity surface. The adjustable air gap has been necessary to allow for an easy way to apply length control. Additional optical loss in the path to the balanced homodyne detector mainly arose due to a Faraday isolator that eliminated parasitic interferences. Finally, the mode missmatch to the BHD as well as its non-perfect quantum efficiency provided additional loss. Inferring the squeeze factor without the BHD detection loss, more than 10\,dB of squeezing are provided by the source. Since 2010 it has been used in basically all observational runs of the GEO\,600 GW detector, see Section~\ref{Sec:6}.

\subsubsection{Generation of two-mode (bi-partite) squeezing} 
`Two-mode squeezed light'  or `bi-partite squeezed light' is light that allows for joint measurements at two locations $A$ and $B$. These joint quadrature measurements reveal correlations and anti-correlations with a remaining uncertainty smaller than the ground-state uncertainty, which certifies the presence of entanglement, cf.~Subsec.\,\ref{ssec:EPR}. Bi-partite squeezed light has been generated by type\,I and by type\,II parametric down-conversion. In case of type\,I, the squeezed fields from two squeezing resonators as described in Subsec.\,\ref{Ssec:sqz} are overlapped on a balanced beam splitter with a 90$^\circ$ phase shift. The two output fields together represent the entangled mode [\cite{Furusawa1998,Bowen2003,Eberle2013}]. 
In case of type\,II, signal and idler fields are non-degenerate regarding polarisation and a single cavity containing an appropriate crystal and a polarising beam splitter are sufficient for the production of bi-partite squeezing. Also in this case, the measurements of the quadrature amplitudes of signal and idler fields show large uncertainties together with bi-partite correlations and anti-correlations that are stronger than the ground state uncertainty of individual subsystems [\cite{Ou1992,Villar2006,Jing2006}]. 

To date, the strongest entanglement of bi-partite squeezed light has been produced based on type\,I parametric down-conversion [\cite{Eberle2013}]. The requirements of producing strong entanglement are identical to those of producing strong squeezing outlined above. The strength of bi-partite entanglement can be given in decibels in full analogy to the squeeze factor. Practically, the strength of bi-partite squeezing is always somewhat smaller than that of single party squeezing, since it requires an additional mode-matching, that results in additional optical loss.\\ 

\FloatBarrier

\subsection{Conclusions} 

The first observation of squeezed light was achieved in 1985 [\cite{Slusher1985}]. Shortly after, cavity-enhanced optical parametric amplification for squeezed-light generation was demonstrated [\cite{Wu1986}], which today enables the observation of up to 15\,dB of squeezing [\cite{Vahlbruch2016}]. 
Quite generally, the maximum squeezing level that is observed does not depend on the strength of the optical nonlinearity. Squeezing cavities can easily be operated at their oscillation threshold, where they should provide infinite squeezing if decoherence is zero. The main limiting factor is optical loss, including that of the photo-electric detection. 

Dedicated experimental research and development towards a squeezed-light source for applications in gravitational-wave detectors can be traced back to 2002 [\cite{McKenzie2002,Bowen2002c}]. Since then a surprising amount of progress has been made, culminating in the first squeezed-light source specifically built for the integration into a gravitational-wave detector. For the future, squeeze factors above 15\,dB will certainly be possible, by further reducing optical loss. This statement is supported by the high degree of matching between experimental data and a theoretical loss model as presented in Fig.\,3 of [\cite{Vahlbruch2016}].\\

%
\section{Quantum noise in laser interferometers} \label{Sec:5}  
\subsection{Interferometric measurements}  \label{Ssec:5.1}

The purpose of a laser interferometer is the precise measurement of small changes of an optical path length with respect to a reference path. 
For this, the interferometer transfers the change of the phase difference between two light fields into an amplitude quadrature change of the interferometer's output light. The latter can easily be detected by a single photo diode. %
Of general interest are differential changes of the optical path length that are much smaller than the laser wavelength, i.e.~differential phase changes that are much smaller than $2 \pi$. In this case, the differential phase change can be described in very good approximation as a differential change of the  \emph{phase quadrature} instead. %

In order to transfer the phase quadrature signal with minimum loss, a high interference contrast at the interferometer's beam splitter is essential. Additionally, instrumental noise in terms of unwanted excitations of the output's amplitude quadrature needs to be reduced to a minimum. Noise arises due to power fluctuations of the input laser light, back-scattered laser light inside the interferometer, thermally driven displacements of mirror surfaces and in many more ways. 
The important measure of the sensitivity of an interferometer obviously is its signal-to-noise-ratio. The most useful measure is given in terms of the \emph{noise spectral density} $S(f\!=\!\Omega/2\pi)$ that is normalized to the physical unit of the signal. $S(f)$ is in fact a `noise-to-signal-ratio' and can be seen as the signal-normalized variance of the photo diode output decomposed into spectral components versus sideband frequency $f$ with the resolution bandwidth of 1\,Hz. As an example, $S(100\,\rm{Hz}) = 10^{-39}$\,m$^2$/Hz means that the instrumental noise in the one hertz band around 100\,Hz equals a signal that would be produced if the mirror of one interferometer arm oscillates with an amplitude of just $\sqrt{10}\cdot10^{-20}$\,m in the very same band. Such small spectral densities are achieved by gravitational-wave detectors [\cite{Abbott2016}].\\

\subsection{Quantum measurement noise and shot noise}  \label{Ssec:5.2}

The most fundamental noise source in laser interferometers is due to the quantum noise of light, which is in fact two-fold [\cite{Caves1980}].
First of all, there is `quantum measurement noise', which arises in the process of photo-electric detection.
For coherent states the quantum measurement noise is the `photon counting noise' of mutually independent photons and usually simply called `shot-noise'. 
Fig.\,\ref{fig:2}\,(b\,i) shows a time series of such noise hiding the actual signal.
The frequency components of the shot noise are well described by the quantum uncertainty of the output field's amplitude quadratures $\hat X_{\Omega,\Delta\Omega}$, see Subsec.\,\ref{Ssec:2.2}. (Recall, this quantity corresponds to the differential phase quadrature $\hat Y_{\Omega,\Delta\Omega}$ of the light beams in the interferometer arms). 
The photon counting noise has a white Fourier spectrum, however, the `shot noise' of an interferometer is usually normalized to the signal, whose transfer function is usually not white, for instance due to the presence of arm cavities or a signal-recycling cavity.

All current and planned gravitational-wave detectors are Michelson-type laser interferometers, with operating points very close to a dark fringe. The light power in the output port is just a couple of tens of mW, which can be handled by a single photo diode. 
In this configuration the signal-to-shot-noise-ratio is actually maximized, which can be shown in three steps [\cite{Bachor2004}]. For the first step we use plane waves to describe the electric field in the output port of a Michelson interferometer. For perfect interference contrast at the balanced beam splitter, i.e.~for perfect mode matching, and for defining $\phi = 0$ as the dark port condition we get
\begin{equation}
\label{eq:Eout}
E_{\rm out} (t,\phi) = \frac{1}{2} E_{0}\,{\rm sin}(\omega t + \phi) - \frac{1}{2} E_{0}\,{\rm sin}(\omega t) \,,
\end{equation}
where $E_{0}$ is the amplitude of the total internal field whose two parts has accumulated a differential phase. It directly follows for the squared fields
\begin{equation}
\label{eq:E2out}
E_{\rm out}^2 (t,\phi) = \left( {\rm sin} \frac{\phi}{2} \right)^2  \left( E_{0} \,{\rm cos}(\omega t + \phi /2) \right)^2 \,.
\end{equation}
We now turn to a light beam with a localized transversal mode that can be focussed onto a photo-electric detector. The photo diode  has perfect quantum efficiency, i.e.~the rate of photo electrons is not only proportional to the rate of output field photons but also has a unity slope efficiency.  %
Since the optical frequency is too high to be resolved we consider the averaged light power
\begin{equation}
\label{eq:Pout}
\overline{P}_{\rm out}(\phi) = \left( {\rm sin} \frac{\phi}{2} \right)^2  \overline{P} \,.
\end{equation}
The next step is a formulation of the signal being the derivative of detected photon number versus phase. 
Let $\overline{n}$ be the average value of the photon number per measuring time interval. Eq.\,(\ref{eq:Pout}) can then be rewritten as 
\begin{equation}
\overline{n}_{\rm out}(\phi) = \left( {\rm sin} \frac{\phi}{2} \right)^2  \overline{n} \,.
\end{equation}
\begin{equation}
\label{eq:signal}
\Rightarrow \; {d} \overline{n}_{\rm out}(\phi) = \overline{n}\, {\rm sin} \frac{\phi}{2}\, {\rm cos} \frac{\phi}{2} \,d \phi  \,.
\end{equation}
The final step is the calculation of the signal-to-shot-noise-ratio. Shot noise refers to coherent states, which have a standard deviation of the photon number of $\sigma (\overline{n}) = \sqrt{\overline{n}}$.
\begin{equation}
\frac{d \overline{n}_{\rm out}(\phi)}{\sigma (\overline{n}_{\rm out})} = \frac{\overline{n}\, {\rm sin} \frac{\phi}{2}\, {\rm cos} \frac{\phi}{2} \,d \phi}{\sqrt{\overline{n}}\, {\rm sin} \frac{\phi}{2}}  \,,
\end{equation}
and find for a signal-to-noise ratio of unity for coherent states and for a non-zero but still small phase difference $\Delta \phi^{\rm Coh} \ll 2\pi$
\begin{equation}
1 = \sqrt{\overline{n}} \, {\rm cos} \frac{\phi}{2} \,\Delta \phi^{\rm Coh}  \,\; {\rm with} \,\; \phi \neq 0.
\end{equation}
In this equation the smallest measurable phase difference is given for $\phi \rightarrow 0$
\begin{equation}
\label{eq:phiSN}
\Delta \phi^{\rm Coh}_{\rm min} = \frac{1}{\sqrt{\overline{n}}} \,.
\end{equation}
This is the well-known shot-noise limit of high-precision phase sensing. $\Delta \phi^{\rm Coh}_{\rm min}$ is the smallest phase shift that can be measured with a signal-to-noise ratio of one when using $\overline{n}$ mutually independent photons per measuring time (those of a coherent state), when the loss of photons is assumed to be zero.\\ 
The typical purpose of a laser interferometer is the continuous sensing (monitoring) of a continuously changing phase. An illustrative example is the phase signal produced by the black hole merger measured by Advanced LIGO on Sept. 14 in 2015 (Fig. 1 in [\cite{Abbott2016}]).
The measuring interval should be short to be able to resolve the time-evolution of the signal. Generally, the measurement of an arbitrary signal that lasts for a finite time thus needs to be understood as $l$ subsequent measurement intervals using $\overline{n}$ photons each.\\ 
It can be shown that Eq.\,(\ref{eq:phiSN}) is also valid for interferometers operated at half fringe, i.e.~when each output port contains the same light power. In this case, photo diodes need to be placed in both output ports and the actual signal is provided by their difference voltage.\\ %

Due to its importance the shot-noise limit deserves some remarks. \\
The phase $\phi$ in Eq.\,(\ref{eq:Eout}) is  the phase difference of two mode-matched fields and might be accumulated by a single pass along the length $L$, such as in a Mach-Zehnder interferometer, or in a double pass, such as in a (simple) Michelson interferometer, or in four passes as realized in a Michelson interferometer with folded arms [\cite{Grote2005}]. The shot-noise limit in Eq.\,(\ref{eq:phiSN}) and its scaling therefore holds independent of the number of passes. The claim in Ref.~[\cite{Higgins2007}] that the scaling according to Eq.\,(\ref{eq:phiSN}) can be surpassed by multiple passes is not justified.\\
The fact that Eq.\,(\ref{eq:phiSN}) is derived by approaching $\phi \rightarrow 0$ correctly describes the actual operation point of gravitational-wave detectors, which is close to, but not exactly at a dark port. In practice a tiny offset from dark port is chosen at which the shot noise is well above the photo diode's electronic dark noise. \\%
Eq.\,(\ref{eq:phiSN}) solely depends on the number of quanta, but not on the light's wavelength $\lambda$. Of course, the shot-noise limit for the change of an \emph{optical path length} $\Delta L$ does depend on the wavelength, and $\Delta \phi^{\rm Coh}_{\rm min}$ needs to be replaced by $\Delta \phi^{\rm Coh}_{\rm min} = 2 \pi \Delta L^{\rm Coh}_{\rm min} / \lambda$.\\
Finally, an essential result of the shot-noise limit is that the ideal precise measurement should use `as much quanta as possible per measuring interval', which translates to `as much light power in the interferometer arms as possible'. Eq.\,(\ref{eq:phiSN}) is indeed the one and only reason why gravitational-wave detectors use high power lasers, power-recycling and arm resonators.
Extending the measuring time for a given light power can also improve the sensitivity, but only if the signal repeats, i.e. is periodic. 
Let us assume that one period of the signal is resolved by $l$ intervals using $\overline{n}$ photons each. In this case, repeating the overall measurement $k$ times improves  Eq.\,(\ref{eq:phiSN}) by $1/\sqrt{k}$. The fundamental statement of Eq.\,(\ref{eq:phiSN}), however, does not change since the actual photon number $\overline{n}$ may then simply incorporate the factor $k$.\\

For a given average photon number, the shot-noise limit in Eq.\,(\ref{eq:phiSN}) can only be surpassed by using photons that are quantum correlated, i.e.~by using nonclassical states of light. How is the shot-noise limit surpassed with the help of squeezed states? 
A Michelson interferometer that is operated close to a dark fringe acts like an almost perfect mirror for both input ports. All the input light is back-reflected towards the laser source. This also accounts for the quantum uncertainty of the input light. The quantum uncertainty that impinges onto the photo diode thus (mainly) enters the interferometer through its (almost) dark port. An interferometer that uses displaced coherent states entering from one port can thus be improved by replacing the ordinary vacuum entering the signal output port by a squeezed vacuum state. This was the proposal by C.M. Caves in 1981 [\cite{Caves1981}], which is labeled here with `CSV'. If the differential phase quadrature of the interferometer is squeezed, Eq.\,(\ref{eq:phiSN}) then, within the limit of large coherent state displacement $ \alpha \gg {\rm sinh}^2 r$, improves to
\begin{equation}
\label{eq:phiSNsqz}
\Delta \phi^{\rm CSV}_{\rm min} \approx \frac{e^{-r}}{\sqrt{\overline{n}}}  \, .
\end{equation}
(The above expression is an approximation since the squeezing operation produces a small number of photons that are not accounted for here.)    
Of course, the mode of the squeezed vacuum needs to be precisely matched to the mode of the interferometer. 
The first experimental demonstrations of squeezed phase measurements used a Mach-Zehnder [\cite{Xiao1987}] and a polarization interferometer [\cite{Grangier1987}].
Fig.\,\ref{fig:2} shows how spatial degeneracy between an externally generated squeezed mode and the signal mode in a Michelson interferometer is achieved using a polarizing beam splitter and a Faraday rotator. Again, the limit in Eq.\,(\ref{eq:phiSNsqz}) can only be achieved if optical loss is zero. Optical loss not only reduces the signal but here also reduces the squeeze parameter, see Eqs.\,(\ref{eq:r}) and ({\ref{eq:loss}}).\\

Let us consider an example. The sensitivity of a laser interferometer that uses coherent states with an excitation of $10^{23}$ photons per second can be improved by a factor of $\sqrt{10}$ by either adding $0.9 \cdot 10^{24}$ photons per second, or by adding about just 2 photons per second and bandwidth in hertz that belong to the 10\,dB squeezed vacuum, confer Eq.\,(\ref{eq:nsqz}). Since the full signal band of ground-based GW detectors covers sideband frequencies up to 10\,kHz, just $2 \cdot 10^{4}$ photons per second are necessary. 
At a wavelength of $\lambda = 1064$\,nm these values correspond to a power increase by 168\,kW and 3.7
\,fW, respectively.\\

The question arises whether a scaling of the sensitivity better than $\propto \sqrt{1/\overline{n}}$ is possible. It was theoretically shown that in principle the scaling can indeed  considerably be improved, yielding the so-called \emph{Heisenberg limit} or \emph{Heisenberg scaling} [\cite{Bondurant1984,Yurke1986,Braunstein1992,Holland1993}]  
\begin{equation}
\label{eq:phiHei}
\Delta \phi^{\rm HL}_{\rm min} \propto \frac{1}{\overline{n}} \,.
\end{equation}
The Heisenberg scaling requires nonclassical states that have a certain number of quanta, similar to Fock states, i.e.~$\overline{n} = n$. The theoretically optimal states describe a superposition of $n$ ($N$) indistinguishable photons in one interferometer arm while having zero ($0$) photons in the second arm, and vice versa, and were named `N00N'-states [\cite{Dowling2008}]. A specific property of these states is `super-resolution'. The output ports of the interferometer show an $n$-times faster oscillation of the interference fringes when changing the phase between the two interferometer arms. Super-resolution corresponds to an $n$-times improved signal transfer function and was demonstrated for instance in Refs.~[\cite{Rarity1990,Kuzmich1998,Mitchell2004,Afek2010}]. The presence of this nonclassical phenomenon, however, does not prove a sensitivity better than the semi-classical bound according to Eq.\,(\ref{eq:phiSN}). Sensitivity is rather related to the signal-to-noise-ratio and needs to take into account all imperfections as well as the probability of a successful detection of the sensing state [\cite{Thomas-Peter2011}]. All experiments so far used post-selection on particular measurement outcomes and neglected the typically large probability that nothing was detected.

Super-resolution was demonstrated with up to $\overline{n} = 5$ [\cite{Afek2010}]. In addition to the fact that super-resolution does not prove a sensitivity better than the semi-classical bound, photon numbers in state-of-the-art super-resolution experiments are extremely small compared to the photon number of about 10$^{23}$ (within a measuring interval of one second) of coherent states used in Ref.~[\cite{Abbott2016}], and of about 10$^{22}$ using coherent states plus squeezed vacuum states used in Ref.~[\cite{LSC2011}](, which did prove a sensitivity better than the semi-classical bound.)\\

Another interesting and related question is, what the smallest phase is that can be estimated in a \emph{single} measurement, again using a given number of quanta. Taking into account that no prior information about the phase shift exists, still a scaling proportional to $1 / \overline{n}$ is possible. In the limit of large $\overline{n}$, however, an additional factor of $\pi$ is required in the nominator of Eq.\,(\ref{eq:phiHei}) [\cite{Sanders1995,Berry2000}] yielding
\begin{equation}
\label{eq:phiHei2}
\Delta \phi^{\rm HL}_{\rm min} \approx \pi / \overline{n} \, .
\end{equation}
The state that can actually achieve this bound is different from the N00N state and was found in [\cite{Summy1990,Luis1996,Berry2000}]. A N00N state is not the optimum state for phase estimation (via a single measurement) since it only provides one bit of information. 
A recent review on generell aspects on phase measurements is given by Ref.~[\cite{Demkowicz-Dobrzanski2015}].\\

It is important to note that Eqs.\,(\ref{eq:phiSN}-\ref{eq:phiHei}) do not consider photon loss. 
Experiments that demonstrated super-resolution and aimed for achieving the scaling in Eq.\,(\ref{eq:phiHei}) were conditioned on zero photon loss.
Let $\eta > 0$ be the average efficiency of detecting (all) photons. Eq.\,(\ref{eq:phiSN}) then reads
\begin{equation}
\label{eq:phiSNloss}
\Delta \phi^{\eta,{\rm Coh}}_{\rm min} = \sqrt{ \frac{1}{\eta\, \overline{n}}} \,,
\end{equation}
Eq.\,(\ref{eq:phiSNsqz}) turns into
\begin{equation}
\label{eq:phiSNsqzloss}
\Delta \phi^{\eta,{\rm CSV}}_{\rm min} \approx \sqrt{ \frac{\eta e^{-2r} + 1 -\eta}{\eta\, \overline{n}}} \, 
\end{equation}
and 
Eq.\,(\ref{eq:phiHei}) turns into [\cite{Demkowicz-Dobrzanski2012,Demkowicz-Dobrzanski2013}]
\begin{equation}
\label{eq:phiHeiloss}
\Delta \phi^{\eta}_{\rm min} = \sqrt{  \frac{1-\eta}{\eta\, \overline{n}}  } \hspace{4mm}  {\rm for \;\;} 0 < \eta < 1 \,.
\end{equation}
For non-zero photon loss, most interestingly, the ultimate sensitivity of a phase measurement for a given photon number also shows a $1\!/\!\sqrt{\overline{n}}$\,-scaling.
The difference between the CSV strategy of using bright coherent states in combination with squeezed vacuum states, which is bounded by Eq.\,(\ref{eq:phiSNsqzloss}) and the strategy of using the optimal nonclassical state, which is bounded by Eq.\,(\ref{eq:phiHeiloss}), is marginal in practice [\cite{Demkowicz-Dobrzanski2013}]. 
For gravitational-wave detectors and for any other laser interferometer using intense light there is no need for an alternative to the CSV strategy.\\
~\\

We now turn back to the shot noise according to Eq.\,(\ref{eq:phiSN}). 
Generally, noise can be decomposed into its spectral contributions.
For a simple Michelson interferometer without arm resonators and without a signal-recycling cavity, the square-root of the single-sided shot-noise spectral density normalized to the differential arm length change $x$  in units of {m/$\!\sqrt{\textrm{Hz}}$} is given by\,[\cite{Saulson1994}]
\begin{equation}
\label{eq:xSN}
\sqrt{S^{\rm MI}_{{\textrm{SN}},x}}=\sqrt{\frac{\hbar c^{2}}{2 \omega P}} \propto \frac{1}{\sqrt{P}} \,,
\end{equation}
where $\omega$ is the optical angular frequency of the quasi-monochromatic carrier light and $P$ the total light power in both arms, including the built-ups from cavities. 
In combination with a squeezed vacuum whose relative phase generates squeezing of the output light's amplitude quadrature, the right hand side reduces according to the factor ${\rm e}^{-r}$. 
Note that the single-sided spectral density is only defined for {positive} sideband frequencies and thus twice as large as the double-sided spectral density.

\begin{figure}[ht]
  \vspace{0mm}
  \includegraphics[width=11.8cm]{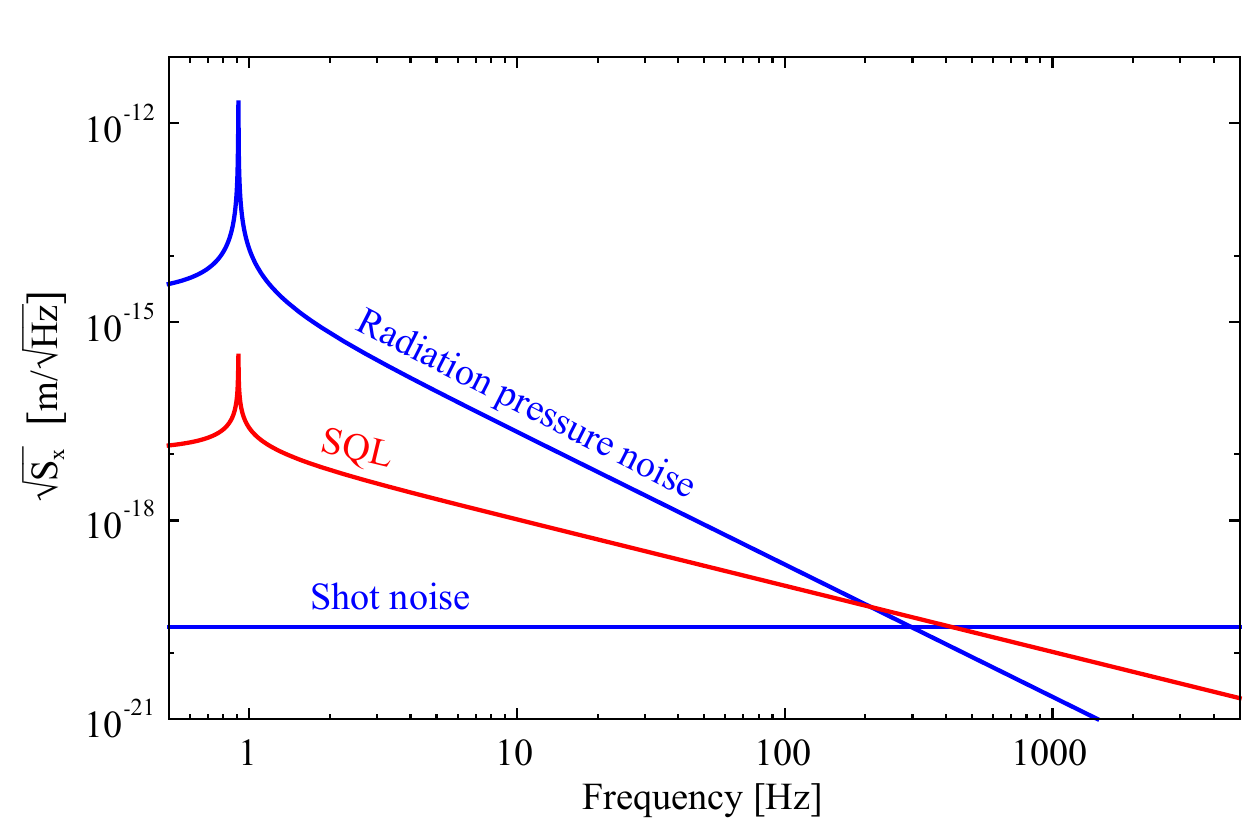}
  \vspace{-1mm}
\caption{\textbf{Displacement-normalized quantum noise spectral densities} -- Considered is a simple Michelson interferometer with neither arm cavities nor signal recycling. The two end mirrors ($m = 100$\,g) of the interferometer arms are suspended as pendula having a resonance frequency of $\Omega_M / 2 \pi= 1$\,Hz and a Q-factor of 10$^7$. The interferometer uses quasi-monochromatic light (in coherent states) with a total power of 4\,kW. Optical loss and the offset from a dark output fringe is assumed to be negligible. Wavelength: $\lambda = 1550$\,nm. The standard quantum limit (SQL) corresponds to the lowest noise achievable at a given sideband frequency when varying the light power without using quantum correlations. 
}
\label{fig:23}
\end{figure}

The spectral density of the measurement of a GW induced strain is given by the same expression but normalized to $h = x/L$. (If the gravitational wave is oriented in an optimal way with respect to the Michelson interferometer, one arm is squeezed while the other is expanded by the same amount of $\Delta L = x/2$ and $h$ then corresponds to the actual gravitational-wave amplitude). The square-root of the single-sided shot-noise spectral density normalized to strain in units $1/\!\sqrt{\textrm{Hz}}$ is given by
\begin{equation}
\label{eq:hSN}
\sqrt{S^{\rm MI}_{{\textrm{SN}},h}}= \sqrt{\frac{\hbar c^{2}}{2 L^2 \omega P}}\,,  %
\end{equation}
Equations (\ref{eq:xSN}) and (\ref{eq:hSN}) show that the smallest measurable signal (corresponding to unity signal-to-\emph{shot}-noise--ratio) is inversely proportional to the square root of the laser power and has a white spectrum for sideband frequencies much smaller than the carrier frequency, see horizontal line in Fig.\,\ref{fig:23}.\\

All first- and second-generation GW detectors use power-recycling and additional cavities to improve their sensitivities. Fabry-Perot arm resonators do not only increase the light power but additionally also increase the signal, for signal frequencies inside the resonator linewidth. 
For lossless Fabry-Perot arm resonators the spectral densities in Equations (\ref{eq:xSN}) and (\ref{eq:hSN}) need to be multiplied by the following factor [\cite{Kimble2001}]
\begin{equation}
\label{eq:HFP}
H_{\textrm{FP}} = \sqrt{ \frac{L^2(\gamma_{\textrm{FP}}^2 + \Omega^2)} {c^2} } \, ,
\end{equation}
where $\gamma_{\textrm{FP}} = c T_{\textrm{FP}} / (4L)$ is the Fabry-Perot arm resonator's half bandwidth and $T_{\textrm{FP}}$ is the light power transmission of the input mirror. The end mirrors are assumed to have perfect reflectivity. 
A similar expression can be derived for describing the improvement due to signal-recycling [\cite{Buonanno2001}].

In summary, shorter laser wavelengths, higher light powers, and squeezing of the amplitude quadrature of the interferometer output reduce shot noise in a broadband way, i.e.~for all signal frequencies. Fabry-Perot arm resonators as well as signal-recycling provide improvements mainly for frequencies inside the resonator linewidths.\\

\subsection{Quantum back-action and quantum radiation pressure noise}   \label{Ssec:5.3}

In laser interferometers, quantum back-action noise results from the uncertainty of the light's radiation pressure force on the interferometer mirrors, and is also called `(quantum) radiation pressure noise' (RPN). Its origin is the quantum uncertainty of the differential amplitude quadrature $X_{\Omega,\Delta\Omega}$ of the fields in the interferometer arms. It results in an uncertain momentum transfer to the mirrors and thus in an position uncertainty of the mirrors at future times with respect to their differential mode of motion [\cite{Caves1980}]. The physical mechanism of radiation pressure corresponds to an intensity dependent phase shift [\cite{Pace1993}].\\  
The higher the light power in the arms of a laser interferometer, the lower is its shot-noise spectral density, see Eq.\,(\ref{eq:xSN}). Unfortunately, the spectral density of  quantum back-action noise increases with light power. 
The single-sided force noise spectral density reads
\begin{equation}
\label{eq:Sff}
\sqrt{S_{{\textrm{RPN},F}}}= \sqrt{\frac{8 \hbar \omega P}{c^{2}}} \,.
\end{equation}
Whereas the force noise of the quantum radiation pressure has a white spectrum, the RPN does not, since the mirror's reaction to external periodic forces depends on frequency.   
The link between the Fourier component of an external force $F(\Omega)$ and the Fourier component of the displacement $x(\Omega)$ is given be the mechanical susceptibility $H_{\rm M}$. It reads for an harmonic oscillator with mass $M$
\begin{equation}
H_{\rm M}(\Omega) =  \frac{1}{M | - \Omega^2 +  \Omega_M^2 + i  \Omega \Omega_M /Q|}  \; ,
\label{eq:HM}
\end{equation}
where  $\Omega_M$ is the oscillator's resonance frequency and $Q$ its quality factor.

The square root of the single-sided spectral density of the RPN normalized to the displacement of an harmonic oscillator with mass $M$ is then given by
\begin{equation} 
\sqrt{S_{{\textrm{RPN}},x}} = H_{\rm M}(\Omega) \sqrt{\frac{8 \hbar \omega P}{c^{2} }}\,.
\end{equation}
In GW detectors, the test mass mirrors are suspended as pendula with high mechanical $Q$-factors and their centre of mass motion corresponds to that of a harmonic oscillator. The resonance frequencies of the pendula are lower than the detection band of interest. The mechanical susceptibility is therefore often approximated for the so-called free-mass regime as $H^{\rm fm}_{\rm M} (\Omega)= (m \Omega^2)^{-1}$. 
 The square root of the single-sided spectral density of the RPN normalized to differential displacement of two mirrors with each of mass $M$ in a simple Michelson interferometer is given by\,[\cite{Saulson1994}]
\begin{equation} 
\label{eq:xRPN}
\sqrt{S^{\rm fmMI}_{{\textrm{RPN}},x}}=\sqrt{\frac{2 \hbar \omega P}{c^{2} m^{2} \Omega^{4} }} \propto \sqrt{P} \,, \hspace{6mm} (\Omega \gg \Omega_M) \, ,
\end{equation}
where $m=M/2$ is the mirrors' reduced mass. In case of a simple Michelson interferometer that is enhanced with arm cavities the spectral density in  Eq.\,(\ref{eq:xRPN}) needs to be multiplied with the expression given in Eq.\,(\ref{eq:HFP}). 
 In combination with a squeezed vacuum whose relative phase generates squeezing of the output light's phase quadrature, the right hand side reduces according to the factor ${\rm e}^{-r}$.  Note if the radiation pressure noise is squeezed, the shot noise must be anti-squeezed, or vice versa.
The radiation pressure noise calibrated to strain of space time is given by the right side of Eq.\,(\ref{eq:xRPN}) divided by the interferometer arm length $L$.\\  
In summary, heavier masses, longer laser wavelengths, lower light powers, and squeezing of the amplitude quadrature in the interferometer arms reduce radiation pressure noise. 

~\\
\subsection{Interferometer total quantum noise and the standard quantum limit}  \label{Ssec:5.4} 

Both, shot noise and radiation pressure noise contribute to the total quantum noise of a given interferometer. If they are not quantum correlated, which is the case for a conventional Michelson interferometer when detecting the output light's amplitude quadrature, their variances add up. (The result is not shown in Fig.\,\ref{fig:23}). It can easily be deduced from the previous sections that changing the laser power will shift the two quantum noise contributions. However, the \emph{total quantum noise} never goes below the {standard quantum limit (SQL)} [\cite{Braginsky1967}]. 

Let us consider Fig.\,\ref{fig:23} for sideband frequencies much greater than the pendulum resonance. Here, the test mass mirrors react as free masses when exerted to external forces.
The SQL in this \emph{free-mass} regime is calculated by minimizing the sum of the squares of Eqs.\,(\ref{eq:xSN}) and (\ref{eq:xRPN}) \,[\cite{Saulson1994}]
\begin{equation}
\label{eq:Stot}
S^{\rm fm}_{{\rm tot},x}=\frac{\hbar c^2}{2 \omega} \left[ \frac{1}{P} + \frac{4 \omega^2}{c^4 m^2 \Omega^4} P       \right] \,.
\end{equation}
Its derivative reads

\begin{equation}
\frac{dS^{\rm fm}_{{\rm tot},x}}{dP} = \frac{-1}{P^2} + \frac{4 \omega^2}{c^4 m^2 \Omega^4} \,. \vspace{2mm}
\end{equation}
Setting the above equation to zero provides the optimum laser power versus sideband frequency in order to achieve the lowest total quantum noise
\begin{equation}
P^{\rm fm}_{\rm opt} = \frac{c^2 m \Omega^2}{2 \omega}  \,.
\end{equation}
Inserting the optimal light power into Eq.\,(\ref{eq:Stot}) provides the square root of the single-sided noise spectral density of the free-mass SQL in m/$\!\sqrt{{\rm Hz}}$
\begin{equation}
\label{eq:SQL}
\sqrt{S^{\rm fm}_{{\textrm{SQL}},x}} = \sqrt{\frac{2 \hbar}{m \Omega^{2}} }\,.
\end{equation}
Again, $m$ is the reduced mass, and dividing by the interferometer arm length $L$ yields normalization to the GW-induced strain $h$. 
Eq.\,(\ref{eq:SQL}) shows that the SQL falls off with sideband frequency. The corresponding equation for a Michelson interferometer that uses arm cavities reads
\begin{equation}
\label{eq:SQLfp}
\sqrt{S^{\rm fmFP}_{{\textrm{SQL}},x}} = \sqrt{\frac{\hbar}{m \Omega^{2}}   \left( \frac{1}{H_{\rm FP}} + {H_{\rm FP}} \right) } \,,
\end{equation}
with $H_{\rm FP}$ according to Eq.\,(\ref{eq:HFP}).

Using the expression for the SQL, the square root of the total quantum noise spectral density of a Michelson interferometer in the free-mass approximation can be written as
\begin{equation}
\label{eq:Stot}
\sqrt{S^{\rm fmFP}_{{\rm tot},x}} = \sqrt{\frac{S^{\rm fmFP}_{{\textrm{SQL}},x}}{2} \left[ \frac{1}{k} + k \right] }\,,
\end{equation}
with the radiation pressure coupling parameter 
\begin{equation}
\label{eq:k}
k(\Omega) = \frac{2 \omega P}{m c^2 \Omega^2}\,.
\end{equation}
For a fixed light power and fixed reduced mass of the mirrors the quantum noise limited interferometer reaches the SQL when $k = 1$, which is realized at the angular sideband frequency $\Omega_{\rm SQL} = \sqrt{2 \omega P /(m c^2)}$.  

Note that neither squeezing the phase quadrature nor squeezing the amplitude quadrature of the interferometer light leads to sub-SQL performance [\cite{Caves1981}], also confer Ref.~[\cite{Schnabel2005}] Fig.~3 (left).  
As we will see in the next sections the standard quantum limit \emph{can} be surpassed if shot and radiation pressure noise are correlated. Then the total quantum noise is \emph{not} given by the sum of the variances, i.e.~the sum of the squares in Eqs.\,(\ref{eq:xSN}) and (\ref{eq:xRPN}).

~\\
\subsection{Squeezed light for surpassing the standard quantum limit} \label{Ssec:5.5}

A measurement with sensitivity better than the standard quantum limit (SQL) is also called a `quantum non-demolition (QND)' measurement [\cite{Braginsky1995,Braginsky1996a,Kimble2001}]. Several QND techniques for laser interferometers were proposed in recent decades [\cite{Jaekel1990,Kimble2001,Purdue2002,Chen2003,McClelland2011,Danilishin2012,Graef2014}]. What they have all in common is they exploit quantum correlations between observable uncertainties. 

Arguably, the most extensive way of introducing quantum correlations and surpassing the SQL is the injection of squeezed states of light [\cite{Jaekel1990}]. If the squeezed quadrature angle of the injected states is neither 0$^\circ$ nor 90$^\circ$, the quantum uncertainties of the amplitude and phase quadrature amplitudes that describe the differential field in the two interferometer arms become correlated.\\ 
Let us consider a very simplified setup that just consists of a quasi-mono\-chromatic light field that is back-reflected from a quasi-free mirror. The light power and the mass be such that reflected light in a coherent state results in a measurement of the mirror position with a noise spectral density at the SQL at sideband angular frequency $\Omega_{\rm SQL}$. At this frequency quantum measurement noise and back-action noise are of the same size, i.e.~the uncertainty in $X$ produces an \emph{equally large, additional} uncertainty in $Y$. Upon reflection the quadrature amplitude variances change from $\Delta^2 \hat X= \Delta^2 \hat Y = 1/4$ to $2 \Delta^2 \hat X = \Delta^2 \hat Y = 1/2$. This result corresponds to the situation in Fig.\,\ref{fig:23} at the crossing frequency of shot noise and radiation pressure noise. The coupling of the uncertainty variances can be described by the matrix ${\bf K} = (1\, -\!k; 0\:\: 1)$, where $k = 1$ at the SQL. If the modulation state at $\Omega_{\rm SQL}$ is the ground state, its variances are transferred according to  
\begin{center}
\vspace{-7mm}
\begin{equation} \label{eq:k1}
{\bf K}^{\rm T}
\parenth{\!\begin{array}{cc}
1	&0	\\
0	&1	
\end{array}\!}
{\bf K}
=
\parenth{\!\begin{array}{cc}
\,\;1	&0	\\
\!\!-1	&1	
\end{array}\!} 
\parenth{\!\begin{array}{cc}
1	&0	\\
0	&1	
\end{array}\!}
\parenth{\!\begin{array}{cc}
1	& \!\!-1	\\
0	& \,\,1	
\end{array}\!}
=
\parenth{\!\begin{array}{cc}
\,\,1	& \!\!-1	\\
\!\!-1	& \,\,2	
\end{array}\!}.
\end{equation}
\end{center}
In accordance with Fig.\,\ref{fig:23} the variance of $\hat Y_{\Omega_{\rm SQL}}$ is twice as large as the vacuum noise variance.\\
Now, let the quantum noise of the light field be 10\,dB squeezed at 45$^\circ$ (Eq.\,(\ref{eq:r1})). %
The projection of the quantum uncertainty onto the $\hat X$-observable produces the radiation pressure noise by being transferred with the coupling factor $k = 1$ at the SQL into the $\hat Y$-observable, in fact with negative sign, since a larger value of $\hat X$ produces a larger optical path length and thus a retardation of the phase. Due to the squeezing at 45$^\circ$ the initial uncertainty in $\hat Y$ cancels with the additional uncertainty that originates from the one in $\hat X$. The following calculation shows that the strength of the cancellation corresponds to the initial  squeezing strength. Upon reflection the quantum uncertainties transform in the following way
\begin{center}
\vspace{-7mm}
\begin{equation}\label{eq:k2}
\parenth{\!\begin{array}{cc}
\,\;1	&0	\\
\!\!-1	&1	
\end{array}\!} 
\parenth{\!\begin{array}{cc}
5.05	& 4.95	\\
4.95	& 5.05	
\end{array}\!}
\parenth{\!\begin{array}{cc}
1	& \!\!-1	\\
0	& \,\,1	
\end{array}\!}
=
\parenth{\!\begin{array}{cc}
\,\,5.05	& \!\!-0.1	\\
\!\!-0.1	& \,\,0.2	
\end{array}\!}.
\end{equation}
\end{center}
The state of light after reflection has a squeezed phase quadrature amplitude. The improvement in comparison to Eq.\,(\ref{eq:k1}) is exactly 10\,dB. The quantum noise improvement corresponds to the input squeeze factor and is also a measure by what factor the SQL is surpassed.
Squeezed vacuum injection thus allows surpassing the SQL upon measuring the conventional $\hat Y$-quadrature (which is realized by a single photo diode in the interferometer's output port), as first realized by [\cite{Unruh1983,Yuen1983a,Jaekel1990}].\\ 
\begin{figure}[h!!!!!!!!!!]
  \vspace{0mm}
  \includegraphics[width=13.7cm]{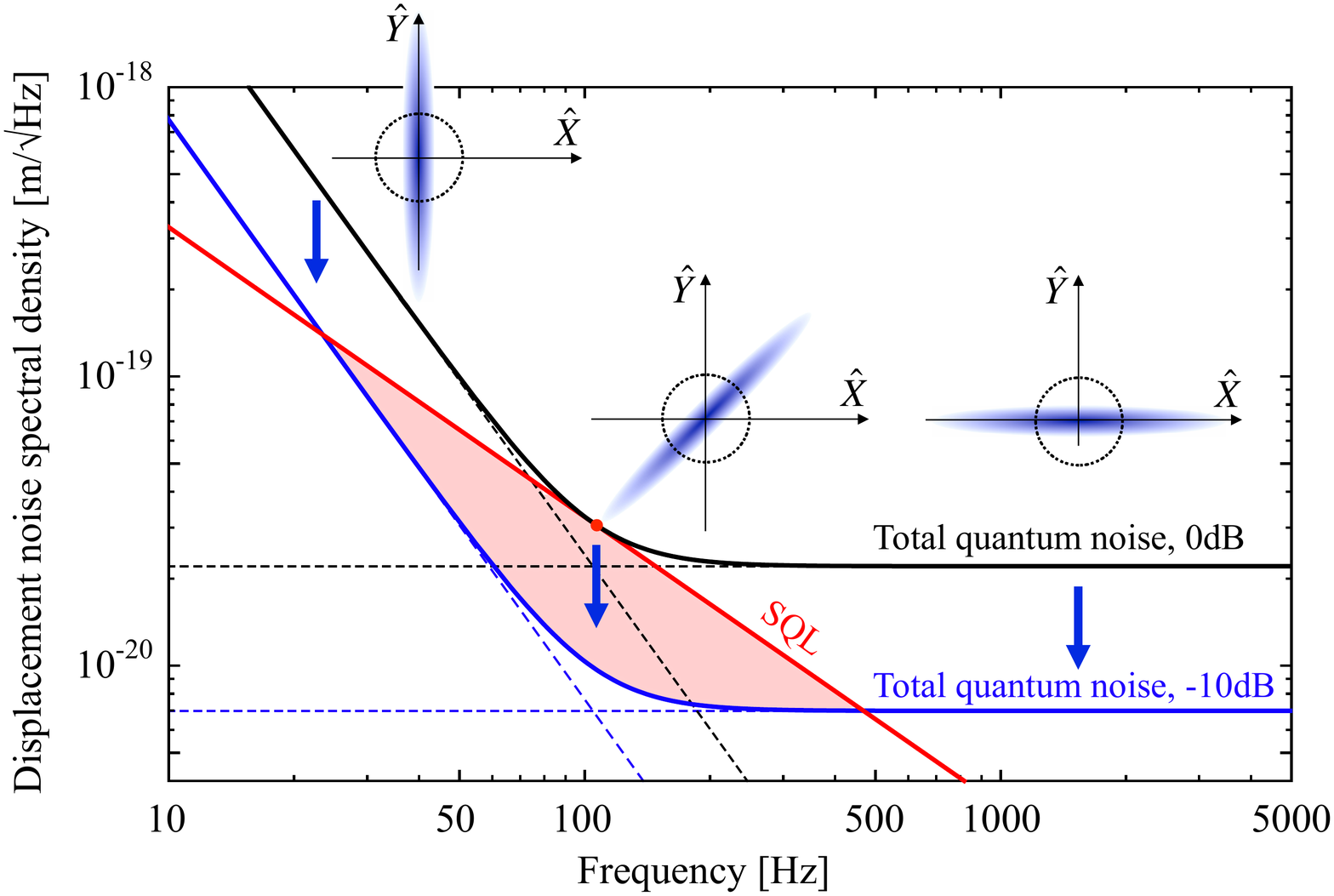}
  \vspace{-1mm}
\caption{\textbf{Surpassing the SQL with squeezed-light injection} -- At shot-noise limited sideband frequencies, squeezing of the $\hat Y$-quadrature amplitude improves the noise spectral density of the interferometer. At radiation-pressure-noise limited sideband frequencies, squeezing of the $\hat X$-quadrature amplitude improves the noise spectral density of the interferometer. If both kinds of quantum noise contribute equally (at the SQL, marked with a dot), a squeeze angle of $45^\circ$ results in surpassing the SQL by the full squeeze factor, see Eq.\,(\ref{eq:k2}). In the graph here, the squeeze angle is optimized for \emph{all} frequencies resulting in a broadband quantum noise reduction [\cite{Jaekel1990}]. Measurement sensitivities beyond the SQL (shaded area) are in the so-called quantum non-demolition (QND) regime [\cite{Kimble2001}]. Dashed horizontal lines represent the (squeezed) shot noise. Dashed straight lines with negative slope represent the (squeezed) radiation pressure noise. The calculations use 10\,dB of squeezing, a conventional Michelson interferometer with neither arm resonators nor signal recycling, a light power at the beam splitter of 1\,MW at a wavelength of $\lambda = 1550$\,nm, and mirror masses of 1\,kg.
}
\label{fig:24}
\end{figure}
In the example above the input squeeze angle is optimized for a single sideband frequency. Injecting a broadband squeezed vacuum field with frequency-independent squeeze angle of $45^\circ$ would result in a rather bad interferometer quantum-noise performance at frequencies far smaller or larger than $\Omega_{\rm SQL}$. Fig.\,\ref{fig:24} shows the quantum-noise performance if the input field has squeeze angles that are optimized for every $k(\Omega)$ as given in Eq.\,(\ref{eq:k}). 
Injected squeezing can thus lead to a broadband sub-SQL performance if the quantum measurement noise (shot noise) and the quantum back-action noise (radiation pressure noise) are correlated in an optimal way. Due to the correlation, shot noise and radiation pressure noise can be squeezed simultaneously.\\[4mm]
\FloatBarrier
\emph{Light with a frequency-dependent squeeze angle} \\[1mm]
The discovery that shot noise and radiation pressure noise can be squeezed simultaneously and thus a broadband reduction of quantum noise beyond the SQL be achieved required the insight that the spectral analysis of light defines a spectrum of many `sideband modulation modes'  that all can be in different quantum states. An ordinary squeezing resonator, which is on resonance for light at twice the pump wavelength, produces a spectrum of  modulation modes that all have the same squeeze angle. 
A frequency-dependent squeeze angle can be introduced by reflecting such a field from a detuned single-ended filter cavity, which was suggested by Kimble and coworkers [\cite{Kimble2001}]. They showed that the optimal frequency dependence that leads to the broadband improvement shown in Fig.\,\ref{fig:24} can be realized by using altogether two filter cavities as shown in Fig.\,\ref{fig:25}.
Motivated by this result, research and development on filter cavities for optimizing the frequency dependence of broadband squeezed fields has been very active in recent years  [\cite{Corbitt2004,Chelkowski2005,Dwyer2013,Kwee2014,Straniero2015,Oelker2016}].

\begin{figure}[h!!!!!!!!]
  \vspace{0mm}
  \includegraphics[width=9.4cm]{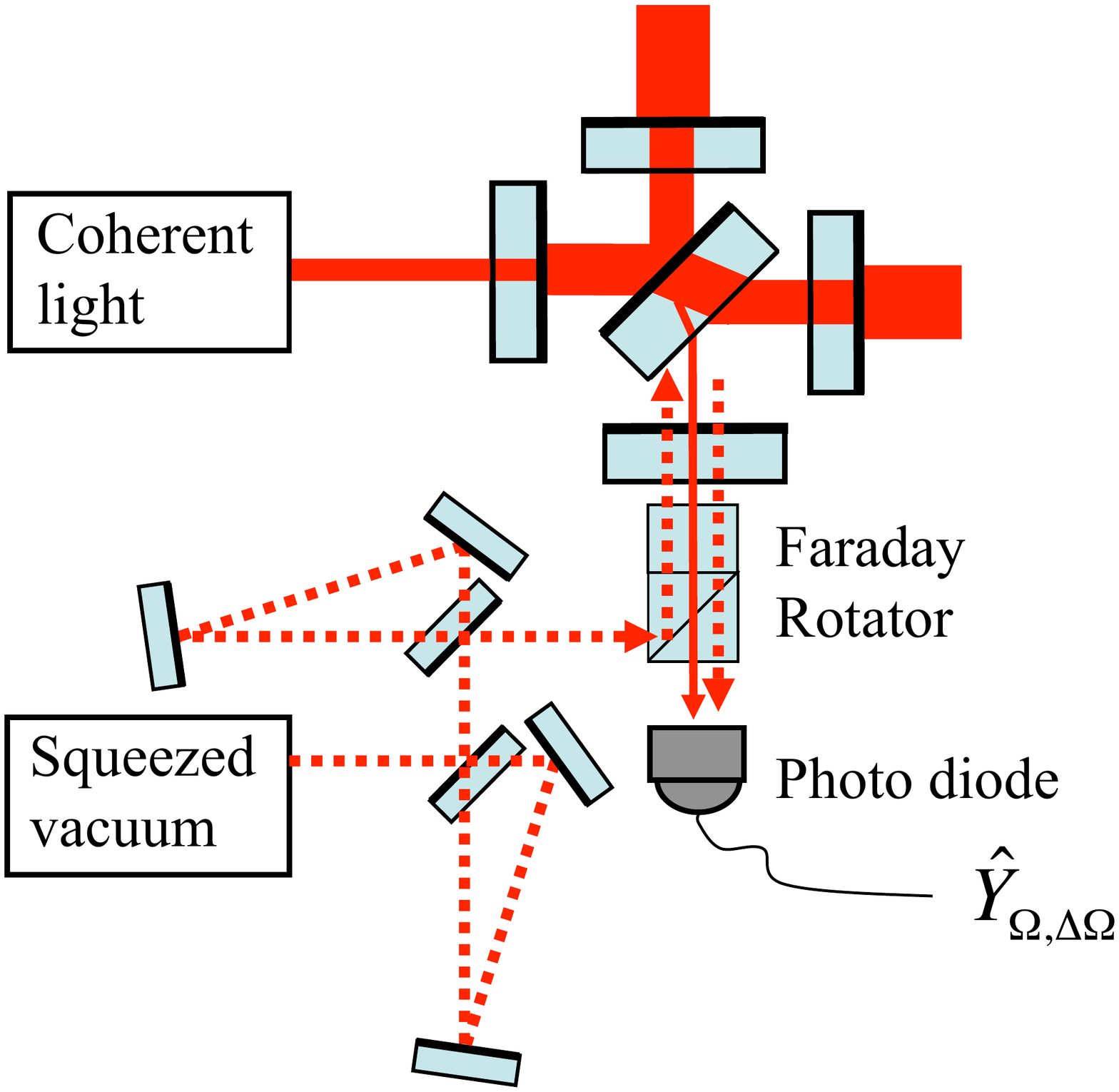}
  \vspace{-1mm}
\caption{\textbf{Frequency dependent squeezing injection} -- A broadband squeezed field with a frequency-dependent squeeze angle that is optimal for gravitational-wave detectors is produced by reflecting off an ordinary broadband squeezed field from two detuned optical filters [\cite{Kimble2001}].
}
\label{fig:25}
\end{figure}

\begin{figure}[h!!!!]
\vspace{-5mm}
\centerline{\includegraphics[width=10.6cm]{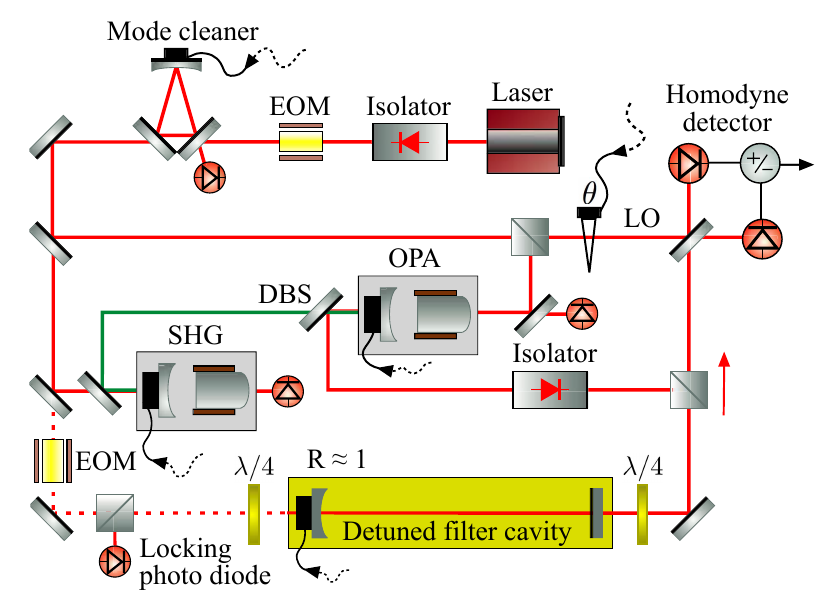}}
  \vspace{0mm}
  \caption{\textbf{Generation of a frequency-dependent squeezing} -- A frequency-dependent orientation of the squeeze ellipse was first demonstrated in Ref.~[\cite{Chelkowski2005}]. Initially, a conventional spectrum of squeezed vacuum states of light was generated in a squeezing resonator (`OPA'). The squeezed vacuum was transmitted through an optical isolator to a detuned filter cavity. After reflection, the squeezed vacuum states were absorbed in a balanced homodyne detector (BHD). The phase of the BHD's local oscillator (LO) was changed for quantum state tomography of the squeezed states in diffe\-rent regions of the spectrum. The result showed a frequency-dependent orientation of the squeeze ellipse, see Fig.\,\ref{fig:27}. SHG: second harmonic generation; EOM: electro-optical modulator for applying phase modulation sidebands for cavity length control; DBS: dichroic beam splitter; R: mirror reflectivity. $\lambda/4$: quarter wave plate for turning linear polarized light into circular polarised light and vice versa.
  }
  \label{fig:26}
\end{figure}

\begin{figure}[h!!!!]
\vspace{-5mm}
\centerline{\includegraphics[width=11.1cm]{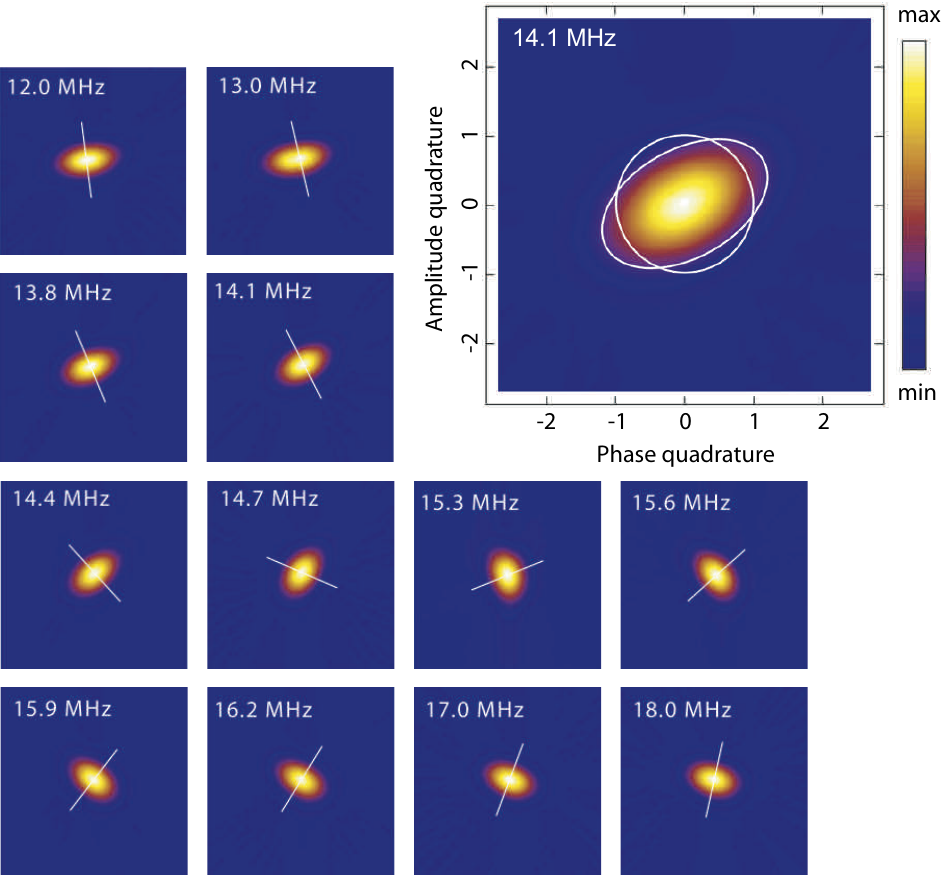}}
  \vspace{0mm}
\caption{\textbf{Frequency-dependent squeezing} -- Picture top right: Reconstructed contour plot of the Wigner function of the sideband modulation at $\Omega / (2 \pi) = 14.1$~MHz after reflection from a 15.15\,MHz detuned filter cavity. The state shows quantum correlations between phase and amplitude quadratures, i.e.~squeezing at an angle of, here,  about 40$^\circ$. The white circle visualizes the standard deviation of the vacuum state uncertainty. The white ellipse represents the standard deviation of the squeezed uncertainty. 
Small pictures:  Measurement results on the same continuous-wave laser beam at various sideband frequencies around $15$~MHz. For each tomographic picture, noise histograms of 100 equidistant quadrature angles were measured. In each case the laser beam was phase locked to a reference beam and the quadrature angle stably controlled and stepwise rotated. The phase reference was given by a phase modulation at $19.8$~MHz [\cite{Chelkowski2005}]. The picture was first published in Ref.~[\cite{Schnabel2005}] (copyright \textcopyright 2007 by Imperial College Press).
}
  \label{fig:27}
\end{figure}

A light field with a frequency-dependent squeeze angle was first demonstrated in Ref.~[\cite{Chelkowski2005}], see Figs.\,\ref{fig:26} and \ref{fig:27}. The experiment consisted of a standing-wave squeezing resonator, which produced an s-polarized, broadband amplitude quadrature squeezed field accompanied by a dim continuous-wave DC control field with a wavelength of $\lambda = 2 \pi c / \omega = 1064$\,nm. The squeeze bandwidth covered sideband frequencies up to about $\Omega / (2 \pi) =  30$\,MHz, which corresponded to the linewidth of the squeezing resonator. 
The optical cavity for producing the frequency dependence of the squeeze angle was a standing-wave cavity composed of a plane incoupling mirror of reflectivity $r_1=\sqrt{0.97}$ and a concave end mirror of reflectivity $r_2=\sqrt{0.9995}$. The cavity length was $L=50$\,cm resulting in a linewidth of $1.47$\,MHz. The squeezed field first passed a Faraday isolator to prevent interference effects between the filter cavity and the squeezing resonator.  A $\lambda/4$-waveplate turned the s-polarized field into a circularly polarized beam which was then mode matched into the detuned cavity. The retro-reflected field was analyzed by a balanced homodyne detector (BHD)
for quantum state tomography. 
The filter cavity was electro-optically controlled to be detuned by 15.15\,MHz with respect to the DC control field. The cavity length control was achieved by the Pound-Drever-Hall (PDH) locking technique utilizing a circularly polarized laser beam that carried $15$~MHz phase modulation sidebands and was coupled into the filter cavity from the back. The output voltage of the BHD was characterized by a spectrum analyser as well as used to perform quantum state tomography.
In the latter case the BHD output voltage was mixed down with an electronic local oscillator at different radio-frequencies around 15\,MHz and low-pass filtered to set the resolution bandwidth to $\Delta \Omega / (2 \pi)= 100$\,kHz. The final electric signal corresponds to a time series of quadrature amplitude measurements $X_{\theta,\Omega_i,\Delta \Omega}$.
Quantum state tomography is a method to reconstruct the phase space quasi-probability distribution (Wigner function) of quadrature amplitudes from sets of measured $X_{\theta,\Omega,\Delta \Omega}$ distributions when varying the angle $\theta$ [\cite{Leonhardt1997}]. For every sideband frequency $\Omega_i$, 100000 quadrature values were measured divided up on 100 equidistant quadrature angles. Each quadrature angle was stably controlled with a precision of $\pm 1^\circ$.
Fig.\,\ref{fig:27} shows the reconstructed Wigner functions, which were all measured on the same laser beam but at different sideband frequencies. For these measurements the detuned filter cavity was locked to the lower sideband at $-15.15$~MHz. The result clearly shows the frequency-dependent orientation of the squeeze ellipse. In a more recent experiment a frequency-dependent squeeze angle was also realized in the kHz regime [\cite{Oelker2016}].

\subsection{Optomechanically induced (ponderomotive) squeezing}   \label{Ssec:5.6} 

The radiation pressure of light, when acting on a movable mirror, results in an intensity dependent phase shift [\cite{Pace1993}]. The coupling produces a so-called `ponderomotive effect' [\cite{Braginsky1967}], which is of third order optical nonlinearity and which transforms a bright coherent state inside an interferometer into a squeezed state of light [\cite{Vyatchanin1993}]. This type of squeezed-light generation is usually called `ponderomotive squeezing' or `optomechanical squeezing'. 
Consequently, even if no squeezed field is \emph{injected} into the interferometer, correlations between the quadrature amplitudes are generated that allow for surpassing the SQL. \\
Ponderomotive squeezing, as produced by the interferometer itself, can only be exploited for evading back-action (radiation pressure noise). It can not be used to squeeze the interferometer shot noise. This is why ponderomotive squeezing is fundamentally less extensive than injecting externally produced squeezed states of light. 
[\cite{Corbitt2006}] suggested an external ponderomotive squeezing source for gravitational-wave detectors. In this case, due to its \emph{external} generation, also the interferometer's shot noise can be squeezed. Recently, ponderomotive squeezing was observed for the first time [\cite{Brooks2012,Purdy2013b}]. The achieved squeeze factors are much smaller than those produced by optical-parametric down-conversion [\cite{Vahlbruch2016}].

Let us have a look again at Eq.\,(\ref{eq:k1}). Rotating the covariance matrix on the right by arctan$(-\sqrt{5/4}-1/2) \approx -58^\circ$ indeed reveals squeezing,
\begin{center}
\vspace{-4mm}
\begin{equation} \label{eq:k3}
\parenth{\!\!\begin{array}{cc}
{\rm cos}\,58^\circ	&\!\!-{\rm sin}\,58^\circ\!	\\
{\rm sin}\,58^\circ	&\,\,{\rm cos}\,58^\circ\!	
\end{array}\!\!} 
\parenth{\!\begin{array}{cc}
\,\,1	& \!\!-1\!	\\
\!\!-1	& \,\,2\!	
\end{array}\!\!}
\parenth{\!\begin{array}{cc}
\,\,{\rm cos}\,58^\circ	&{\rm sin}\,58^\circ\!	\\
\!\!-{\rm sin}\,58^\circ	&{\rm cos}\,58^\circ\!	
\end{array}\!\!} 
\approx
\parenth{\!\begin{array}{cc}
\!2.62	& \!0	\!\\
\!0	& \!0.38\!	
\end{array}\!}.
\end{equation}
\vspace{1mm}
\end{center}
The vacuum-noise normalized variance of 0.38 corresponds to about 4.2\,dB of ponderomotive squeezing. This is the general value that is  produced at the angular sideband frequency $\Omega_{\rm SQL}$. At higher frequencies the squeeze factor gets smaller, at lower frequencies higher. The squeezing strength of 4.2\,dB can be observed if the photo diode in the interferometer output port is replaced by a balanced homodyne detector using a local oscillator phase of about  $-58^\circ$. 
It can be shown, however, that the optimal signal-to-quantum-noise-ratio at the SQL is achieved for a local oscillator phase of exactly 45$^\circ$. At this angle back-action is fully evaded.

Full evasion of radiation pressure noise at all frequencies requires an optimized frequency dependence of the relative local oscillator phase. This can be achieved by reflecting off the interferometer output field from two detuned filter cavities [\cite{Kimble2001}]. The scheme was called `variational output'. In the case of zero optical loss this scheme can fully evade radiation pressure noise, just leaving the shot noise as the only quantum noise contribution.

The variational-output scheme can be used to enhance the frequency-dependent squeezed input scheme.  The right site of Eq.\,(\ref{eq:k2}) shows that the output state's squeezing is not optimally detected in the $Y$-quadrature. Rather than with a single photo diode, the detection should be done with a balanced homodyne detector with optimized phase of its local oscillator. In this case, the output light's quantum noise is solely given by squeezed shot noise. The total quantum noise in Fig.\,\ref{fig:24} would then be given by the lowest (dashed) horizontal line.
This combined scheme was called `squeezed variational' [\cite{Kimble2001}]. It can be realized by reflecting off the interferometer output light from in total two optical filter cavities placed in front of the balanced homodyne detector.

\subsection{Conclusions} 

The highest quantum-noise-limited sensitivities of high-precision laser interferometers are achieved by employing a large number of quanta, to maximize the signal strength, in combination with strongly squeezed states, to minimize the quantum noise. 
From this perspective it is clear that the quantum-noise-limited sensitivity of future gravitational-wave detectors will be further improved -- by increasing the light power \emph{and} the squeeze factor. To be able to do so, the optical loss in these devices needs to be reduced.

In principle, the optical loss in laser interferometers can be made small, but never zero. Recent theoretical research has shown that for any non-zero loss the sensitivity scales proportional to $1/\sqrt{\overline{n}}$ at best, where $\overline{n}$ is the average photon number per measurement. This scaling is efficiently achieved by combining strongly displaced coherent states with squeezed vacuum states of light.

If a repeated measurement is not only limited by quantum measurement noise but also by quantum back-action noise, squeezed states of light can be used to simultaneously reduce both, i.e.~in the case of an interferometer, shot noise and radiation pressure noise.\\

\section{The first application of squeezed light in an operating gravita\-tional-wave detector} \label{Sec:6} 

Squeezed states of light have been successfully used to improve the sensitivity of the gravitational-wave detector GEO\,600 from 2010 up to the point when this Review was written [\cite{LSC2011,Grote2013}]. After decades of proof-of-principle experiments [\cite{Xiao1987,Grangier1987,McKenzie2002,McKenzie2004,Vahlbruch2005,Vahlbruch2006,Vahlbruch2007,Vahlbruch2008,Goda2008}] the implementation of a squeezed-light source in GEO\,600 has resulted in the first sensitivity improvement beyond shot noise of a measurement device that targets new observations in nature.   
The implementation of squeezed states in GEO\,600 was not done to provide another proof-of-principle demonstration, but was realized because it offered a relatively cheap way of further improving the measurement sensitivity. Of course, the sensitivity of GEO\,600 can also be further increased by purely classical means, however, the  implementation of arm resonators to enable higher light powers without increasing the thermal load on the beam splitter, or even the realization of longer interferometer arms are much more expensive.
In this respect, the sensitivity improvement of GEO\,600 with squeezed light can arguably be regarded as the first `true' application that developed out of the field of `nonclassical (quantum) metrology'.
(Note that the term 'quantum metrology' is currently defined in different ways [\cite{Giovannetti2006,Goebel2015}], and the term 'nonclassical', referring to a nonclassical P-function, gives a distinct description.)

\subsection{Gravitational waves} 

Einstein's General Theory of Relativity  [\cite{Einstein1916}], or simply `General Relativity' (GR), predicts that accelerating mass distributions produce gravitational radiation, analogous to electromagnetic radiation from accelerating charges. Experimental evidence of their existence is given by the observation of the slow spiraling together of two neutron stars, caused by the loss of orbital energy to gravitational waves. The inspiral rate exactly matches the predictions of Einstein's theory\,[\cite{Weisberg2005}]. 
Recently Advanced LIGO observed gravitational waves for the first time [\cite{Abbott2016}], thereby giving the go-ahead for gravitational-wave astronomy. 
The gravitational-wave source was the final inspiraling and the merger of two black holes 1.3 billion light years away from earth. 

\begin{figure}[ht]
  \vspace{-6mm}
  \includegraphics[width=12.0cm]{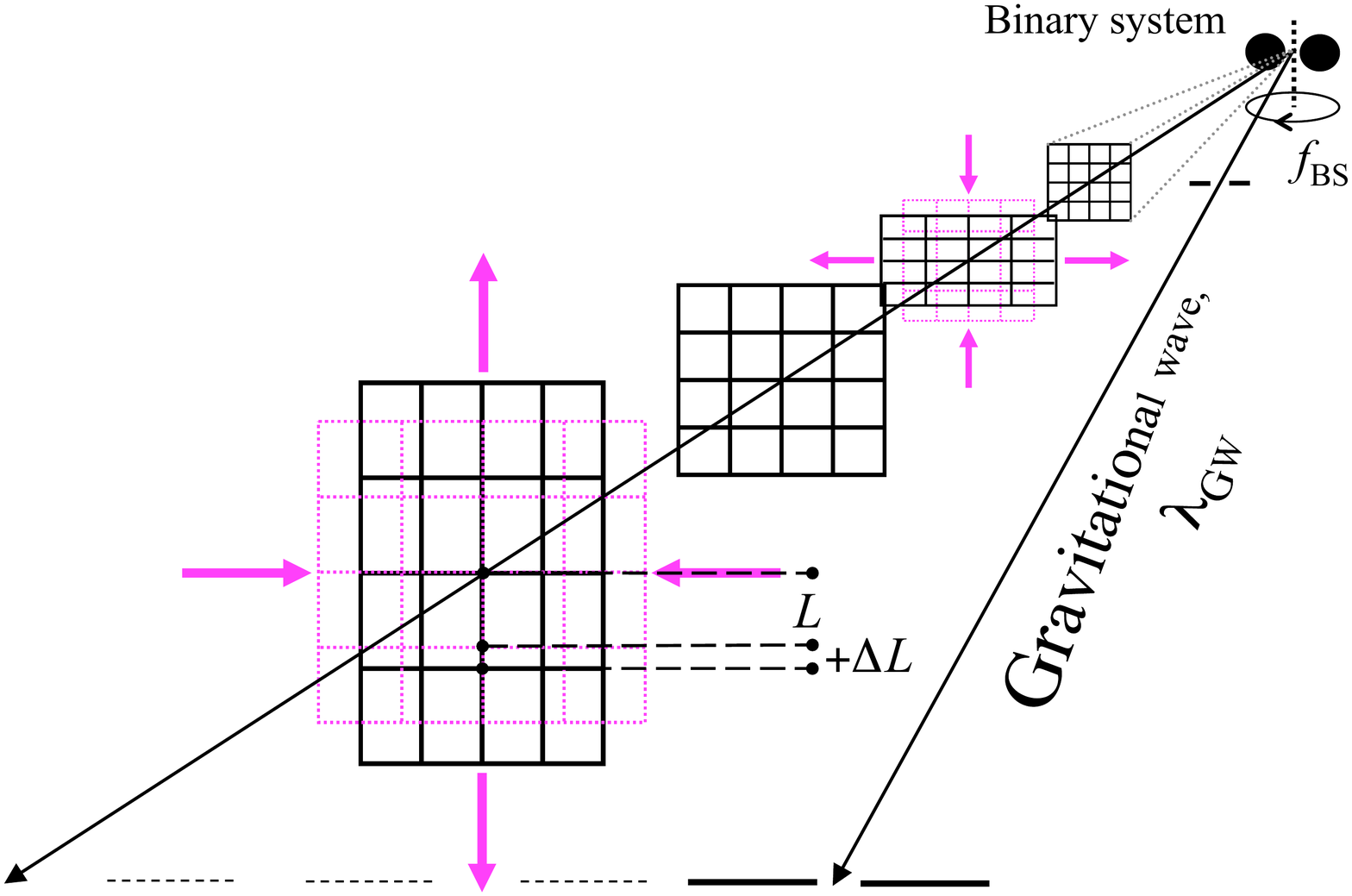}
 \vspace{-2mm}
\caption{
\textbf{Space-time oscillation} --  
Gravitational waves are dynamical deformations of space-time that form in the plane perpendicular to the direction of wave propagation. As a result, distances between free-falling test masses in a transverse plane will change with a strain $h = \Delta L / L$. For black hole or neutron star binary systems with orbital frequency $f_{\rm BS}$,  
distances will oscillate at frequency $f_{\rm GW}= 2 f_{\rm BS}$. The wavelength of this oscillation is given by $\lambda_{\rm GW} =  c / f_{\rm GW}$, where $c$ is the speed of light. The wave of orthogonal polarization with respect to the one shown is rotated by 45$^\circ$ around the propagation axis.
}
\label{fig:28}
\end{figure}

Gravitational-waves evolve in the far field of the source, propagate with the speed of light, and are measurable on earth with laser interferometers.  Fig.\,\ref{fig:28} displays a gravitational wave propagating along a certain direction. Gravitational waves are dynamical changes of space-time. They are transversal and quadrupolar in nature, and have two polarization states.

A variety of known astrophysical and cosmological sources are predicted to emit gravitational radiation that should reach the Earth with a measurable strength [\cite{Sathyaprakash2009}]. The first gravitational wave event detected was produced by two black holes of 36 and 29 solar masses. During the final 0.2 seconds of their inspiraling they produced a peak gravitational strain in our solar system of $10^{-21}$, covering frequencies up to 250\,Hz [\cite{Abbott2016}]. 
Other predicted sources are mergers of neutron stars, super novae and background signals from the Big Bang. 
According to GR, GWs from complex astrophysical sources carry a plethora of information that will have a major impact on gravitational physics, astrophysics and cosmology.

\subsection{Interferometric detection of gravitational waves}

Current gravitational wave detectors are kilometre-scale laser interferometers [\cite{Dooley2016,Aasi2015,Acernese2015,Aso2013}]. 
Continuous-wave laser light is split into two beams traveling in orthogonal directions. Both beams are reflected back towards the central beam splitter where they interfere. Gravitational waves change the optical path length difference, and thus the light power directed towards the photo-diode that is positioned in the signal output port of the beam splitter. A gravitational wave at frequency $f_{\rm GW} = \Omega_{\rm GW} / (2 \pi)$ reveals itself as a light-power modulation at the same frequency. 
The spectral decomposition of the output signal is described by a spectrum of the quadrature amplitude $\hat Y_{\Omega,\Delta \Omega}$ introduced in Sec.\,\ref{Sec:3}. It corresponds to the amplitude quadrature amplitude of the output light and  relates to the differential phase quadrature of the interferometer arms. 
\begin{figure}[ht]
  \vspace{-8mm}
  \includegraphics[width=12.2cm]{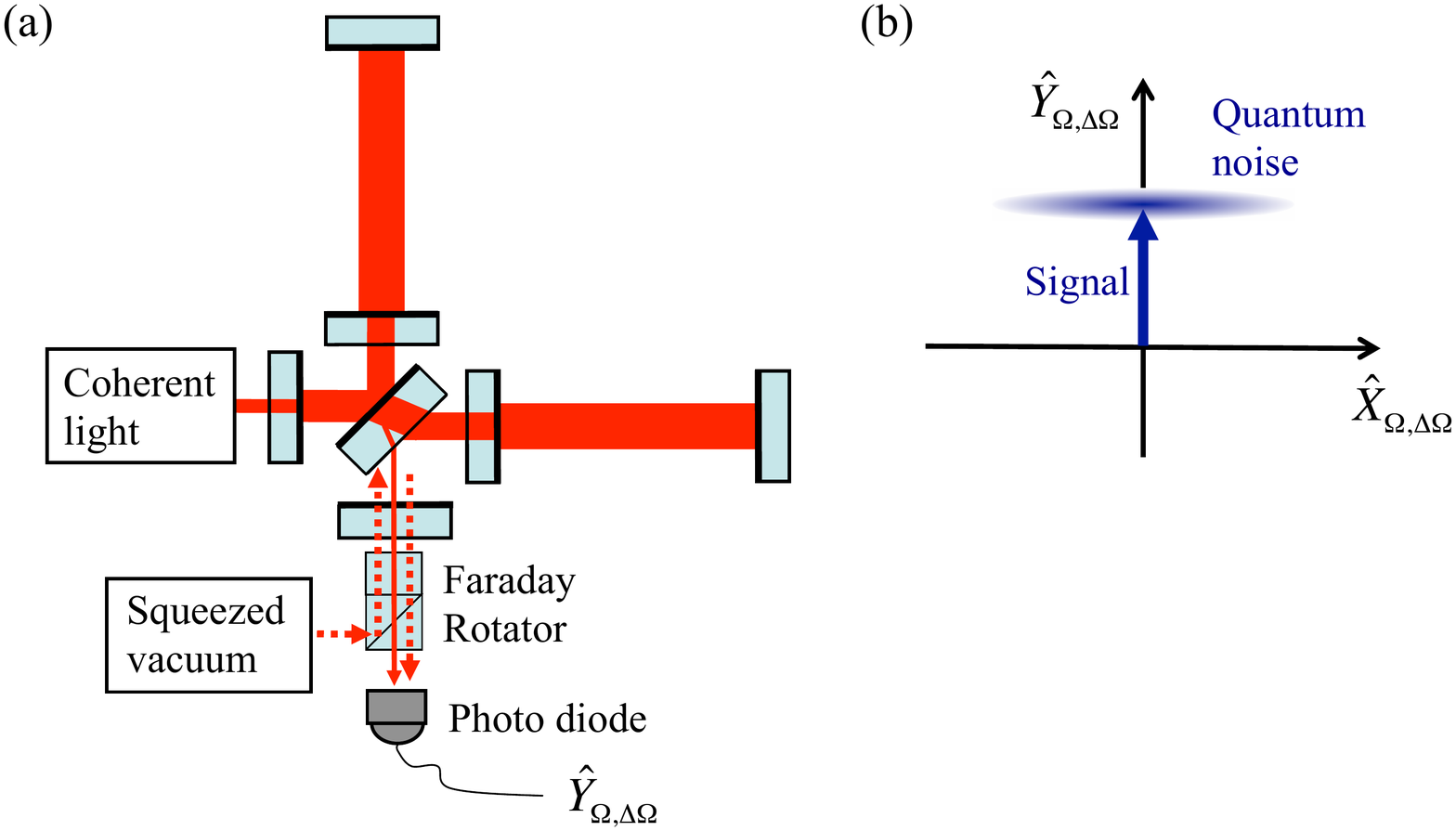}
  \vspace{-8mm}
\caption{\textbf{Squeezed-light-enhanced interferometric measurement} -- (a) Michelson interferometer with arm cavities, power recycling, and signal recycling (see main text for explanation). The interferometer is operated close to a dark fringe such that the quantum noise entering from the dark port is back-reflected. The squeezed field is mode-matched to the signal output field. (b) Phase space diagram of the gravitational-wave signal output at sideband frequency $\Omega / (2 \pi)$. The quantum noise is squeezed below the ground state uncertainty and thus the signal to quantum noise ratio improved.
}
\label{fig:29}
\end{figure}

The first key ingredient of an interferometric gravitational-wave detector are suspended, heavy mirrors that can be regarded as quasi-free in the direction of laser light propagation, thereby acting as test masses that probe spacetime. Being on ground, current detectors are located in rather noisy environments that allow the realization of undisturbed, quasi-free mirrors only above a sideband frequency of the order of 10\,Hz. Since sufficiently strong GW signals are expected up to a frequency of 10\,kHz, today's gravitational wave detectors target at signals in the acoustic band from 10\,Hz  to 10\,kHz.
The quasi-free motion of the test mass mirrors in this frequency regime is achieved by suspending the mirrors as sophisticated multiple-stage pendula in vacuum chambers [\cite{Aasi2015}]. Far above the pendula's resonant frequencies, which are typically around 1\,Hz, the centre of masses of the mirrors are isolated from vibrations of the ground and they react on frequency components of small external forces approximately as free masses. The mirrors and their suspensions are built from materials having exquisitely high mechanical quality factors. This helps to concentrate the thermal energy that causes displacements of the mirror surface into well-defined vibrational frequency modes. At these particular, very sharp frequencies, no gravitational waves can be detected. 

The second key ingredient of an interferometric gravitational-wave detector is laser light with a power of up to hundreds of kilowatts or even megawatts. The light is quasi-monochromatic and needs to show very low amplitude (quadrature) noise and phase (quadrature) noise at sideband frequencies within the detection band. Low amplitude noise is necessary to provide a shot noise limited output field. It is also necessary to avoid classical radiation pressure noise, which becomes an issue if the light power or the mirror masses in the two arms are not identical. 
Low phase noise is required if the storage time of the light in the two arms is not identical. This might accidentally occur due to different linewidths of the arm cavities or might be part of the interferometer design to allow for the length control scheme proposed by Schnupp [\cite{Heinzel1998}]. To maximize the light power inside the interferometer's cavities, it should be produced in an almost perfect transversal spatial distribution of a Gaussian TEM$_{00}$ mode.

Light sources of gravitational-wave detectors are ultra-stable Nd:YAG master-slave systems that provide up to 200\,W of light at 1064\,nm [\cite{Winkelmann2011,Kwee2012}].
The high power in the interferometer arms is achieved by cavity built-ups in the so-called power-recycling cavity 
and in the arm cavities. Power recycling uses a partially reflective mirror that is located between the light source and the interferometer beam splitter. Its surface is matched to the light's wave front and forms an optical cavity together with the rest of the interferometer. Since gravitational-wave detectors are operated close to a dark fringe, large power built-ups can be achieved. The highest power built-up is achieved for a mirror transmission equal to the (given) interferometer round trip loss. In this case an impedance-matched cavity is achieved. The power-recycling cavity as well as the arm cavities are stabilized on resonance for the input light. The difference between their functionality is that the power-recycling cavity does not limit the detection bandwidth of the interferometer. GEO\,600 as well as Advanced LIGO employ a third type of cavity, the so-called signal-recycling cavity.  
Similarly to power recycling, a partially reflecting mirror that is placed between the output port of the beam splitter and the photodiode is used to resonantly enhance the GW signal [\cite{Meers1988}]. The signal-recycling cavity resonantly enhances the signal modulation fields within its linewidth, without further enhancing the carrier light power. In combination with low linewidth Fabry-Perot arm resonators it can also be used to \emph{extract} the signal by reducing the effective finesse of the arm resonators for the signal sidebands. This scheme is called resonant sideband extraction [\cite{Heinzel1996}]. The signal-recycling cavity has also been tested in a detuned setting, in which just the upper or lower sideband is extracted or resonantly enhanced, respectively [\cite{Heinzel2002}]. Current gravitational-wave detectors, however, use carrier-tuned signal recycling.

All these techniques are `classical' approaches for maximizing the signal-to-shot-noise ratio. At frequencies above a few hundred Hertz, however, shot-noise is still the limiting noise source in gravitational-wave detectors.
Future gravitational-wave detectors will therefore use even higher light powers, but further increasing the light power becomes more and more challenging. Optical materials with less light absorption need to be found to counteract an increasing thermal load inside the mirrors. Mirror masses need to be further increased to counteract the increasing radiation pressure noise. 
Nonclassical approaches are superior and become more and more attractive the farther classical approaches are pushed to the extremes. Nonclassical approaches allow for simultaneously increasing the signal-to-shot-noise ratio and the signal-to-radiation-pressure-noise ratio \emph{without} changing light power or mirror masses, see Fig.\,\ref{fig:24}. They also allow for a complete evasion of radiation pressure noise [\cite{Braginsky1995,Braginsky1996a,Kimble2001}], see Subsec.\,\ref{Ssec:5.6}.\\

\subsection{Squeezed-light enhancement of the gravitational-wave detector GEO\,600}

\begin{figure}[hb!!!!!!!!]
  \vspace{2mm}
  \includegraphics[width=12.4cm]{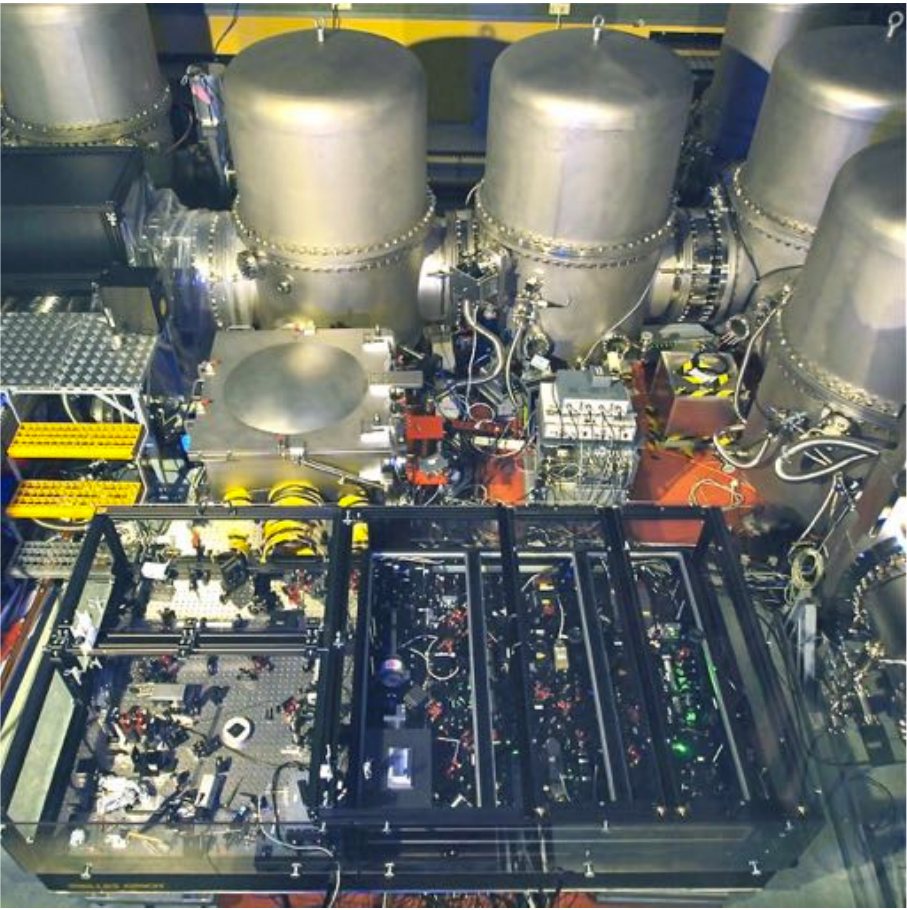}
  \vspace{-1mm}
\caption{
\textbf{GEO\,600 } --  View into the central building of the British-German GW detector located close to Hannover, Germany. The vacuum chambers contain the suspended beam splitter, power- and signal recycling mirrors, additional input and output optics as well as mirrors to realize a double pass of the laser light through the 600\,m long interferometer arms. By courtesy of the AEI.
}
\label{fig:30}
\end{figure}

In 2010, GEO\,600 was equipped with the squeezed-light source shown in Fig.\,\ref{fig:18}. The location of the squeezed-light source close to the output port is shown in Fig.\,\ref{fig:30}. It was known that GEO\,600 was shot-noise limited at sideband frequencies above about 700\,Hz. In this frequency regime the replacement of the ordinary vacuum states that entered the interferometer from the output port by a spectrum of squeezed vacuum states was expected to reduce the noise spectral density into the nonclassical regime. It was not precisely clear, what squeezing factors could be expected since the optical loss upon mode-matching an external field into the output port, propagation along the arms and the final photo-electric detection was not determined.

Critical components were the quantum efficiency of the photo-diode as well as the optical loss of Faraday rotator for overlapping the squeezed field with the output mode. Also the transversal output mode of GEO\,600 was an issue, since it contained significant contributions from higher order modes, to which a good mode matching of the squeezed mode was not possible. The implementation of the squeezed-light source thus was accompanied with a new custom made InGaAs photo-diode with 3\,mm diameter. The goal was a quantum efficiency of greater 99\% [\cite{Vahlbruch2016}]. Also the Faraday rotator was custom-made and optimized for lowest optical loss, which involved a rather precise rotation of the polarisation of 45$^\circ \pm 0.5^\circ$ over an aperture of more than 15\,mm. Finally a ring cavity (output mode cleaner) was placed in front of the photo-diode, which acted as a passive filter for higher transversal modes. Since GEO\,600 was not limited by radiation pressure noise and since it used a carrier-tuned signal-recycling cavity, a frequency independent orientation of the squeezing angle was optimum.
\begin{figure}[ht]
  \vspace{2mm}
  \includegraphics[width=13.6cm]{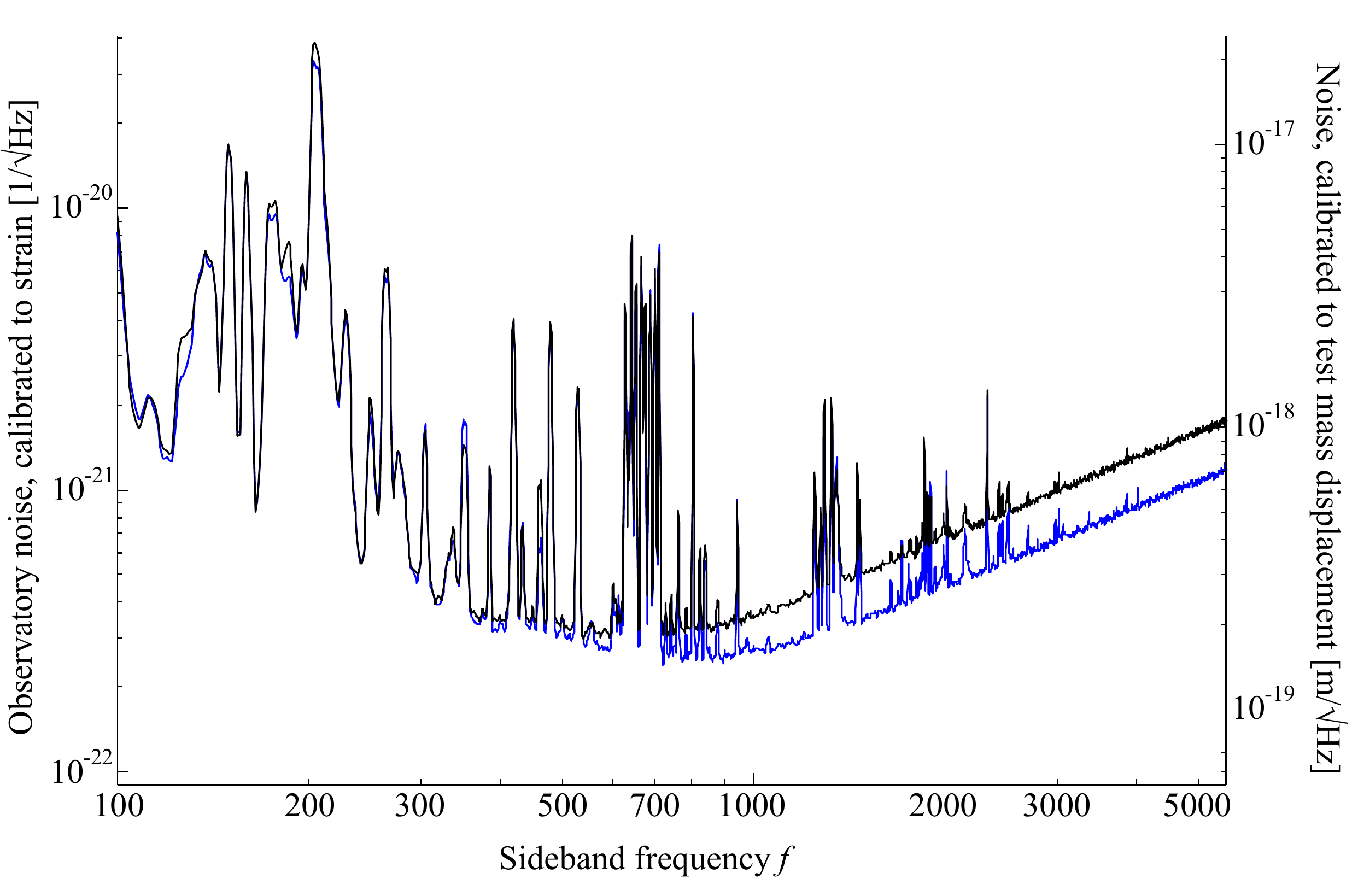}
  \vspace{-1mm}
\caption{
\textbf{Nonclassical reduction of the GEO\,600 instrumental noise} --  Shown are the square roots of the noise spectral densities without squeezed vacuum states (top) and with squeezed vacuum states (bottom) [\cite{LSC2011}]. Their calibrations [\cite{Affeldt2014}] to gravitational-wave strain and differential arm length change are shown on the left and right y-axes, respectively. Note that both traces increase towards higher frequencies due to the frequency-dependent signal enhancement of the signal-recycling cavity. The injection of squeezed vacuum states  leads to a broadband noise reduction of up to 3.5\,dB at shot-noise limited frequencies. The spectral features are for instance caused by excited violin modes of the mirror suspensions (600--700 Hz and harmonics). Data by courtesy of the AEI.
}
\label{fig:31}
\end{figure}
After several months the combination of the squeezed-light source and the gravitational-wave detector succeeded. The main laser of the  squeezed-light source was phase locked to the main laser of GEO\,600 and a stable mode matching between the squeezed field and the interferometer output field was achieved [\cite{LSC2011}]. The effect on GEO\,600's (strain normalized) noise spectral density is shown in Fig.\,\ref{fig:31}. At frequencies above about 700\,Hz the square root of noise spectral density was reduced by up to one third. This value corresponds to a quantum noise variance of 0.45 ($-3.5$\,dB), with the shot noise variance normalized to unity. For isotropically distributed gravitational-wave sources this factor produces a detection rate increase by a factor of $1.5^3 \approx 3.4$.
After its integration into GEO\,600 the squeezed-light source was used in all scientific runs seeking for gravitational waves, for instance in the observational run S6e/VSR4 that was undertaken from June 3$^{\rm rd}$ to September 5$^{\rm th}$ in 2011 [\cite{Grote2013}].

Towards the end of 2011, right before the start of the detector upgrade to Advanced LIGO, a nonclassical sensitivity improvement was also demonstrated in one of the LIGO detectors [\cite{Aasi2013}]. In this experiment a nonclassical sensitivity improvement corresponding to up to 2.15\,dB above frequencies of about 150\,Hz was achieved. The successful test is a strong motivation for a squeezed-light upgrade of Advanced LIGO. Note that the design of Advanced LIGO was completed in 1999, and squeezed-light sources were not mature at those times.   

In the past years the squeezing enhanced GEO\,600 detector was not only used for observations, but also was the control of the injected squeezed mode further improved. Stabilizing the overlap between squeezed mode and bright mode of the interferometer to close to perfect is necessary to reduce the effective optical loss and to maximize the measurable squeezing factor. Recently superior methods for stabilizing the longitudinal phase of squeezed vacuum mode were found [\cite{Dooley2015}] and the first automatic alignment system for stabilizing and optimizing the transversal mode overlap was demonstrated [\cite{Schreiber2016}].

\FloatBarrier
\subsection{Are squeezed states the optimal nonclassical resource in \\ gravitational-wave detectors?} \label{Optimality}

For a given number of photons, Eq.\,(\ref{eq:phiHei2}) quotes the ultimately smallest phase change that can be measured with a signal-to-noise-ratio of one. The scaling with number of photons per measuring time of this Heisenberg limit seems appealing compared to the scaling achievable with coherent states or squeezed states according to Eqs.\,(\ref{eq:phiSN}) and (\ref{eq:phiSNsqz}), respectively. The Heisenberg limit, however, is only valid for precisely zero photon loss. Since the nonclassical states required to achieve Eqs.\,(\ref{eq:phiHei}) and (\ref{eq:phiHei2}) show an exponentially increasing sensitiveness to loss when increasing the photon number, the actual scaling can not be deduced from  Eq.\,(\ref{eq:phiHei}).
Proposals to use Fock states and the so-called N00N states for optimizing interferometer sensitivities [\cite{Holland1993, Dowling1998, Mitchell2004, Afek2010}] are thus only applicable when the experiment is conditioned on zero photon loss. As discussed in recent publications, the correct expression for the fundamental sensitivity limit needs to consider not only the total photon number inside the interferometer but also the total photon loss [\cite{Dorner2009,Kolodynski2010,Knysh2011,Escher2011,Demkowicz-Dobrzanski2012}].

Based on these earlier works, Ref.\,[\cite{Demkowicz-Dobrzanski2013}] proved that the nonclassical sensitivity enhancement of GEO\,600 reported in Ref.\,[\cite{LSC2011}] has been exceedingly close to fundamental quantum interferometry bound under given energy constraints and photon loss levels. 
More than that, it was generally proven that  the approach of combining displaced coherent states and squeezed vacuum states is optimal for gravitational-wave detectors.

In Ref.\,[\cite{LSC2011}] the gravitational-wave detector GEO\,600 used an effective number of photons per second of approximately $\overline{n} = 2 \cdot 10^{22}$, which corresponded to a total optical power inside the interferometer arms of $P \approx 3.7$\,kW at a wavelength of 1064\,nm. The total optical loss was $1 - \eta \approx 0.38$. The injected squeezing factor was $e^{-2r} \approx 0.1$. For these numbers, the ratio of Eqs.\,(\ref{eq:phiSNsqzloss}) and (\ref{eq:phiHeiloss}) is calculated to
\begin{equation}
\label{eq:CSVgen}
\frac{\Delta \phi^{\rm CSV}_{\rm min} } {\Delta \phi^{\rm gen}_{\rm min} }  \approx \sqrt{ \frac{\eta e^{-2r} + 1 -\eta}{1-\eta }}   \approx \, 1.08 \, ,
\end{equation}
which is a good approximation within the limit of large coherent state displacements $ \alpha \gg {\rm sinh}^2 r$. 
The quantum noise of GEO\,600 including the squeezed-light source was just 8\% above the fundamental quantum interferometry bound. 
An increased squeezing strength of 16\,dB ($e^{-2r} \approx 0.025$), which is in reach, would bring the approach based on coherent states and squeezed vacuum states to within just 2\% above the fundamental bound.

Future GW detectors will have significantly reduced optical loss values $(1-\eta)$. `Loss' includes scattering and absorption at mirrors, non-perfect fringe contrasts, and the non-perfect quantum efficiency of the photo detector. 
Optical loss reduction is important for at least four reasons. First, it leads to an increased signal, second, it leads to a reduced quantum noise when employing squeezed states, third, less absorption reduces the thermal load on the test mass mirrors, and fourth, less scattering reduces the probability of back-scattered light, which produces disturbance signals [\cite{Billing1979,Vahlbruch2007,Punturo2014}].
The higher the finesse values of the arm and signal-recycling cavities are, the more significant is optical loss at mirror test masses, the beam splitter and the signal-recycling mirror. The finesse value of the power-recycling cavity and the loss of mirrors and lenses that guide the output field to the photo-diode are less critical. Suitable photo detectors of 99.5\% quantum efficiency are available today [\cite{Vahlbruch2016}], but achieving a total optical loss of 10\% is still challenging. The reason for that is that first of all a measurement device aiming for best absolute sensitivity should use as much quanta (photons) as possible. High finesse values for the enhancement cavities are thus essential, but results in an unavoidable scaling-up of the effect of mirror losses. A realistic example of future gravitational wave detectors thus considers $\eta = 0.9$ with a squeezing factor of 20\,dB ($e^{-2r} = 0.01$). In this case, the quantum noise will be about 4\% above the ultimate fundamental bound for a given photon number.

From Eq.\,(\ref{eq:CSVgen}), it can be concluded that there is no need for any more sophisticated nonclassical states than squeezed states. In particular nonclassical states with a defined photon number, such as N00N states, are not required. Within the approximation quoted, this result is independent of the photon number. This result is also independent of the physical system used for interferometric phase estimation and can also be made for quantum-enhanced atomic clock calibration in the presence of dephasing. Here, theoretical results indicate that the precision of Ramsey interferometry with spin-squeezed states is close to the optimal one in the asymptotic regime of a large number of atoms [\cite{Huelga1997,Ulam-Orgikh2001,Escher2011}] as already stated in Ref.\,[\cite{Demkowicz-Dobrzanski2013}].
More sophisticated nonclassical states with fixed number of $n$ quanta might still be useful for the exceptional case when the absorption of  \emph{one} quantum already results in zero measurement sensitivity anyway. An example is an ensemble measurement where the absorption of a single photon demolishes the source of the phase change to be characterized. A typically used approach of conditioning the measurement result on $n$ clicks of $n$ single photon counters conditions on precisely zero loss and is thus able to use the advantage of Eq.\,(\ref{eq:phiHei}) over Eq.\,(\ref{eq:phiSNsqz}).

\FloatBarrier

\subsection{Conclusions}

Squeezed states of light will contribute to realizing gravitational-wave observatories with much higher sensitivities than existing or planned ones. To benefit from squeezed states in a most efficient way, optical loss in terms of absorption and scattering must be minimized. In particular the optical loss of mirror coatings and mirror substrates need to be minimized. The relevant mirrors include the test masses, the balanced beam splitter, the signal recycling/extraction mirror and all optical components between the latter and the photo diode. Excellent spatial mode matching between the bright interferometer field and the squeezed vacuum field is also of great importance. Achieving this requires further improvement of the surface figures of all reflective optical components of the interferometer, as well as improved homogeneity of all  optical components that the light passes through. 

The quantum noise reduction achieved in a gravitational-wave detector is of course always smaller than the highest squeeze factor provided by the squeezed-light source. As an example, let us consider the observation of 15\,dB of nonclassical noise suppression directly at the source. If the squeezed field senses an additional loss of 5\% when propagating through the interferometer, which is a very challenging number from today's point of view, the remaining squeezing level is about 11\,dB, see Eq.\,(\ref{eq:loss}).\\

\section{The application of 2-mode-squeezed light in laser interferometers} \label{Sec:7} 
\subsection{Quantum Dense Metrology}
\begin{figure}[hb!!!!!!!]
  \vspace{-1mm}
  \includegraphics[width=12.4cm]{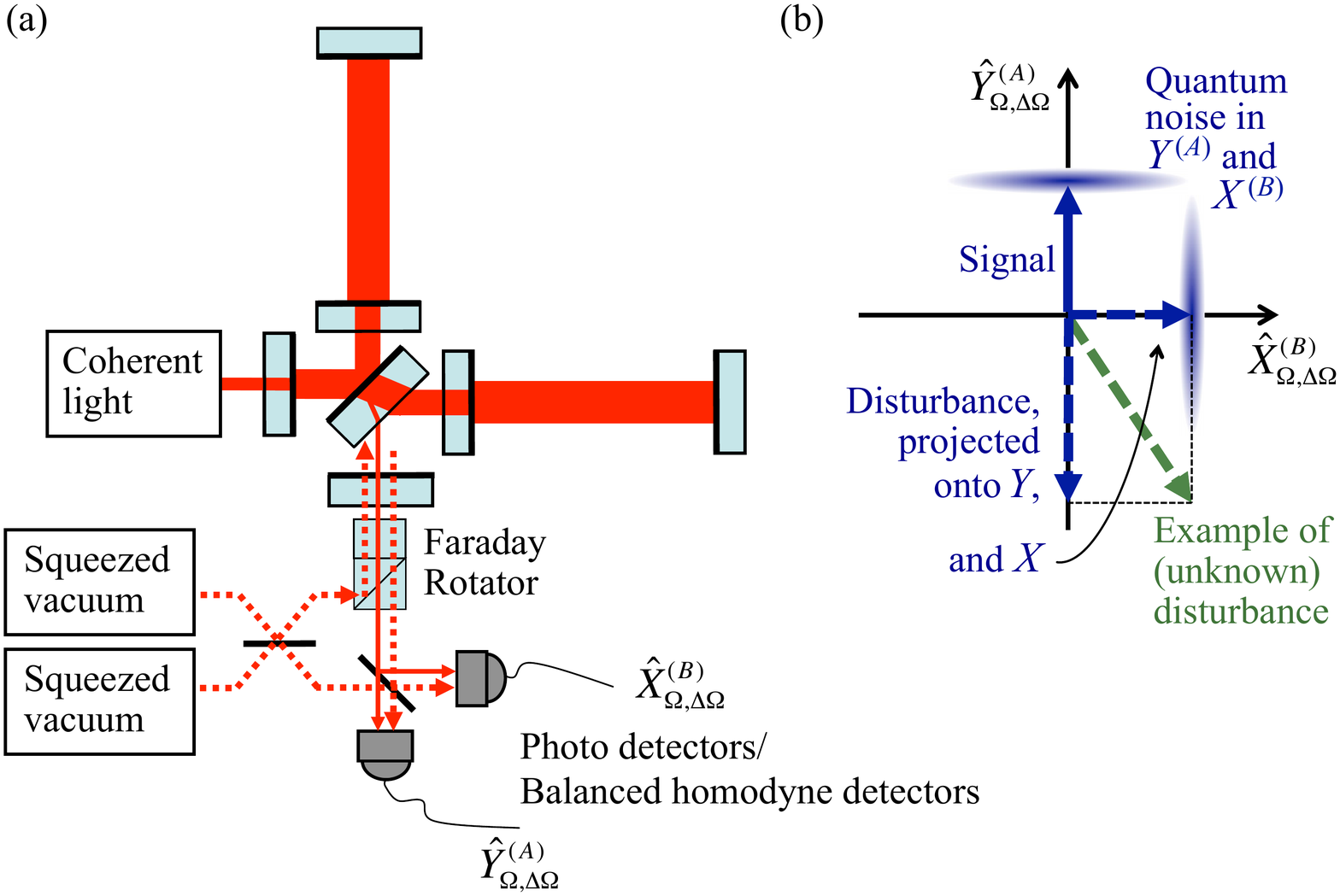}
  \vspace{0mm}
\caption{\textbf{Bi-partite-squeezed-light-enhanced measurement} -- (a) Setup for the application of bi-partite (two-mode) squeezed light in a laser interferometer on the basis of QDM. Two squeezed vacuum fields are overlapped on a balanced beam splitter with programmable squeeze angles, for instance with a relative angle of 90$^\circ$, which produces a bi-partite state as shown in Fig.~\ref{fig:12}. The beam splitter outputs are entangled for any relative angle greater than zero. One part is matched to the interferometer mode. The second part is kept outside as a reference beam. The interference of the interferometer output and the reference beam is arranged with such a phase difference that it reproduces the two squeezed inputs on the photo detectors. The two squeezed beams are photo-electrically detected measuring the respective squeezed quadrature (using balanced homodyne detectors). Both beams carry half of all interferometer induced modulations, which include signals as well as disturbances. A single readout as shown in Fig.~\ref{fig:21} cannot distinguish between the two kinds. The double readout shown here provides additional information and allows for recognition of the disturbance [\cite{Steinlechner2013b}] as well as in principle a modeling of the disturbance and, with a correct model, an improvement of the noise spectral density of the interferometer [\cite{AstM2016}]. (b) Phase space diagram describing phase quadrature readout $A$ as well as amplitude quadrature readout $B$. Both show squeezed quantum noise. The amplitude quadrature readout does not contain any gravitational-wave signal, i.e.~any feature in this channel must be due to disturbances. This information can be used to improve the interferometer.
}\vspace{0mm}
\label{fig:32}
\end{figure}
At first glance, the application of bi-partite (two-mode) squeezed states to a device whose goal is measuring a \emph{single} observable seems meaningless. Squeezing the uncertainty of that observable should be the optimum one can do. This is indeed true when concerning just quantum noise, but recently it was discovered that in the presence of classical disturbances, bi-partite squeezing can improve such measuring devices [\cite{Steinlechner2013b}]. 
The concept was named \emph{quantum dense metrology} (QDM).
The potential improvement of a gravitational-wave detector with bi-partite squeezed states is shown in Fig.~\ref{fig:32}\,(a). A description is given in the caption.
The pre-condition for a potential improvement can be best understood within a phase space diagram. Fig.~\ref{fig:32}\,(b) contains two different kinds of `signals'. The first is the actual signal, which always shows up as a phase space displacement along the $Y$ axis. The second is a disturbance signal that can produce a displacement in arbitrary direction in phase space. A prominent example for such a disturbance is parasitic interference due to back-scattered laser light [\cite{Vahlbruch2007}]. Back-scattering is a limiting noise at low signal frequencies of gravitational-wave detectors [\cite{Billing1979,Vinet1997,Hild2007Diss,Ottaway2012,Punturo2014}]. Note that all noise that couples in via unwanted motions of the test mass mirrors, so-called `displacement noise', always produces a phase space displacement along the $Y$ axis and cannot be tackled with QDM.

\begin{figure}[t!!!!!!!!!!!!!!!!!!!!!!!!!!!!!!!!!!!!!!!!!!!!]
  \vspace{-6mm}
  \includegraphics[width=9.1cm]{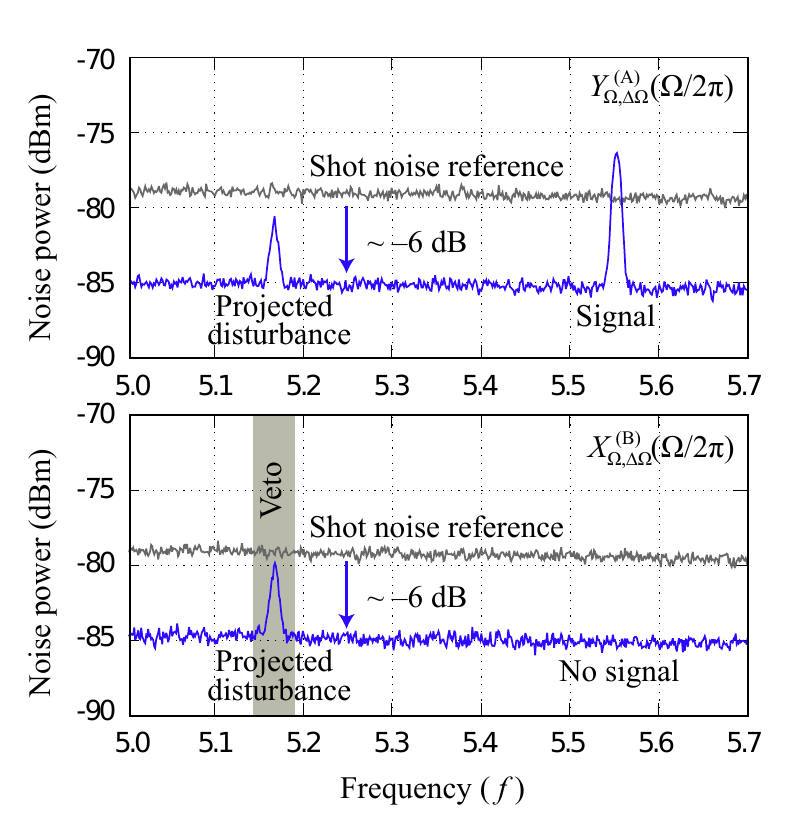}
  \vspace{-2mm}
\caption{\textbf{Bi-partite-squeezed-light-enhanced measurement} -- The result was achieved in a table-top setup [\cite{Steinlechner2013b}]. In the two panels, the lower (blue) traces show the squeezed quadrature noise-power spectra $\Delta^2 {\hat Y^{\rm (A)}_{\Omega,\Delta \Omega}}(\Omega/2\pi)$ (top) and $\Delta^2 {\hat X^{\rm (B)}_{\Omega,\Delta \Omega}}(\Omega/2\pi)$ (bottom) as simultaneously measured with balanced homodyne detectors `A' and `B', respectively. The conventional $Y$-measurement (top) cannot distinguish between signal and disturbances. The additional $X$-measurement (bottom) does not detect any phase quadrature signal, thus any feature in this measurement is a parasitic signal due to a disturbance. The respective projection onto the $Y$-measurement can thus be `vetoed'. In a more sophisticated approach, the $X$-data might be used to model and then to eliminate the disturbance as well as its projections on both quadrature measurements.  The result is a reduced spectral density of the actual phase quadrature measurement [\cite{AstM2016}].  
Traces shown here are slightly sloped due to the decreasing transfer functions of the balanced homodyne detectors. The resolution bandwidth was $\Delta\Omega / (2 \pi)= 10$\,kHz, the video bandwidth was 100\,Hz. All traces were averaged three times. 
}
\label{fig:33}
\end{figure}

Fig.~\ref{fig:33} shows measurement results obtained in Ref.~[\cite{Steinlechner2013b}]. In a table-top experiment, one part of a bi-partite squeezed state of a continuous-wave mode at 1064\,nm was mode-matched into the output port of a Michelson laser interferometer operated at its dark fringe, in full analogy to Fig.~\ref{fig:32}(a). A `signal' was produced by driving the piezo behind one of the end mirrors at a frequency of 5.55\,MHz. The `disturbance' was introduced by re-injecting a small amount of light that leaked through the second end mirror with an additional piezo-mounted mirror. The piezo was driven at a frequency of 5.17\,MHz to produce a phase modulation. An additional DC voltage defined an arbitrary and unknown optical path length of the light before being re-injected and as such the phase space orientation of the disturbance signal. 
This mechanism of a parasitic interference is realized naturally in any interferometric device due to back-scattering of quanta from moving surfaces in the environment.

The interferometer output consisted of the signal as well as the disturbance, with a quantum uncertainty given by one subsystem of the bi-partite entanglement. It was overlapped with the second subsystem of the entangled state on a balanced beam splitter and the two outputs were analysed with balanced homodyne detectors. The phases of the bi-partite entanglement and the BHD local oscillators were controlled to resemble Fig.~\ref{fig:32}(b), i.e.~both BHDs measured a squeezed uncertainty, regardless of the phase of the (generally unknown) disturbance. 

The beam splitter that combines interferometer output and the entangled reference beam, unavoidably splits the signal as well as the disturbance into two  paths. For a balanced beam splitter, this generally reduces the signal and disturbance power by 3\,dB for both quadrature measurements.
Fig.~\ref{fig:33} shows, however, that both BHDs performed about 6\,dB below shot noise, which demonstrates the usefulness of the scheme. The squeeze factor can in principle be infinite, which thus qualifies the `3\,dB penalty'. In the above figure the additional information from the second BHD output was used to recognize the parasitic interference in the first BHD output providing a `veto' signal to trigger its removal from the data stream.

The question arose whether the additional information can be used to reduce the actual noise spectral density of the first measurement, i.e.~to re-cover signals that were buried by parasitic interferences. 
Very recently it was shown that this is indeed possible. Ref.~[\cite{AstM2016}] reports a table-top proof-of-principle experiment, in which the additional information in the QDM approach could be used for improving the sensitivity of an interferometer. The measurement sensitivity was improved from above-shot-noise to sub-shot-noise (sub-Poissonian) performance. 
This result was possible not because the way the parasitic interference arose was known, but because the additional information provided by QDM allowed for fitting a model of the excess noise to the readout data. 

\FloatBarrier

\emph{Quantum dense metrology} (QDM) as shown in Fig.~\ref{fig:32} improves a measurement by simultaneously reading out two conjugate observables. Both readout observables show a squeezed quantum noise and act as estimators of independent physical quantities. This situation was recently described as `quantum-mechanics free' [\cite{Tsang2012}]. QDM is based on an Einstein-Podolsky-Rosen (EPR) entangled [\cite{Einstein1935}] bi-partite system as described in Subsec.~\ref{ssec:EPR}. EPR entanglement was previously considered for the quantum-informational task of \emph{dense coding}, which doubles the capacity of quantum communication channels [\cite{Bennett1992,Braunstein2000}]. The application of EPR entanglement in metrology was first proposed by D'Ariano \emph{et al.} [\cite{DAriano2001}].

\subsection{Conclusions}

A single beam that carries an optimized spectrum of squeezed vacuum states and that is injected into the interferometer's dark port provides the most efficient and practically optimal approach to reduce the quantum noise in laser interferometers by means of nonclassical states (see Section\,\ref{Sec:5}).
The conclusion of the section here is that two \emph{entangled} beams provide a superior approach if the interferometer's sensitivity is limited by classical noise that is not exclusively restricted to the actual observable, which is the phase quadrature amplitude $\hat Y$.
Parasitic interferences due to laser light that is backscattered from vibrating surfaces are an example. 
Current gravitational-wave detectors use light fluxes of about $10^{24}$ photons per second [\cite{Abbott2016}].
Just a single photon per second and hertz which leaves the main light beam and is backscattered from a vibrating surface, and in this way gets frequency shifted into the detection band, produces a significant disturbance signal. 
The `quantum-dense metrology' approach might provide a powerful technique to tackle this problem.

Very recently it turned out that QDM is  not the only technique that may exploit EPR entanglement to improve phase measurements.
Ref.~[\cite{Ma2016}] proposes to use EPR entanglement to simultaneously suppress shot noise and radiation pressure noise in a gravitational-wave detector without the need for an additional filter cavity (confer subsection~\ref{Ssec:5.5}). In this case, EPR entanglement is exploited that is carried by one broadband squeezed beam and that is present between quadrature amplitudes defined with respect to different optical frequencies $\omega$ and $\omega'$ as investigated in Ref.~[\cite{Hage2010}]. 
Such `frequency multiplexed' EPR entanglement might result in considerably lower costs of building a gravitational-wave detector with a broadband simultaneous squeezing of shot noise and radiation pressure noise. 
Also this proposal does not lead to a fundamentally lower quantum noise but rather improves on classical aspects of an interferometer.\\


\FloatBarrier
\section{Summary and Outlook} \label{Sec:8} 

In many cases, experiments that involve interference of quantum states can be described in a semi-classical way. This description uses the classical wave picture for the interference part of the experiment and subsequently the classical particle picture when the states transfer their energy to a detector, or more generally, to a thermal bath. 
This semi-classical description is not possible when using the specific class of `nonclassical' states. Squeezed states of light are a prominent example of these. Squeezed states and other nonclassical states allow for observations that made Einstein, Podolsky, and Rosen formulate their critical and seminal paper on quantum theory [\cite{Einstein1935}].

In the review here it is argued that, after many successful proof-of-principle experiments with nonclassical states in the past decades, 
the routine use of squeezed-light in observational runs of the gravitational-wave detector GEO\,600 goes beyond proof-of-principle and is a true application of nonclassical light. Since 2010 the squeezed-light source has improved the measurement sensitivity of GEO\,600 in basically every observational run  [\cite{LSC2011,Grote2013,Dooley2016}]. At quantum noise limited frequencies, i.e.~above a few hundreds of hertz, the sensitivity has been improved corresponding to a squeezing strength in the noise spectral density
of up to 3.7\,dB, which corresponds to an increase of the average gravitational-wave detection rate by a factor of $0.43^{-3/2} = 3.6$.
This success is a strong motivation to also equip the Advanced LIGO, Virgo and Kagra gravitational-wave detectors with squeezed light. Similar improvement factors, even down to lower signal frequencies are expected [\cite{LSC2013}]. The achievable improvement factors are mainly limited by the optical loss on the squeezed states, and much higher factors are achievable in principle.

Up to now, squeezed states have not been used to reduce the radiation-pressure noise in gravitational-wave detectors. The reason is that so far other noise sources are larger than radiation pressure noise and such an effect cannot be observed. It is expected, however, that  future gravitational-wave detectors will eventually be partly limited by radiation pressure noise. From this point on, squeezed light will be used to simultaneously reduce shot noise and radiation pressure noise. \\
Squeezed states are the optimum nonclassical states for gravitational-wave detectors, or more generally for all laser interferometers operating with large average photon numbers per measuring interval [\cite{Demkowicz-Dobrzanski2013}].  
In addition to using higher light powers and heavier test mass mirrors, higher squeeze factors will thus contribute to mitigate the light's quantum noise in laser interferometers.

Two-mode (bi-partite) squeezed light has not been used in gravitational-wave detectors so far. They are not capable of further reducing the quantum noise in laser interferometers, but they can be used to mitigate classical noise that originates from fluctuating phase space displacements. A well-known such noise source is back-scattered light. Proof-of-principle experiments were performed recently [\cite{Steinlechner2013b,AstM2016}]. This new technique could turn out to be valuable in next generations of gravitational-wave detectors, in particular in those targeting high sensitivities at low, sub-audio signal frequencies and using high light powers. Such an implementation in gravitational-wave detectors does not require any new technology. Compared to a squeezed-light enhanced interferometer, just a second squeezed-light source is required. 

It is certainly remarkable that those quantum states that made Einstein, Podolsky and Rosen falsely think quantum theory incomplete, are now exploited as new technologies in measurement devices targeting new observations in nature.

~\\
{\bf Acknowledgements}\\[4mm] 
R.S. thanks \, M. Ast, J. Bauchrowitz, C. Baune, S. Chelkowski, J. DiGugliel\-mo, A. Franzen, B. Hage, J. Harms, A. Khalaidovski, L. Kleybolte, N. Lastzka, M. Mehmet, S. Steinlechner, and H. Vahlbruch  for their contributions, many fruitful discussions and their support with the figures, and J. Fiur\'{a}\v{s}ek for many valuable comments on the manuscript. Thanks are also due to Y. Chen, F. Khalili, and H. Miao for fruitful discussions within the quantum noise working group of the LIGO Scientific Collaboration (LSC). Special thanks are due to H. Vahlbruch and H. Grote together with the GEO\,600 team for their pioneering work on the squeezed-light implementation in GEO\,600.
R.S. is supported by the Deutsche Forschungsgemeinschaft (Grant No. SCHN 757-6) and by the European Research Council (ERC) project `MassQ' (Grant No. 339897).




~\\
\section*{References}

\bibliographystyle{elsarticle-harv} 
\bibliography{RS-library}

%
%
%
\end{document}